\documentclass[showpacs,a4paper,aps]{revtex4}

\usepackage{amsmath}
\usepackage{amssymb}
\usepackage{graphicx}
\usepackage{color}

\DeclareGraphicsExtensions{.eps}

\begin{document}

\title{Statistical distributions in the folding of elastic structures}
\author{Mokhtar Adda-Bedia$^1$, Arezki Boudaoud$^{1,2}$, Laurent Bou\'e$^{1,3}$, and St\'ephanie Deb\oe{}uf$^{1,4}$}
\affiliation{$^1$Laboratoire de Physique Statistique, Ecole Normale Sup\'erieure, UPMC Paris 06, Universit\'e Paris Diderot, CNRS, 24 rue Lhomond, 75005 Paris, France\\
$^2$RDP, ENS Lyon, 46 all\'ee d'Italie, 69007 Lyon, France\\
$^3$Department of Chemical Physics, The Weizmann Institute of Science, Rehovot 76100, Israel\\
$^4$Department of Physics, McGill University, 3600 University, Montr\'eal (Québec) H3A 2T8, Canada}

\date{\today}

\begin{abstract}
The behaviour of elastic structures undergoing large deformations is the result of the competition between confining conditions, self-avoidance and elasticity.  This combination of multiple phenomena creates a geometrical frustration that leads to complex fold patterns.  By studying the case of a rod confined isotropically into a disk, we show that the emergence of the complexity is associated with a well defined underlying statistical measure that determines the energy distribution of sub-elements,``branches'', of the rod.   This result suggests that branches act as the ``microscopic'' degrees of freedom laying the foundations for a statistical mechanical theory of this athermal and amorphous system.
\end{abstract}


\maketitle

\section{Introduction} 

Classical equilibrium statistical mechanics stands as a major cornerstone of modern physics. Tools issued from this theory have been instrumental in rationalizing a huge number of seemingly unrelated physical situations ranging from phase transitions in atomistic systems to the behaviour of polymers.  This is possible because the size of the fundamental components of those systems is sufficiently small so that thermal fluctuations allow the degrees of freedom to span all the possible configurations through an ergodic exploration of the energy landscape. For macroscopic systems thermal agitation becomes negligible and, while those systems may be mechanically stable, they fall out of equilibrium with respect to thermodynamics.  A usual example where gravitational energy completely dominates thermal effects concerns the physics and mechanics of granular assemblies~\cite{jaeger96}. Because of this mismatch of energy scales and of the presence of dissipative interactions between the grains, it had long been believed that those systems could never be reconciliated with the basic principles underlying classical statistical physics.  However, it has been recognized in the last decade that granular materials do in fact share many properties with thermal, although very slowly evolving, systems such as molecular glasses~\cite{bouchaud98,debenedetti01}. Indeed, analogies between the slow relaxation of granular materials and the glassy phenomenology of supercooled liquids has led to the now famous ``Jamming diagram'' proposed by Liu and Nagel~\cite{liu98}.  While this analogy is not free from criticism, it does open up the possibility that macroscopic systems could still be described using a combination of the tools of statistical physics and some new out-of-equilibrium concepts. In fact it has recently been proposed that the rheology of athermal amorphous systems is characterized by well defined statistical distributions that are amenable to a probabilistic treatment similar to that of classical statistical mechanics~\cite{edwards89,barrat00,makse02,dean03,bertin06,aste07}. An important new concept often encountered in those descriptions is the presence of an ``effective'' temperature that is different from the bath temperature and that depends on the driving mechanism or on the details of the coupling between the thermostat and the system~\cite{cugliandolo97a,ono02,srebro04,boue10}. 

Here we are interested in another kind of athermal macroscopic systems: elastic structures.  Due to the geometrical frustration created by the interaction between confinement, self-avoidance and elasticity, it is expected that confined elastic structures should display a complex behaviour. Indeed crumpling a sheet of paper is a typical everyday illustration of the phenomena that we want to study. A rapid visual inspection of the numerous valleys and mountains present after unfolding a piece of crumpled paper often leaves one with the impression of fascinating albeit extremely complex fold patterns~\cite{witten07}. Accordingly, the folding phenomena are associated with a rich class of crumpling phenomena which belong to a wider class of interfacial deformation phenomena. Both deterministic and random folding of thin materials are of noteworthy importance to many branches of science and industry. Examples range from DNA packaging inside virus capsids~\cite{purohit05} and polymerized membranes at the microscale~\cite{bourdieu,saintjalmes} to folded engineering materials and geological formations at the macroscale, including insect wings~\cite{brackenbury94} or leaves in buds~\cite{kobayashi98,couturier09}.  They usually consist of thin sheets or rods constrained to undergo large deformations. Because of their biological and technological importance, the properties of randomly folded thin materials are now the subject of increasingly growing attention.  Because self-avoidance and nonlinear deformations make the description of fully crumpled or folded materials very difficult, earlier studies focused on the identification of the elementary generic features displayed by an elastic surface that has to accommodate a geometrical mismatch reducing its accessible volume. It was found that those individual structures consist of sharp vertices (or developable cones)~\cite{benamar97,chaieb98,cerda98,cerda99,liang05} and linear ridges~\cite{witten,lobkovsky95,lobkovsky96,lobkovsky97,didonna02} and their properties are now rather well established.  These singularities are conceptually similar to the dislocations, cracks, necks or shear zones that are created when a continuum media is forced at the large scales and localizes the stress in those tiny regions that dissipate the energy at the smaller length scales.  In general, these singularities can either be moving or be quenched depending on their type and on the mechanical properties of the material~\cite{boudaoud00,mora}.  Their interactions are usually carried through a long ranged elastic field either instantaneously or with retardation via wave propagation.  In the case of the packing of flexible structures, almost nothing is known about how these interactions lead to the complex fold patterns that are observed at the surface of a piece of crumpled paper.  This situation is similar to the one encountered in statistical physics where, given the knowledge of the microscopic interactions, one tries to bridge the gap and rationalize the macroscopic behaviours~\cite{katzav06,boue07}.

The question that motivated the study presented here is: {\it ``Are there any general statistical properties associated with the crumpling and folding phenomenology that is observed when an elastic structure is confined in an environment smaller than its rest size?''}  So far there has been some degree of incompatibility between crumpling experiments and numerical simulations.  For one thing, most simulated sheets have been fully elastic~\cite{kramer97,sultan06,vliegenthart06,lin08b} whereas the ones used in experiments have been made of elasto-plastic materials such as Mylar, paper, aluminium foil and even layers of cream~\cite{gomes90,plouraboue,matan02,blair05,balankin06,andresen07,balankin07a,balankin07b,gomes08,lin08,balankin08,lin09,lin09b}.
It is not clear to what extent these analyses could make the difficult distinction between elastic and elasto-plastic behaviours~\cite{tallinen09}.  On the other hand, the packing of elastic objects clearly depends on the dimensionality of the object ($d$) and of the container ($D>d$). The configurational properties are totally different if the dimension of the container satisfies $D\ge d+2$ from the case $D=d+1$ due to the rigid constraint of the inter-penetrability.

Here we provide a case study in order to answer our original question.  We are studying the statistical properties of an elastic rod ($d=1$) that is confined in a disk ($D=2$)~\cite{donato03,donato07,boue06,stoop08,lin08b,deboeuf09}.
In this geometry, there are no singularities and the most important constraint is of geometrical origin.  Using a comparison between two widely different implementations of rod compaction we suggest that there exists an underlying statistical measure that describes the folding process.  After describing the systems that we studied (sheet pulling experiment in section~\ref{sec_exp} and minimal numerical simulations in section~\ref{sec_num}) and pointing out their different confining and geometrical conditions, we show a compared statistical analysis of several quantities.  Similarities between stochastic observables can be detected by comparing their probability density function (pdf) and this is the method that we employed here.  The particular pdf's that we observed during this study are presented and put in a more general context in section~\ref{sec_stat}.  Our main result is that while the topological (section~\ref{sec_topo}) and the geometrical (section~\ref{sec_geometro}) properties are indeed different, the energy of some sub-parts of the rod (referred to as ``branches'' afterwards) display identical statistical properties (section~\ref{sec_energo}).  The energy of the branches turns out to be distributed according to a Gamma law that reduces to a Boltzmann distribution for high energies.  In section~\ref{sec_discu}, we argue that branches correspond to the relevant ``microscopic'' degrees of freedom that define the statistical mechanics of folding.  Finally we propose that the spiral configurations that define the ground state of the system may be viewed as a condensed phase with all the branches lying on top of each other.

\section{Statistical distributions}
\label{sec_stat}

Since we will be extensively using probability density functions (pdf) in the remaining part of the article, it is worth reminding the reader of the expressions of the pdf's that will come up later.  In order to be general, we denote by~$x$ the random variable in this section but that will of course be replaced by some physical observables such as curvature or energy when we come to the description of our results.  In addition to defining our notation for the parameters of the pdf's, this short list also allows us to summarize results from other related studies.
\begin{itemize}
\item Exponential distribution:
\begin{eqnarray}
\rho_E^{\,\mu}(x) = \frac{1}{\mu}\exp{\left(-\frac{x}{\mu}\right)}   \mathrm{ ;}  \label{eq_exp} 
\end{eqnarray}
where $\mu$ is the mean of $x$.
Such pdf's have been found before in measurements of local 3d curvatures~\cite{blair05} and number of intersecting ridges~\cite{andresen07} in unfolded sheets of paper.  They also appeared to describe the lengths of folds in a 3d crumpled sheet~\cite{vliegenthart06}.   Exponential distributions are usually associated with the presence of uncorrelated events that are distributed randomly.
\item Log-normal distribution: 
\begin{eqnarray}
\rho_\mathrm{LN}^{\,\mu,\sigma}(x) =  \frac{1}{\sqrt{2\pi}\,\sigma x}\exp\left(-\frac{(\ln(x)-\mu)^2}{2\sigma^2}\right)  \mathrm{ ;} \label{eq_logn} 
\end{eqnarray}
where $\mu$ and $\sigma^2$ are respectively the mean and the variance of $\ln x$.
Such pdf's have been found to describe the length of plastic linelike ridges in experiments~\cite{blair05,andresen07} as well as in numerical simulations of crumpled sheets~\cite{sultan06,vliegenthart06}.  Log-normal distributions are usually associated with random events that are occurring hierarchically.
\item Gamma distribution: 
\begin{eqnarray}
\rho_G^{\alpha,\chi}(x) = \frac{x^{\alpha-1}}{\Gamma(\alpha)\chi^{-\alpha}} \exp{\left(-\frac{x}{\chi}\right)}    \mathrm{ ;} \label{eq_gam}
\end{eqnarray}
where $\chi$ is the mean of $x$ and $\alpha$ is the exponent of the Gamma distribution. Such pdf's are known to describe the lengths of bent segments of highly confined sheets~\cite{sultan06}.  Gamma distributions are usually associated with the presence of correlations in otherwise randomly distributed events.  When~$\alpha < 1$, there is a power-law divergence for small values of~$x$. Notice that for large values of~$x$, a Gamma distribution reduces to a simple exponential distribution.
\end{itemize}

We will keep the same notation for all the parameters throughout the rest of the article.  Also, we will not re-write the expressions for the pdf's and refer the reader to this section whenever one of these pdf's is encountered in the following sections.  The goal of our study is to identify what is the relevant observable~$x$ leading to a universal pdf in the context of confined elastic rods.  The robustness of the pdf's is tested by comparing between experimental work (set-up described in section~\ref{sec_exp}) and ``model'' numerical simulations (algorithm described in section~\ref{sec_num}).

\section{Sheet pulling experiments}
\label{sec_exp}

\begin{figure*}  
\includegraphics[width=0.3\linewidth]{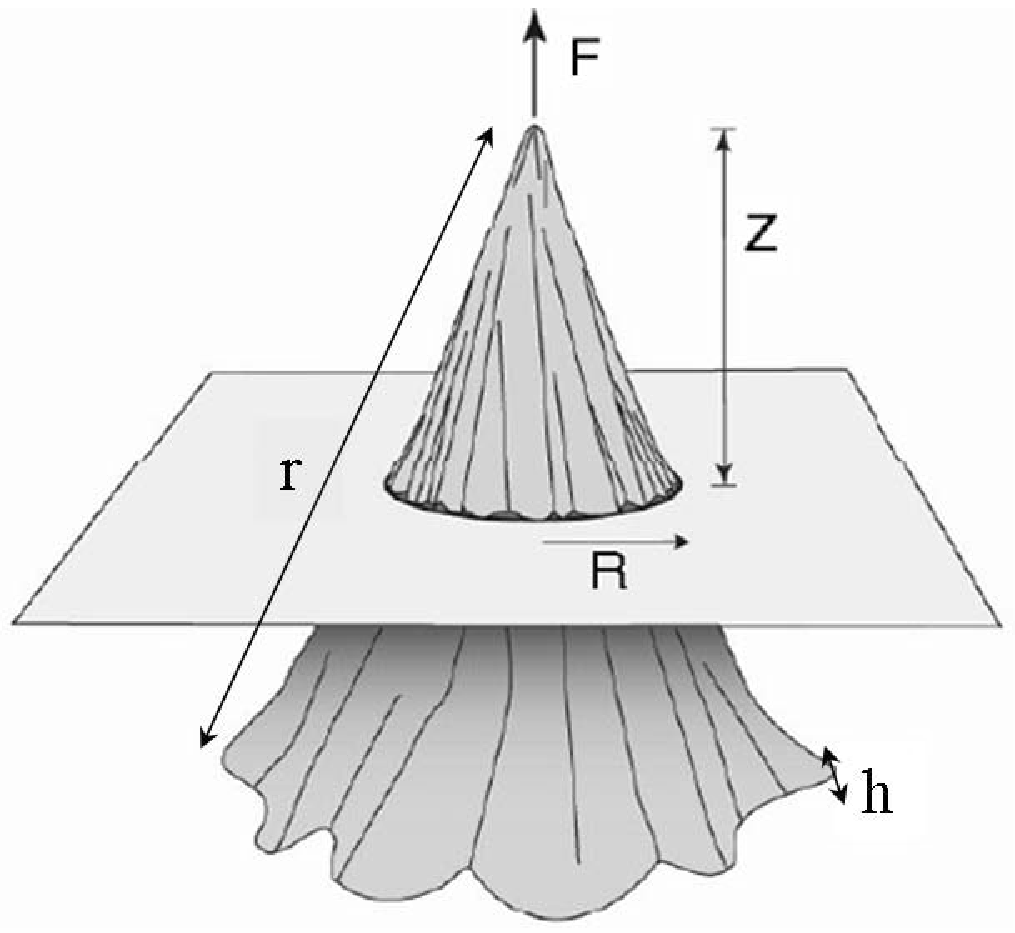} \hfill
\begin{minipage}[b]{0.7\linewidth}
\includegraphics[width=0.24\linewidth]{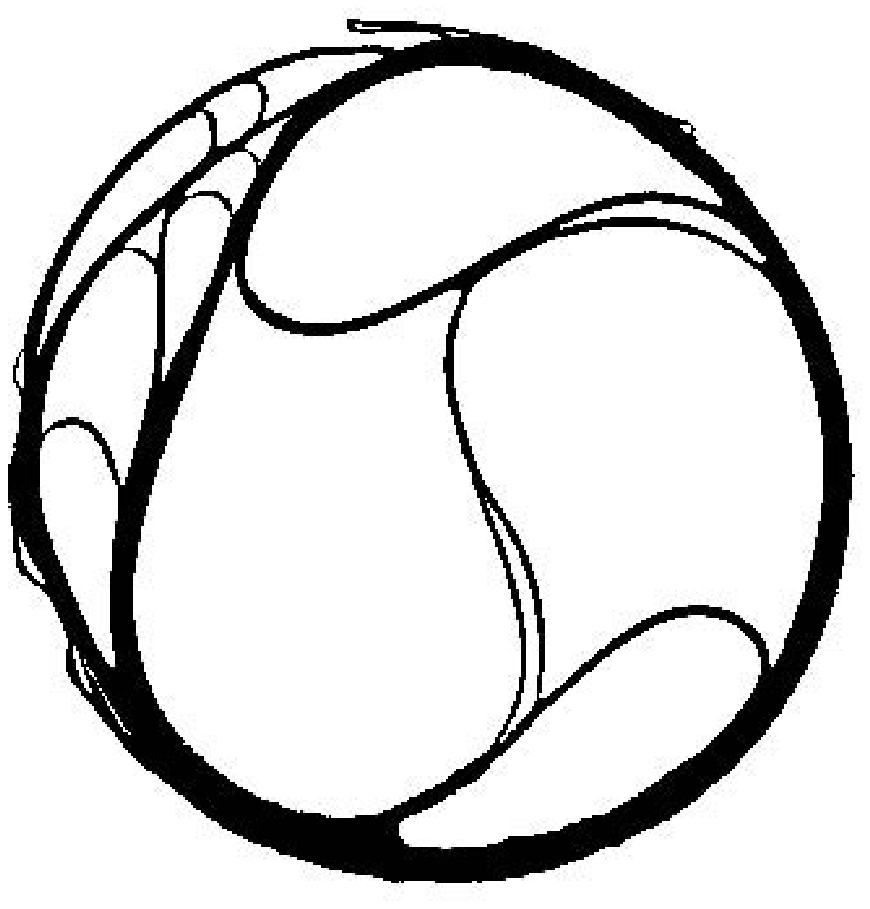}  \hfill 
\includegraphics[width=0.24\linewidth]{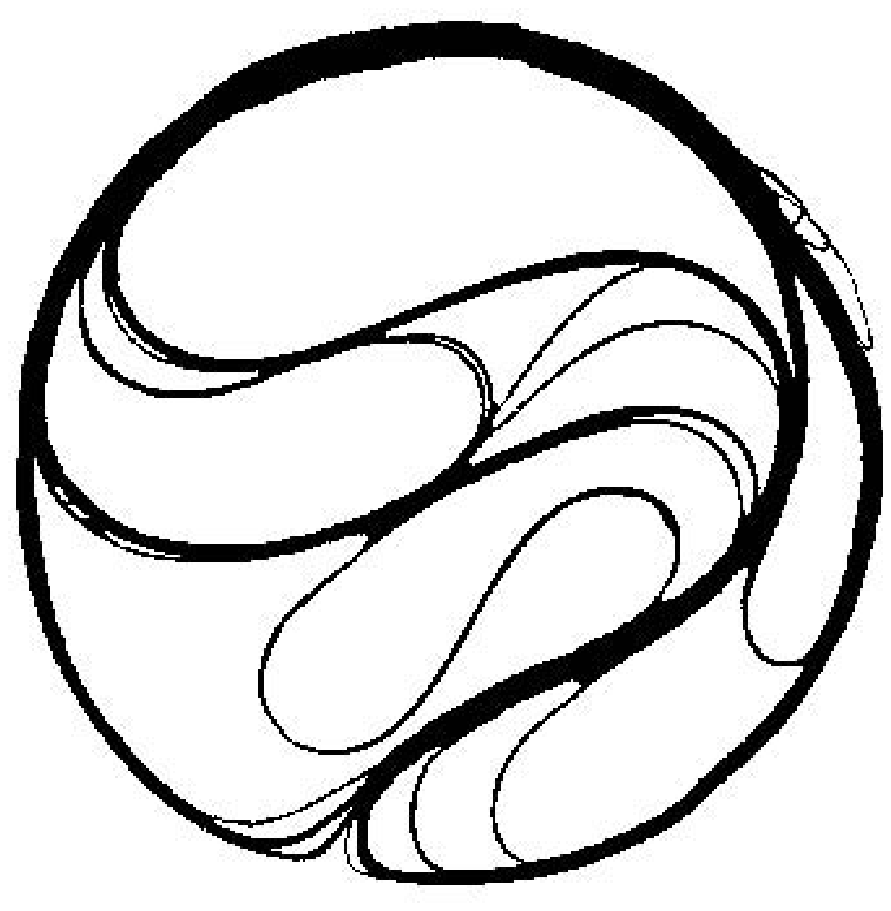} \hfill
\includegraphics[width=0.24\linewidth]{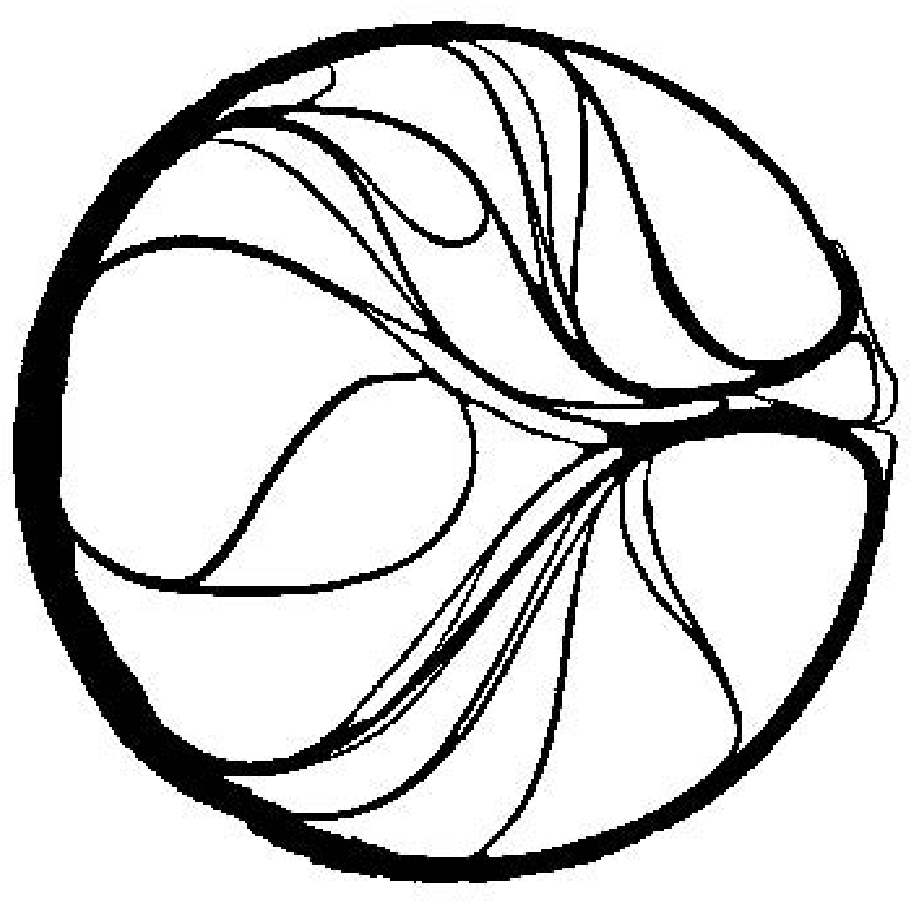} \hfill
\includegraphics[width=0.24\linewidth]{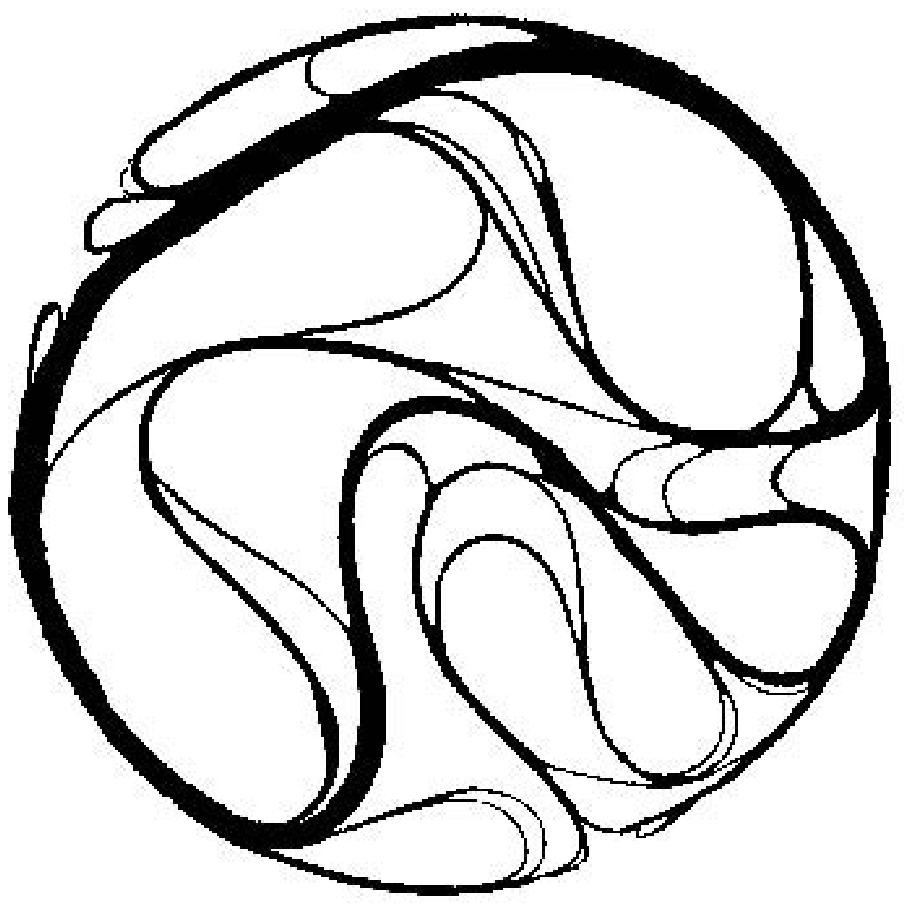} \\
\includegraphics[width=0.24\linewidth]{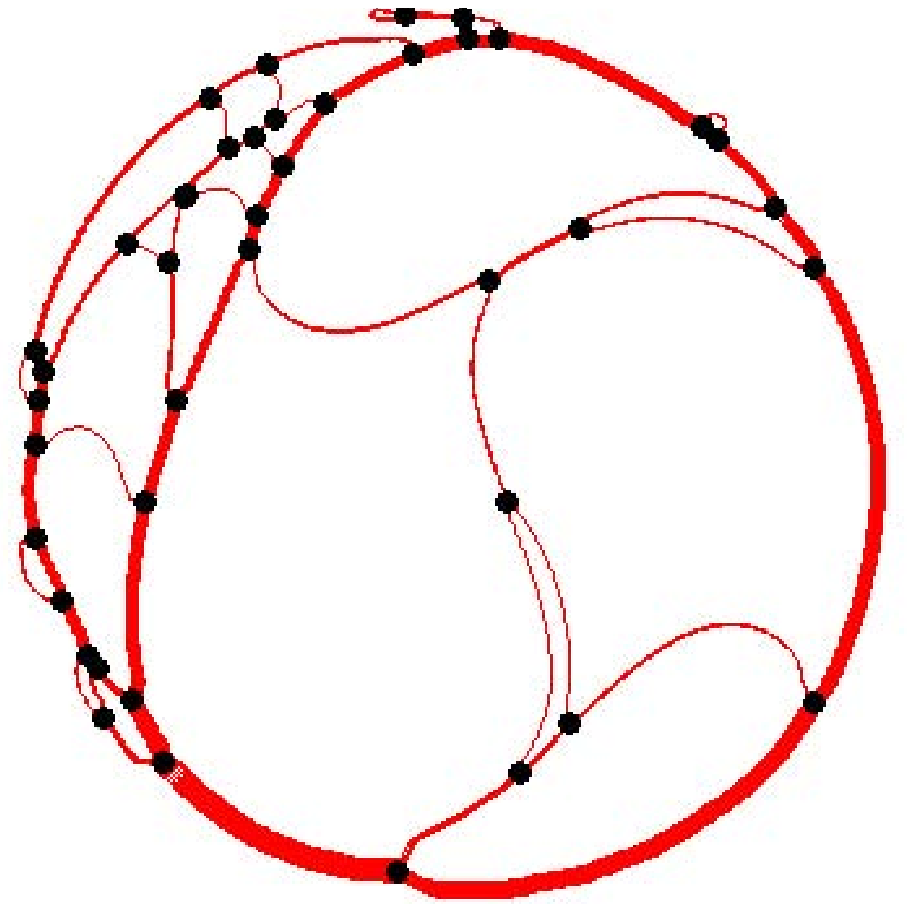}  \hfill 
\includegraphics[width=0.24\linewidth]{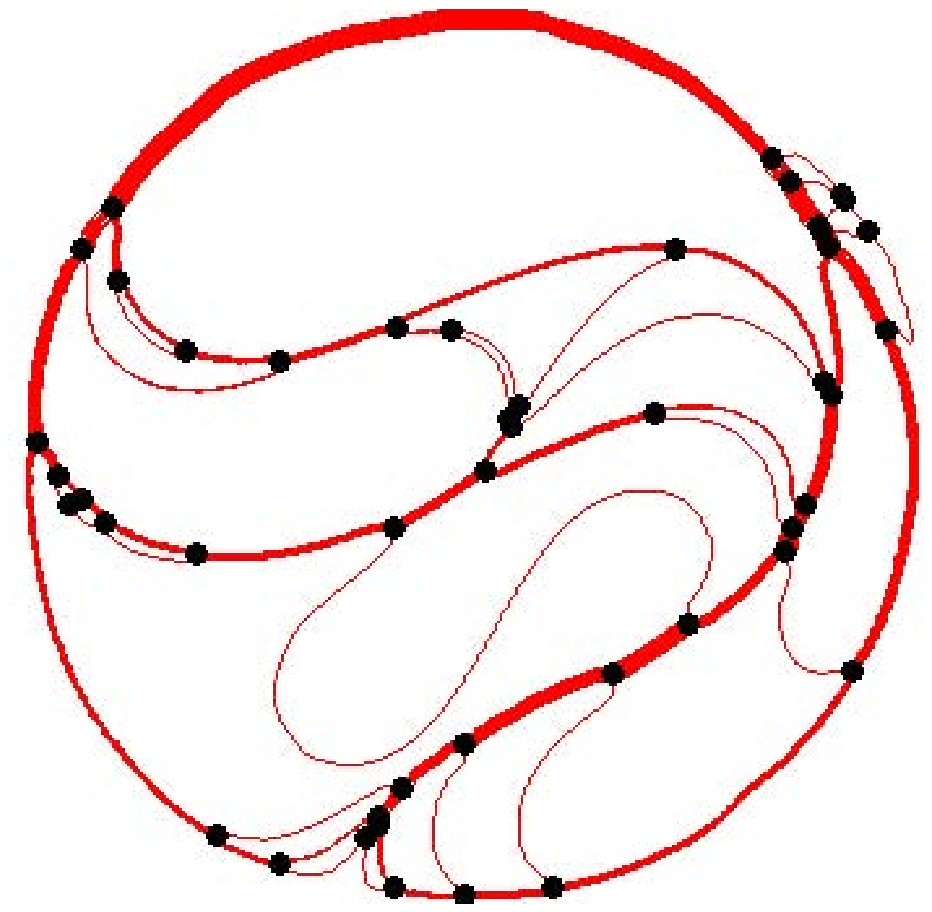} \hfill
\includegraphics[width=0.24\linewidth]{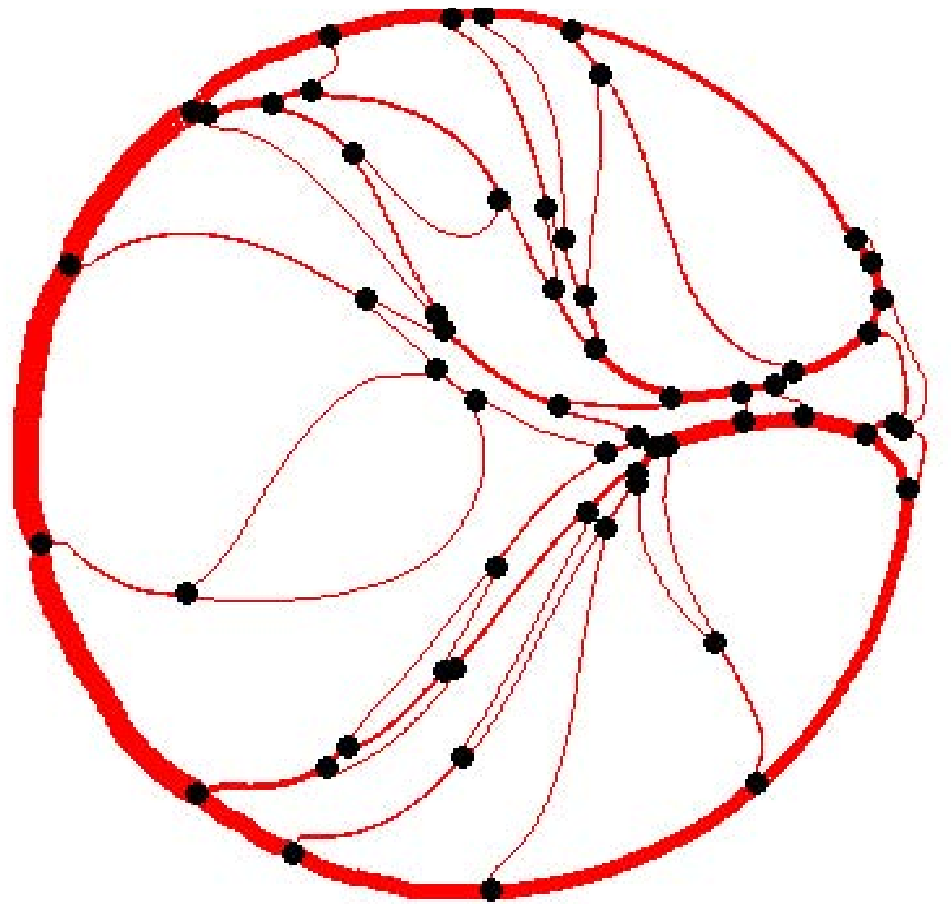} \hfill
\includegraphics[width=0.24\linewidth]{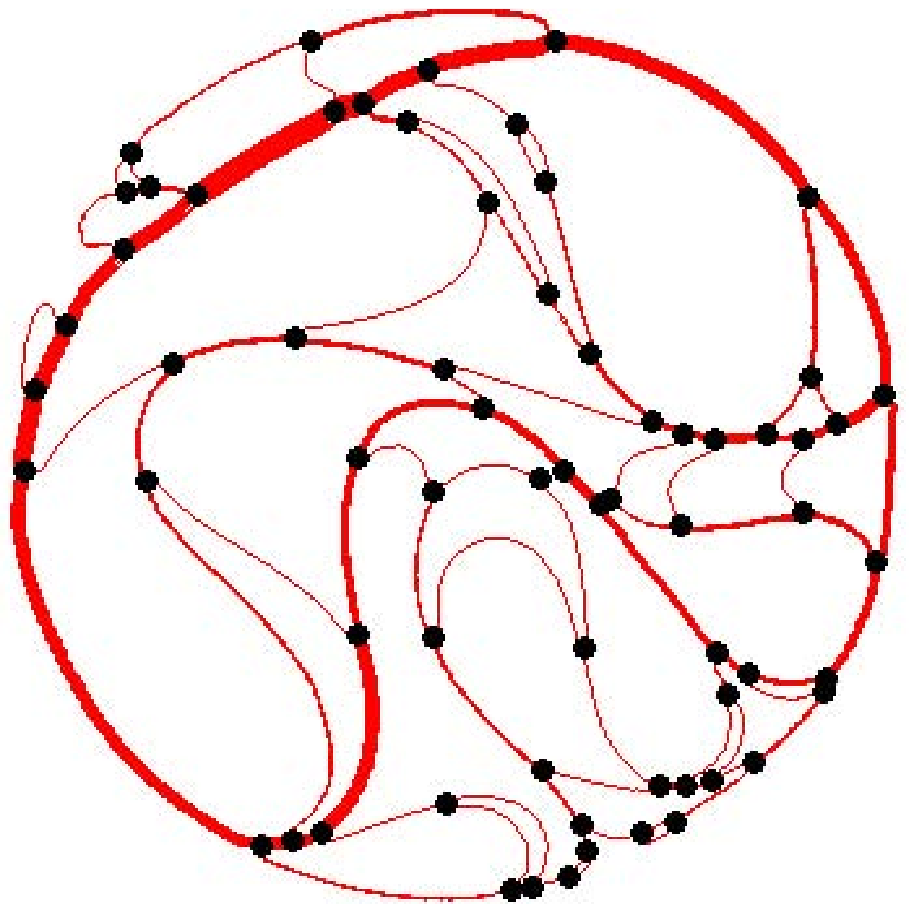} 
\end{minipage}
\caption{{\it {\bf a)}~Schematic representation of the sheet pulling experiment, showing the radius of the sheet~$r$, its thickness~$h$, the radius of the hole~$R$ and the control parameter~$Z$. {\bf b)}~Typical configurations observed after a cross-sectional cut right below the hole.  Note that these very different shapes were obtained by following the same protocol (compaction rate~$\varepsilon = 19$). {\bf c)}~Corresponding configurations after the image analysis procedure (see subsection~\ref{subsec_analim}).  Black circles represent the Y-shaped junction points   where multi-branched stacks converge (merge or separate). Branches in self-contact within a stack make the thickness of these stacks proportional to the number of individual branches they carry.}}
\label{fig_setexp}
\end{figure*}

\subsection{Experimental set-up}

The experiment has been already described in~\cite{deboeuf09,boue06}.  A circular polyester sheet of radius~$r \sim 30$~cm and thickness~$h \sim 0.1$~mm is pulled by its center through a smaller circular rigid hole of radius~$R \sim 2$~cm, as illustrated in Fig.~\ref{fig_setexp}.  The Young's modulus of the sheets has been measured as $E=5$~GPa, its density is $1.4$~g/cm$^3$. The bending rigidity has been measured as $B=7\,10^{-5}$~J for sets of experiments $1$, $2$, $4$ and  $B=1\,10^{-3}$~J for set $3$.  In this specific set-up, the sheet undergoes preferentially bending deformation rather than stretching, in order to minimize its elastic energy.  Thus a  self-affine conical shape is expected and observed, so that the whole 3d shape of the sheet is prescribed by  the shape of one cross-section.  Any cross-section at a distance $z$ from the cone tip, draws a virtual rod of thickness $h$ and length $L\simeq2\pi z$, compacted in a circle of surface $S=\pi (Rz/Z)^2$, $Z$~being the pulling distance between the cone tip and the plane hole (Fig.~\ref{fig_setexp}). In this article, we are specifically interested in the compaction of such a virtual rod, rather than the compaction of the whole sheet.  Lengths, surfaces and volumes measured in a cross-section obviously depend on its position $z$, but become independent on $z$, when  their units are non-dimensionalized by using the total length  of the rod $L$.  In practice, the cross-section is observed in the hole plane $z=Z$, and only once during the pulling experiment for $Z=Z_m\sim$~30~cm (see Table~\ref{tab_paramexp}). For each  experiment, a rod of non-dimensionalized length~$1$ is compacted in a circle of non-dimensionalized surface $S/L^2$:  
 \begin{eqnarray}
 S/L^2=R^2/4/Z_m^2=\varepsilon^2/4 \mathrm{ ,}
\end{eqnarray}
defining the compaction rate $\varepsilon$, the ratio of rod length over hole perimeter: 
\begin{eqnarray}
\varepsilon=L/(2\pi R) \mathrm{ .} \label{eq_epsexp}
\end{eqnarray}     
A second parameter characterizing the compaction rate, but in terms of higher dimensionality, is defined as the ratio of the sheet volume over its conical envelope~:
\begin{eqnarray}
C=3h\varepsilon^2/r \mathrm{ .} 
\end{eqnarray}
Note that the sheet dimensions ($h$, $r$) appear in the expression of the 3d-compaction rate~$C$, but not of the 2d-compaction rate~$\varepsilon$. In this article, the experiments are typically characterized by compaction rates~$\varepsilon$ of the order of $10$ and $C$ of $10\%$.  The values of experimental parameters $L$, $h$, $r$, $R$, $Z_m$, $\varepsilon$, $C$, $E$ and $B$ are reported in Table~\ref{tab_paramexp}.

The compaction mechanism in the present experiment is geometrically controlled by imposing the total available surface $S$ for a rod of length $L$. The use of a rigid hole constraint can be viewed as a hard-wall repulsion potential $V(r)$ acting on the rod such that $V(r)=0$ for $r<R$ and $V(r)=\infty$ elsewhere.
For fixed control parameters ($E$, $h$, $\varepsilon$ and $C$), several pulling experiments have been performed in order to obtain various configurations corresponding to different energy minima and to analyse them in a statistical way.  We realized four sets of experiments at constant control parameters (Table~\ref{tab_paramexp}). Some typical examples of folded rod obtained are shown in Fig.~\ref{fig_setexp}.

\begin{table*}  
\begin{center}  
\begin{tabular}{l||l||l|l|l|l||l|l||l}
& $L (m)$& $h/L$& $r/L$ & $R/L$& $Z_m/L$& $\varepsilon$& $C$&  $B (J)$  \\ \hline
$1$& 1.9& 2.6\,10$^{-5}$& 8.9\,10$^{-3}$& 0.18& 0.16& 19& 15$\%$& 7\,10$^{-5}$ \\  
$2$& 1.9& 2.6\,10$^{-5}$& 1.2\,10$^{-2}$& 0.17& 0.16& 14& 8$\%$& 7\,10$^{-5}$ \\  
$3$& 1.2& 2.1\,10$^{-4}$& 2.2\,10$^{-2}$& 0.18& 0.16& 8& 9$\%$ & 1\,10$^{-3}$ \\    
$4$& 1.0& 5.0\,10$^{-5}$& 1.6\,10$^{-2}$& 0.33& 0.16& 10& 4$\%$ & 7\,10$^{-5}$    
\end{tabular}  
\end{center}  
\caption{{\it {\bf Experimental parameters:} dimensions, compaction rates and material properties of the polyester sheets.}}
\label{tab_paramexp} 
\end{table*}  

\begin{figure*}  
\includegraphics[width=.18\linewidth]{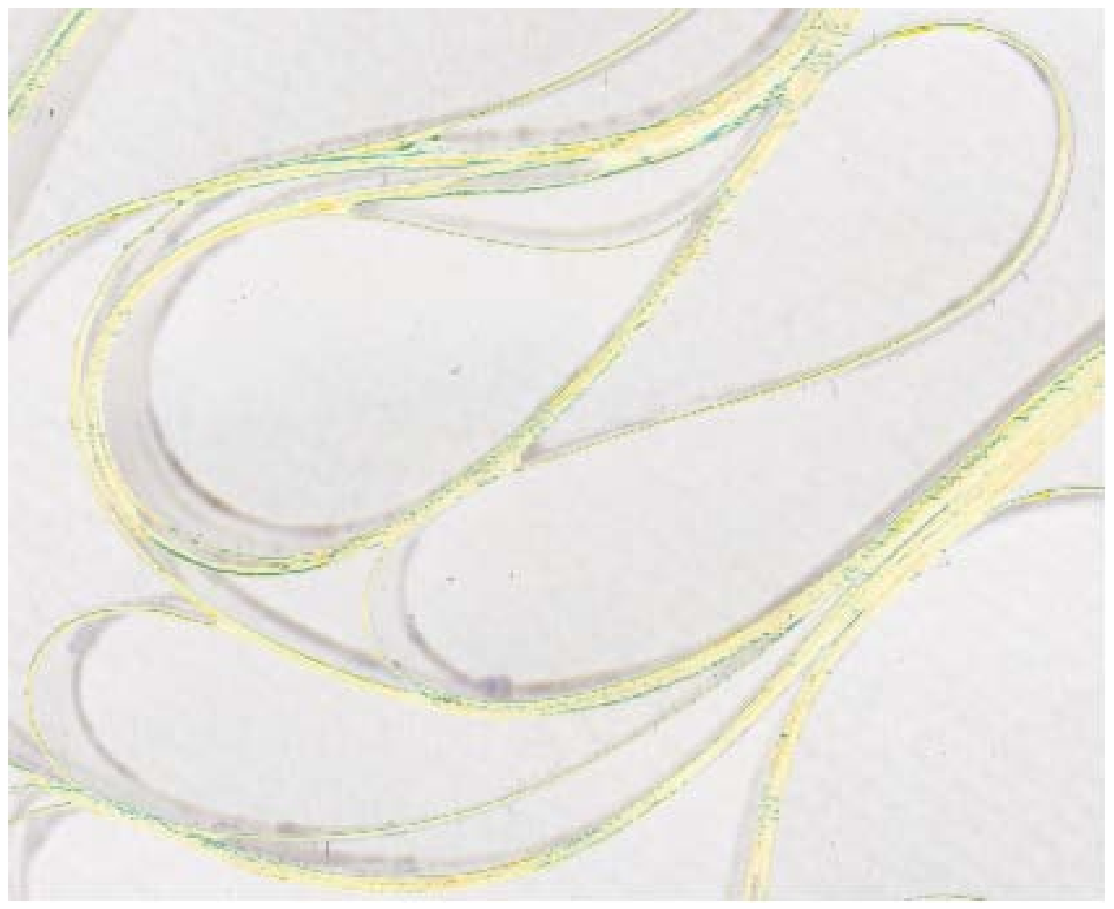} \hfill
\includegraphics[width=.18\linewidth]{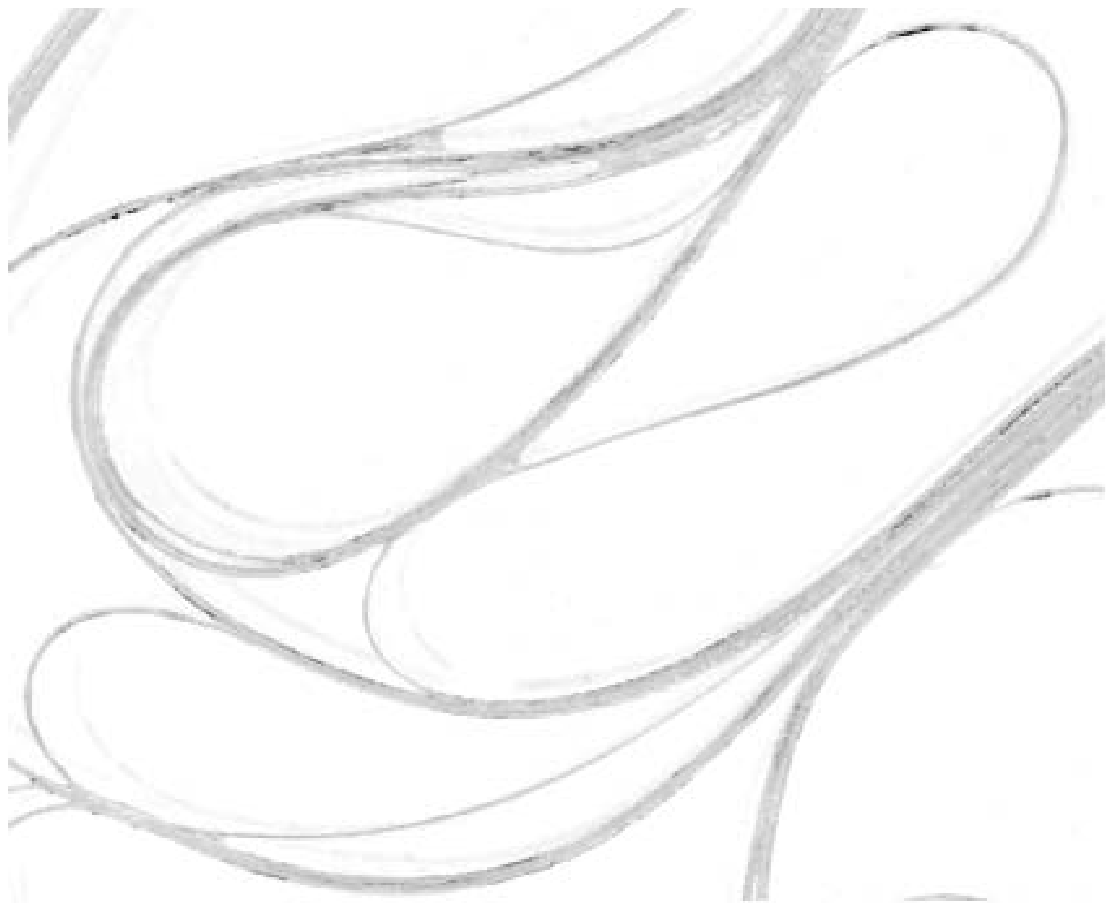} \hfill
\includegraphics[width=.18\linewidth]{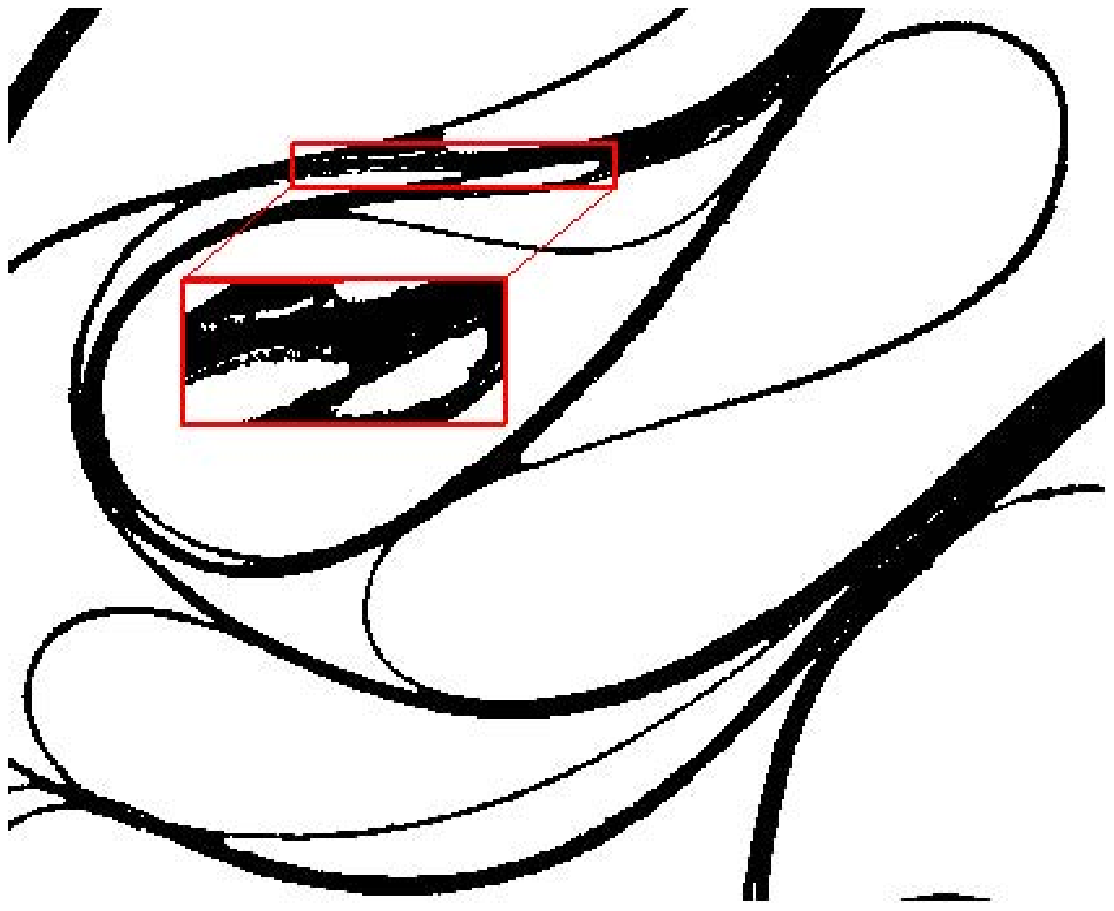} \hfill
\includegraphics[width=.18\linewidth]{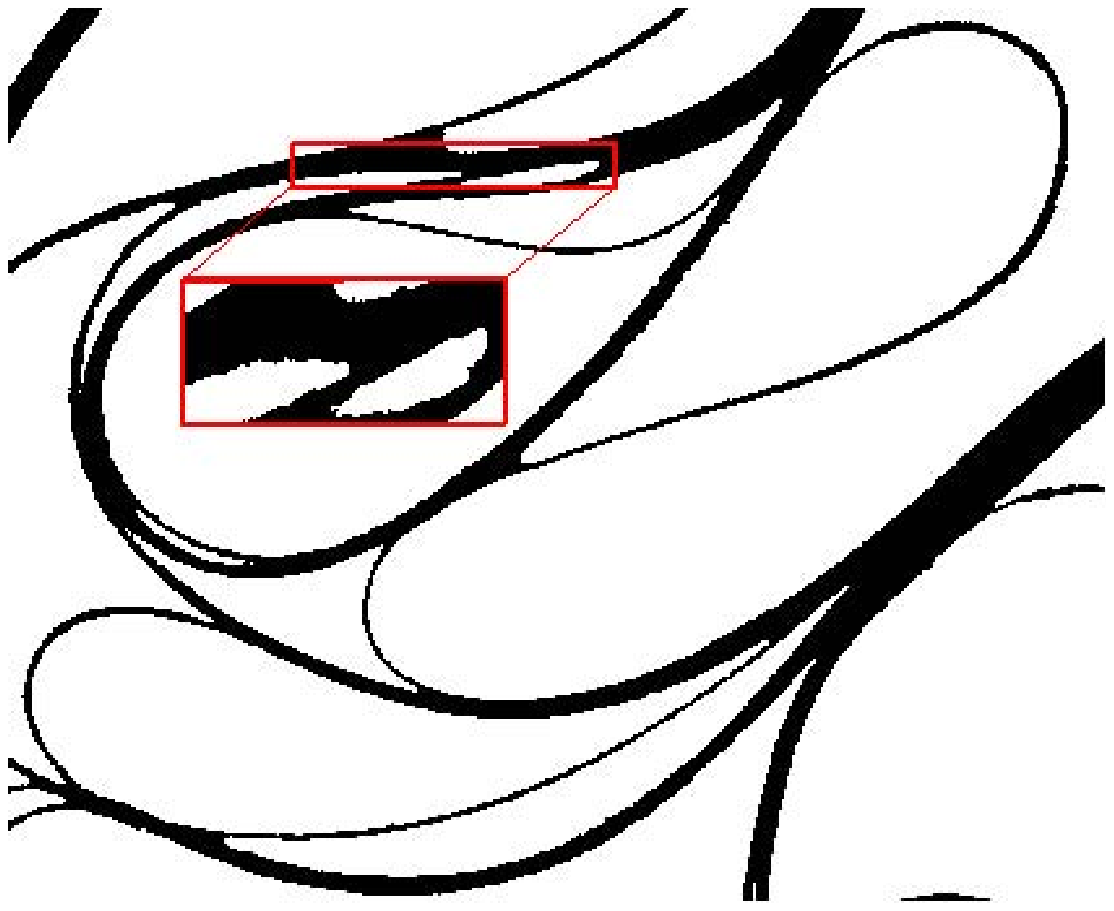} \hfill
\includegraphics[width=.18\linewidth]{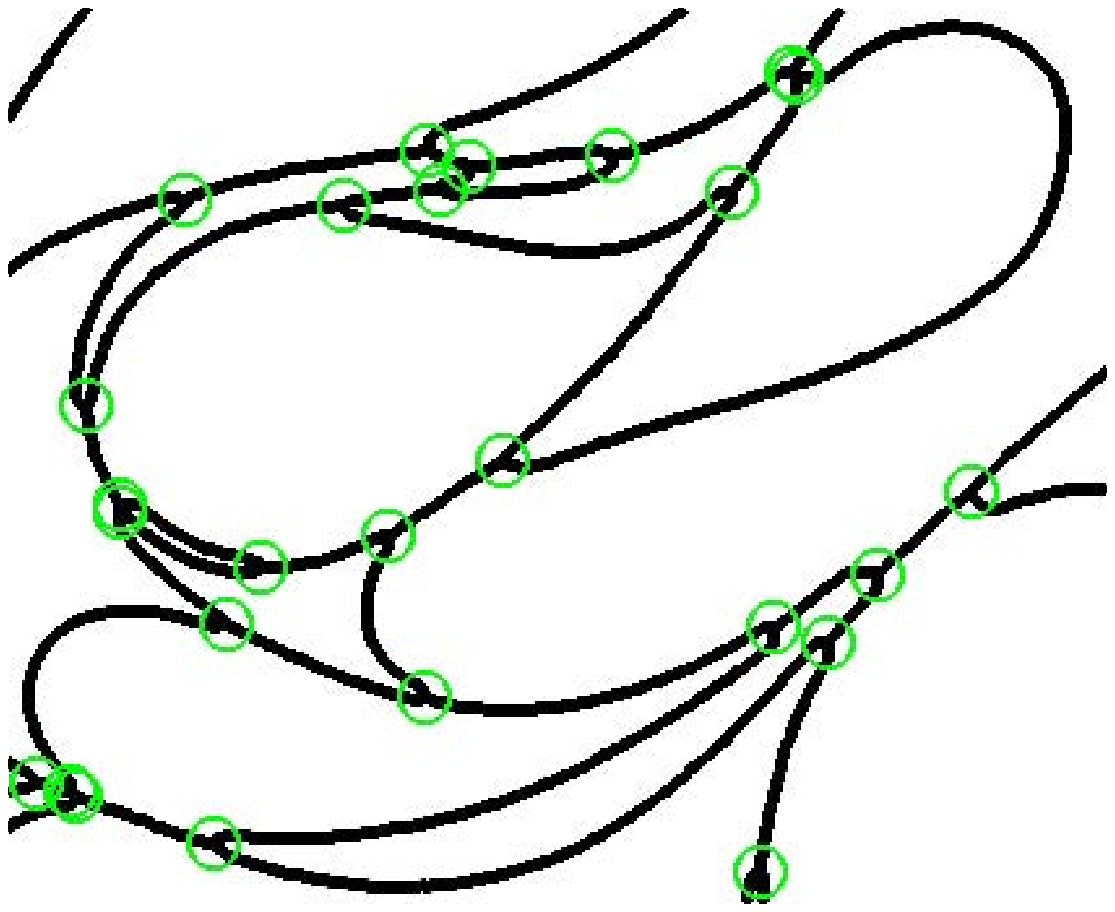}   
\caption{{\it Zoom on a typical configuration detailing  the successive steps of image processing, from left to right. Raw image. Pixels of blue values larger than a given threshold but of red and green values smaller than another given threshold are kept. Thresholded image. Empty spaces of area larger than $s_{min}$ are kept.  Skeleton of the binary image and junction points.}}
\label{fig_analim}
\end{figure*}

\subsection{Image analysis procedure}
\label{subsec_analim}

Because of technical difficulties \cite{deboeuf09}, we resorted to a hot wire cutting tool to obtain cross-sections in the hole plane.  With great care, one obtains neat cuts without perturbing the configuration, and inks them in blue to reinforce contrast of the cut edge with back receding surfaces. The cross-section is digitized with a scanner at a resolution of 50 pixels/mm, which yields 5 pixels for the thickness of the thinner sheets (Fig.~\ref{fig_analim}).  A thresholding based on RGB values results in a binary image: only pixels of blue values larger than a given threshold but of red and green values smaller than another given threshold are kept, allowing to distinguish the cut edge from back receding surfaces and to remove noise from the raw image  (Fig.~\ref{fig_analim}). Then, empty spaces of surface area larger than $s_{min}=(10h)^2$ are kept, allowing to remove light noise from the binary image (Fig.~\ref{fig_analim}). Finally, the binary image is skeletonized --reduced to a one pixel thick skeleton, without redundant kinks-- (Fig.~\ref{fig_analim}). Junction points are then defined as pixels with at least $3$ neighbours (Fig.~\ref{fig_setexp} and Fig.~\ref{fig_analim}). Two neighbouring junction points delimit a stack of branches in close contact, separated by distances smaller than the image resolution.

The next step is to determine the number of branches $n_{br}$ in each of the $M$ stacks. The conservation of the number of branches at each junction point yields $2M/3$ equations, because $3$ stacks intersect at each junction point. This  requires to distinguish converging stacks in 2 opposite packs: merging and separating stacks.  To this aim, the vectors tangent to the stacks locally to the junction point are determined. Then, the straight line perpendicular to one of these vectors is taken as a reference axis, that delimits two semi-planes:  stacks located in the same (resp. opposite) semi-plane are in the same (resp. opposite) pack. The result is systematically checked to be independent on the choice of the reference axis. The remaining $M/3$ equations are found from the thickness of the stacks in the binary image as follows. The heating by the cutting tool thickens a stack nonlinearly, which was calibrated by separately cutting stacks of sheets. We keep the $M/3$ stacks with the best estimation of the thickness as given by the calibration (in general, thin stacks). The solution of the linear $M\times M$ system yields the number of branches in each stack. We reopened a few configurations (5 per set of experiments) and checked by counting the number of branches in each stack: we found no error for sets $2$, $3$ and $4$ (corresponding to $\varepsilon\leq14$, $C\leq9\%$), and an error of $\pm 1$ for a part of the thicker stacks ($20\%$ of the stacks) in the more compact set $1$ (corresponding to $\varepsilon=19$, $C=15\%$). These errors are small thanks to the fact that the number of branches is an integer, and to the successive steps of image processing.  The central position of stacks is thus deduced from the skeleton and smoothed out by slide-averages along the stack trajectory. The local curvature is measured through a parabolic fit on size $5\,n_{br}\,h$, after coordinates translation and rotation in the tangent frame.  Image processing as described above allow to detect branches of length $\ell\geq\ell_{min}=1~mm$ and voids of surface $s\geq s_{min}=(10h)^2$.

\section{Numerical simulations}
\label{sec_num}

\subsection{Energy functional }

We consider an elastic rod of bending rigidity~$B$ and total length~$L$.  Its configurations are represented by a~2d vector~${\bf R}(s)$ parametrized by the arclength~$s \in [0,L]$.  Contrary to the experiment described in section~\ref{sec_exp}, the global constraint of compaction is introduced by plunging the rod into an external quadratic potential.  In that case, instead of a hard-wall geometry, the external field acts as a body force attracting the rod towards the minimum of the potential located at the origin~${\bf R}={\bf 0}$.  Note that imposing a stronger constraint such as~${\bf R}^{a}$ with exponent~$a \gg 2$ (which would be closer to the experimental compaction potential) makes the minimization algorithm significantly slower.  We will see later that this is not a limitation as we do not intend to precisely mimic the experiment in the numerical simulations but rather to extract some robust observations common to both systems.  The strength of this confining potential can be varied through a control parameter~$\lambda$.  Once embedded in this confining force-field, the shape of the rod is prescribed by the competition between two effects: While its bending rigidity tends to keep it straight, the rod responds to the external confining force by buckling and developing folds.  Since we will consider infinitely thin rods in the simulations, the resulting deformations can only be of pure bending and the rod is unstretchable: Its total length~$L$ remains a conserved quantity.  Elasticity theory shows that the energetical cost for bending deformations is proportional to the square of the rod curvature.  In addition to the bending energy, there is another source of energy associated with the confinement.  The total energy of the confined rod can be written as:
\begin{equation}
E = \frac{B}{2} \int_{0}^{L} \left( \frac{\mbox{d}^2 {\bf R}}{\mbox{d}s^2} \right)^2 \mbox{d}s  + \frac{\lambda}{2} \int_{0}^{L} {\bf R}^2 \mbox{d}s + \mbox{``Hard-core repulsion''}. \label{eq_enum}
\end{equation}
The physical constraint of self-avoidance is implemented through a discontinuous hard-core interaction. In the following, the spatial variables~${\bf R}$ and~$s$ are rescaled (now denoted~$\tilde{\bf R}$ and $\tilde{s}$) by the rod length~$L$ and the total energy $E$ by $B/2L$.  This yields the total dimensionless energy~$\tilde{E}$:
\begin{equation}
\tilde{E} = \int_{0}^{1} \left( \frac{\mbox{d}^2 {\bf \tilde{R}}}{\mbox{d}\tilde{s}^2} \right)^2 \mbox{d}{\tilde s} + \Lambda \int_{0}^{1} \tilde{\bf R}^2 \mbox{d}{\tilde s} + \mbox{``Hard-core repulsion''} \mathrm{,}
\label{modreduced}
\end{equation}
with the dimensionless control parameter~$\Lambda =  \lambda L^4 /B $.  In order to compute this energy numerically, the rod is discretized into~$N$ segments of constant length~$L/N$.  Derivatives for the bending energy are determined via finite-differences and integrals are computed with usual trapezoidal rules.  Our goal is to explore the energetical landscape by minimizing Eq.~(\ref{modreduced}).  An obvious constraint that has to be satisfied during the minimization is self-avoidance (physical objects cannot cross themselves).  This constraint is encoded in the ``hard-core repulsion'' term of Eq.~(\ref{modreduced}).  Whenever a configuration contains at least one self-intersection, its energy is set to infinity (a very large number in practice). Otherwise only the first two terms (bending and confinement) contribute to the total energy for configurations free of any self-intersections.

Avoiding self-intersections in the energy minimization is a delicate operation because it is a non-local interaction.  Regions of the rod that are far away in the rest state (straight rod) become very close when the available area decreases and folds start to appear.  Since the location and the nature (localized or extended) of these contact regions cannot be known beforehand, the detection and treatment of self-contact areas is numerically expensive.  Because the rod is discretized into a connected polyline of~$N$ segments, looking for self-intersections is a procedure that usually involves testing each pair of segment and therefore grows as $N^2$.  However here, we can take advantage of the particular (connected) geometry of the problem and lighten this procedure. Instead of testing each pair of segments, the idea is to restrict our search to a limited set of segments that are more likely to contain self-intersections.  This is possible by designing a variation of the brute force~$N^2$ method that keeps track of the distance between segments.  Once a pair of segment has been tested we also determine their distance in units of segment length~$L/N$.  This information is then used to determine how many of the following segments can be ignored.  These segments are skipped because they are too far to generate an intersection with the tested segment (even in the worst case scenario where they would  come straight back towards it).  Therefore the actual number of tested segments depends on the input curve.  We found this input sensitive algorithm to run much faster than the simple~$N^2$ method and as fast as more elaborate techniques (sweep-line for example) in computational geometry, at least for the moderate values of~$N \approx 300$ used here. 

\subsection{Minimization procedure}

\begin{figure*}  
\centerline{
\includegraphics[width=0.32\linewidth]{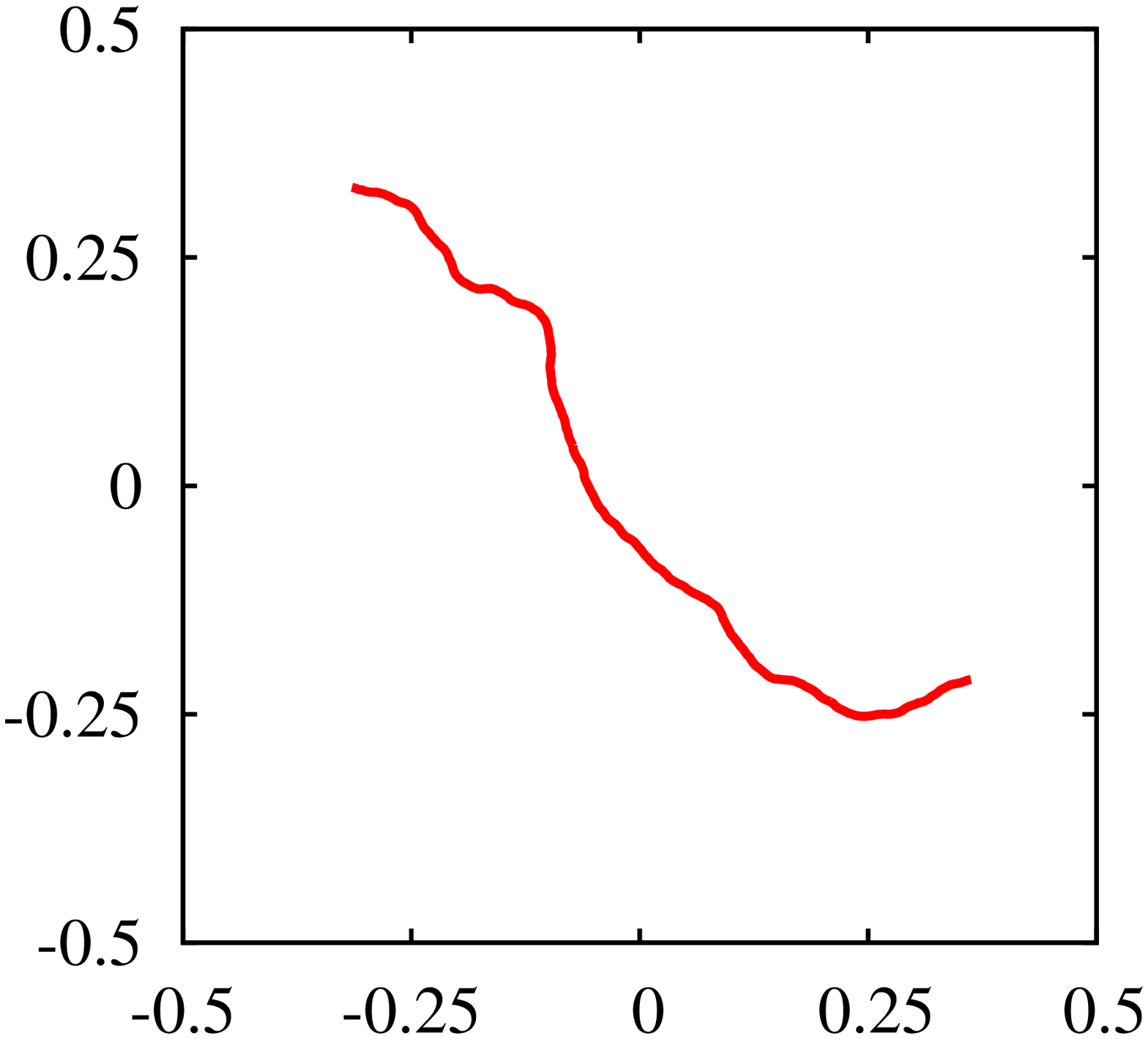} \hfill
\includegraphics[width=0.32\linewidth]{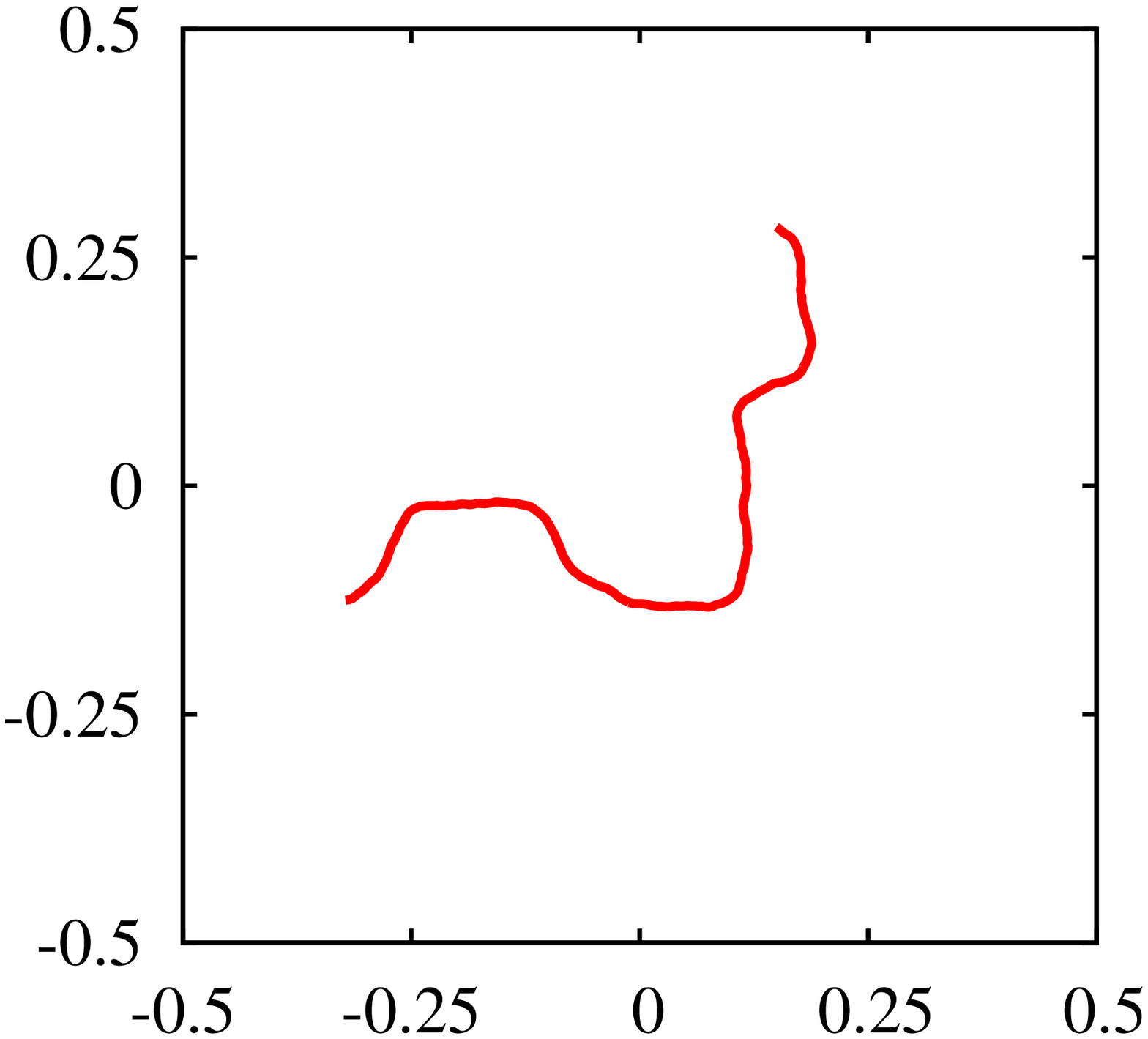} \hfill
\includegraphics[width=0.32\linewidth]{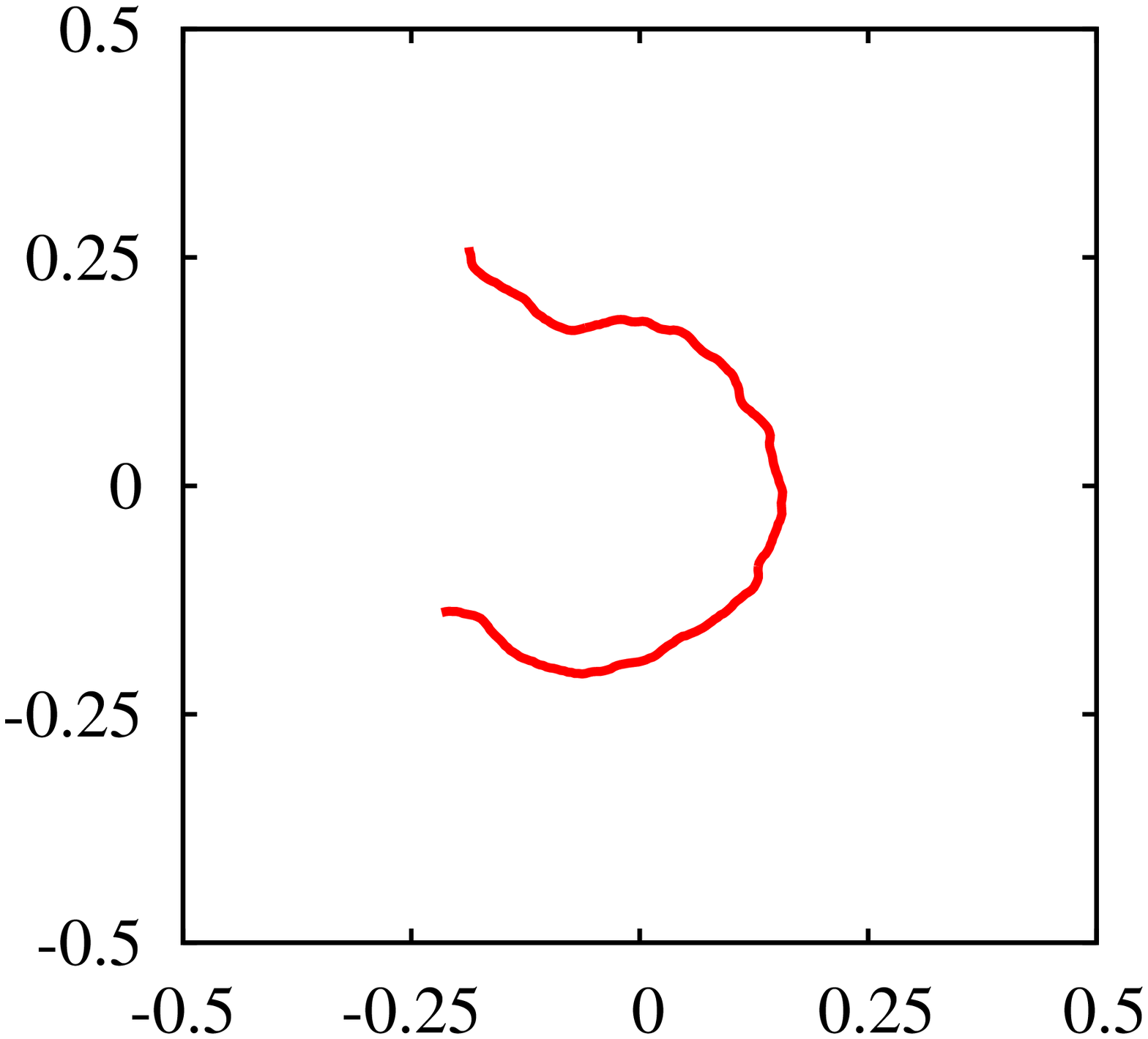} 
}  
 \caption{{\it {\bf Simulations.} Typical random initial configurations used as input in the numerical simulations.  These are generated by imposing a white noise on their curvature according to Eq.~(\ref{white_noise}), with noise intensity~$D=750$.}}
 \label{typical_ini}
\end{figure*}

As mentioned above, the self-avoidance property introduces discontinuities in parameter space by building up infinite barriers between different attraction basins.  In general, the question of finding extrema of discontinuous functions is very difficult, because one cannot use familiar procedures such as gradient methods.  We found Powell's algorithm to be convenient in our minimization problem.  It is a derivative-free procedure whose search directions in parameter space are updated after each iteration, finally generating non-interfering (or conjugate) search directions.  The numerical protocol is now described step by step:
\begin{enumerate}
\item{{\bf Initial configurations.}} A catalog of random initial conditions is constructed by the introduction of a white noise of amplitude $D$ on the curvature of the rod such that~:
\begin{eqnarray}
 \left\langle {\bf R}'' (s)\right\rangle  &=& 0, \\
 \left\langle {\bf R}''(s_1) {\bf R}''(s_2) \right\rangle &=& 2 D\, \delta(s_1 - s_2), \label{white_noise}
\end{eqnarray}
where the average is taken over each realization of the rod and $\delta$ is the Dirac distribution function.  Once a value for~$D$ has been set, we can iteratively generate multiple initial shapes.  The configurations' center of gravity is translated back to the origin.  Some typical configurations used as random initial shapes are shown in~Fig.~\ref{typical_ini}. The largest accessible value of~$D$ is the one which does not generate self-intersections.
\item{{\bf Introduction of the confinement.}} The parameter~$\Lambda$ controls the strength of the confining potential. Going instantaneously from~$\Lambda=0$ for the construction of initial configurations, to a non-zero value to take into account the confinement, can be seen as a quenching mechanism.  This is because the confinement constraint is instantaneously turned on to its desired value without taking intermediate steps.  All of the results presented here were obtained by following this procedure.   We also tested an annealing process by increasing~$\Lambda$ very slowly, about which we will say few words in the conclusion.
\item{{\bf Search for local minima.}}  In practice we ran Powell's algorithm 10 times for each initial configuration.  We verify at each iteration that configurations do not contain any self-intersections.  If they do, we set the energy of such configurations to a very large number, so that these configurations are immediately rejected by Powell's algorithm.  The set of search directions is regularly re-initialized to a set of random directions in order to optimize our span of local minima.  Running the minimization procedure many times also improves numerical convergence to these local minima. One should note that while the hard core repulsion procedure ensures self-avoidance, it raises problems when parts of the rod are aligned. Indeed our numerical procedure does not allow sliding of self-contact areas.  Therefore we (temporarily) modeled self-contacts as a nematic interaction of the form~$E_{\mbox{\tiny self-contact}} = u \sin^2 \alpha$ where~$u$ is a dimensionless parameter and~$\alpha$ is the angle between touching segments~\cite{katzav06,boue07}.  Typically, we used~$u$ in the range~50-150 but its precise value did not affect the resulting configurations.  This geometric self-interaction has a physical interpretation as a self-excluded volume resembling what was introduced by Onsager in the context of polymers.  In our case, it destabilises  tightly clamped configurations by allowing branches in self-contact to self-align and slide along one another.   This nematic term is included only in 2 early algorithm runs (out of~10) and ignored otherwise (and in particular in the last run).
\end{enumerate}

\begin{figure*}  
\centerline{
\includegraphics[width=0.32\linewidth]{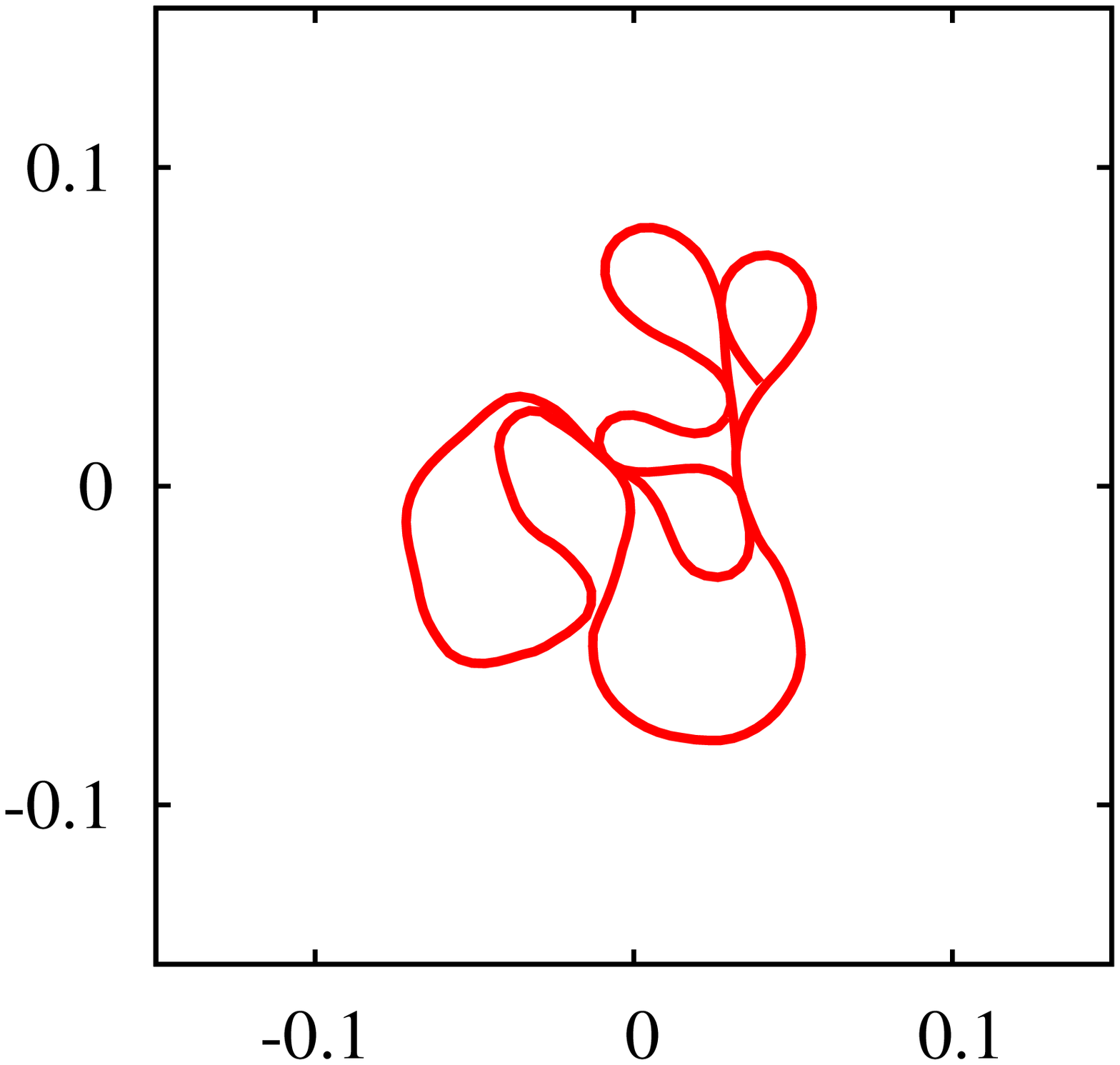} \hfill
\includegraphics[width=0.32\linewidth]{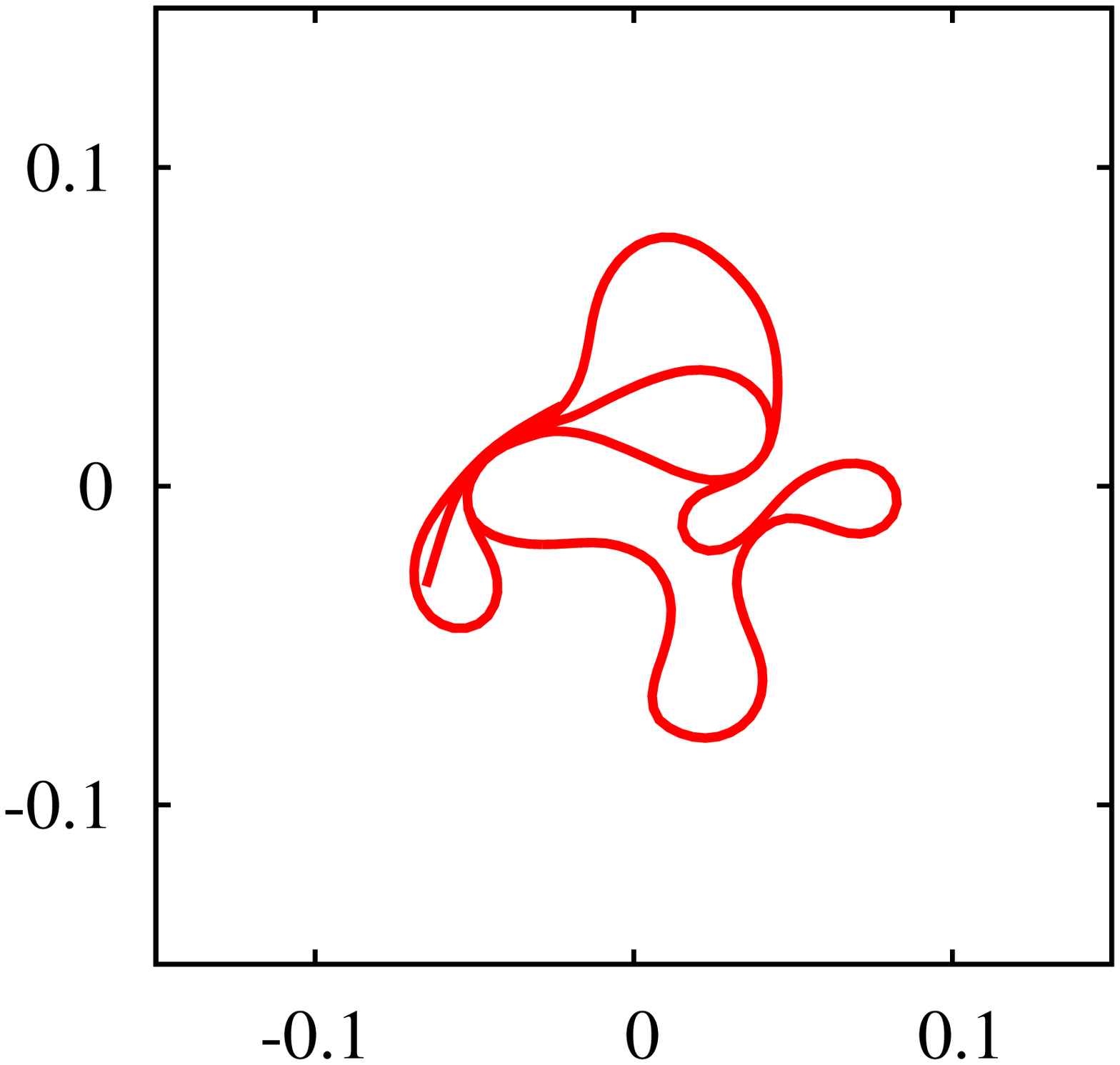} \hfill
\includegraphics[width=0.32\linewidth]{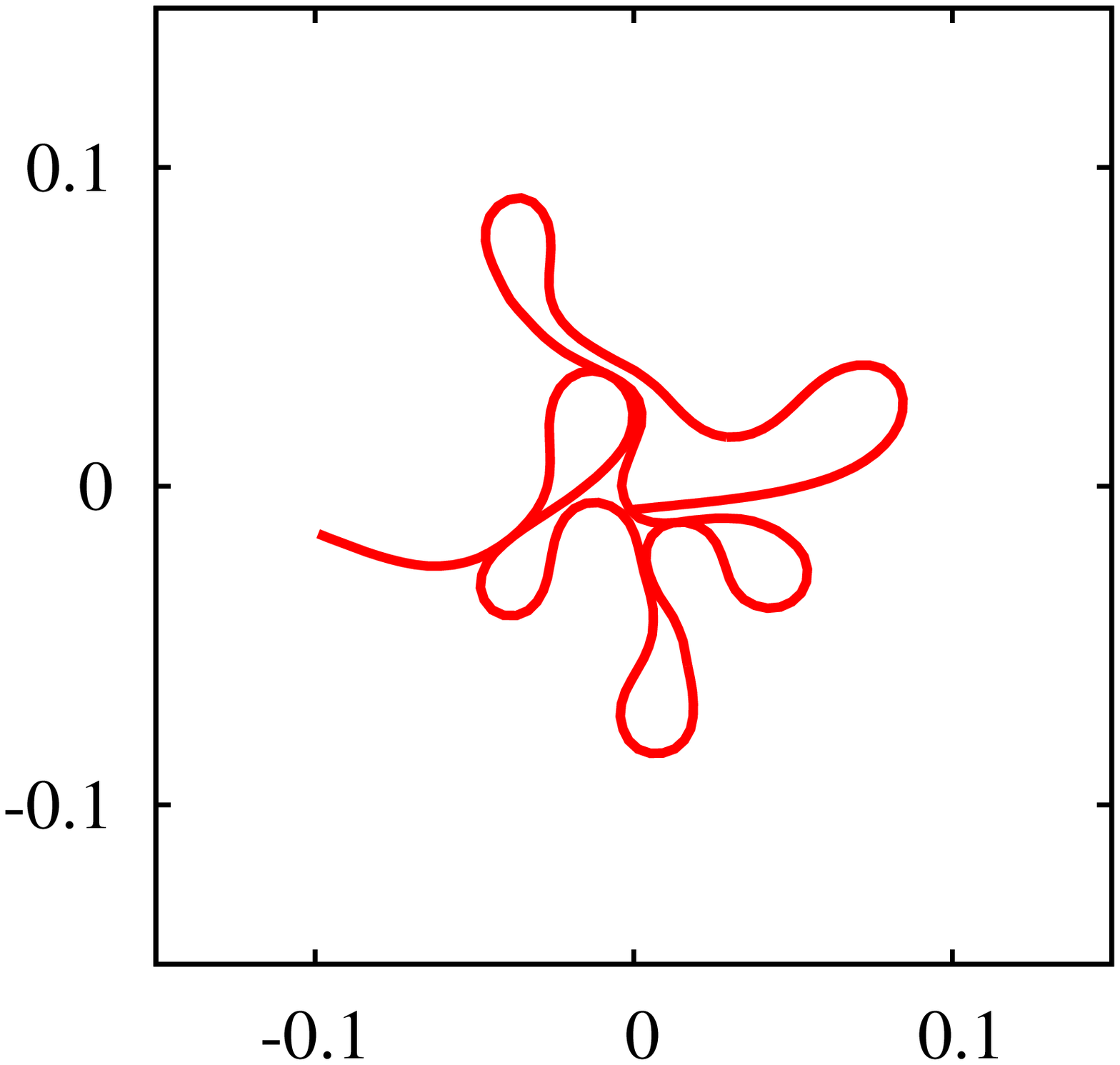} 
}  
\caption{{\it {\bf Simulations.} Typical local minima obtained with~$\Lambda = 7\,10^5$ and~$D=750$. The corresponding mean compaction ratio is~$\left\langle \varepsilon\right\rangle=2.3$.}}
 \label{fig_simu}
\end{figure*}

After going through this procedure a number of times with different initial conditions but under the same compaction conditions (noise intensity~$D$ and control parameter~$\Lambda$), a wide variety of different possible configurations is found, some typical examples of which are shown in Fig.~\ref{fig_simu}.  The results presented here are with~$\Lambda = 7\,10^5$ and  $D=750$ for a total of over~250 different realizations.

\subsection{Simulations vs experiments }

\begin{figure*} 
\centerline{
 \includegraphics[width=0.32\linewidth]{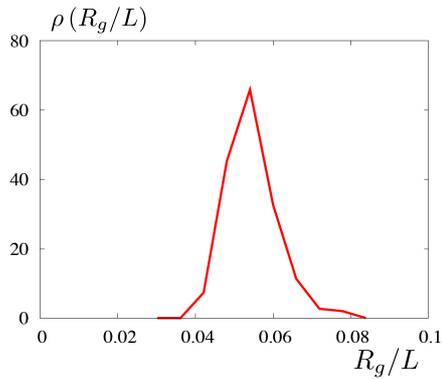} 
}
 \caption{{\it {\bf Simulations.} Distribution of the radius of gyration $R_{\mbox{\tiny g}}$ non-dimensionalized by the rod length $L$,  of the 252 configurations with~$\Lambda=7\,10^5$ and~$D=750$.}}
 \label{gyra}
\end{figure*}

As previously noticed, the global compaction constraint is of different nature in simulations and experiments: compaction is mechanically controlled in  simulations through $\lambda$, the intensity of the quadratic potential  exerted on the rod,  whereas it is geometrically controlled in experiments, through the available size $R$ for the rod.  The pressure exerted on the quasi-1d rod in experiments is related to the force necessary to pull the 2d sheet through the hole. This raises the question of the parameter common to experiments and simulations, relevant for the description of the compaction strength.  On one hand, pressure is not trivially accessible in experiments;  on the other hand, the size of the occupied surface can be easily characterized for each configuration in simulations by its radius of gyration~$R_{\mbox{\tiny g}}$: 
\begin{equation}
R_{\mbox{\tiny g}} = \sqrt{ \frac{1}{L} \int_0^L {\bf R}^2 \mbox{d}s } \mathrm{ ,  } \label{eq_raygir}
\end{equation} 
allowing to compute the compaction rate $\varepsilon$, written in Eq.~(\ref{eq_epsexp}), by replacing $R$ by $R_{\mbox{\tiny g}}$.   The radius of gyration~$R_{\mbox{\tiny g}}$ has been computed for all the numerical configurations.  Its probability density, presented in Fig.~\ref{gyra}, features a sharp peak at~$\left\langle R_{\mbox{\tiny g}}\right\rangle \simeq 0.05 L$, showing that imposing a confining potential and noise amplitude in the simulations, results to indirectly imposing the size of the surface occupied by the rod.  In these early stages of folding that are accessible numerically, we define the geometric compaction rate as $\varepsilon=L/\left(2\pi R_{\mbox{\tiny g}}\right)$, and thus have on average for all configurations~$\left\langle \varepsilon\right\rangle=2.3$.

To summarize, the values of~$\varepsilon$ studied in this paper are collected for experiments and simulations respectively.  Four sets of experiments are realized at fixed compaction rate $\varepsilon=8$, $10$, $14$ and $19$ (see Table~\ref{tab_paramexp}).  Even if the compaction rate is not a priori controlled in the simulations, it is indirectly imposed to the mean value $\left\langle \varepsilon\right\rangle=2.3$.   Despite the different nature of compaction, geometrically and  mechanically  controlled in experiments and simulations respectively, implying a different expression of the rod energy (as it has been suggested previously and it will be shown later), and despite the different values of compaction rates $\varepsilon\sim10$ in experiments and $\varepsilon\sim2$ in simulations (visible when comparing Fig.~\ref{fig_setexp} and Fig.~\ref{fig_simu}), both systems share the property that these are governed by elasticity and self-avoidance. The main issue of this study, allowed by the investigation of these two systems, is to point out the general characteristics of folding and to identify the ones that are system dependent.  Two types of averages will be used in this study:~$\overline{x}$ refers to the mean of the variable $x$ over one configuration, whereas $\left\langle x\right\rangle$ refers to the ensemble average of $x$ over the whole set of configurations realized at constant control parameters.  Due to better statistical averaging, most of the results presented here were obtained by ensemble statistics.

\begin{figure*}  
\centerline{
\includegraphics[width=.25\linewidth]{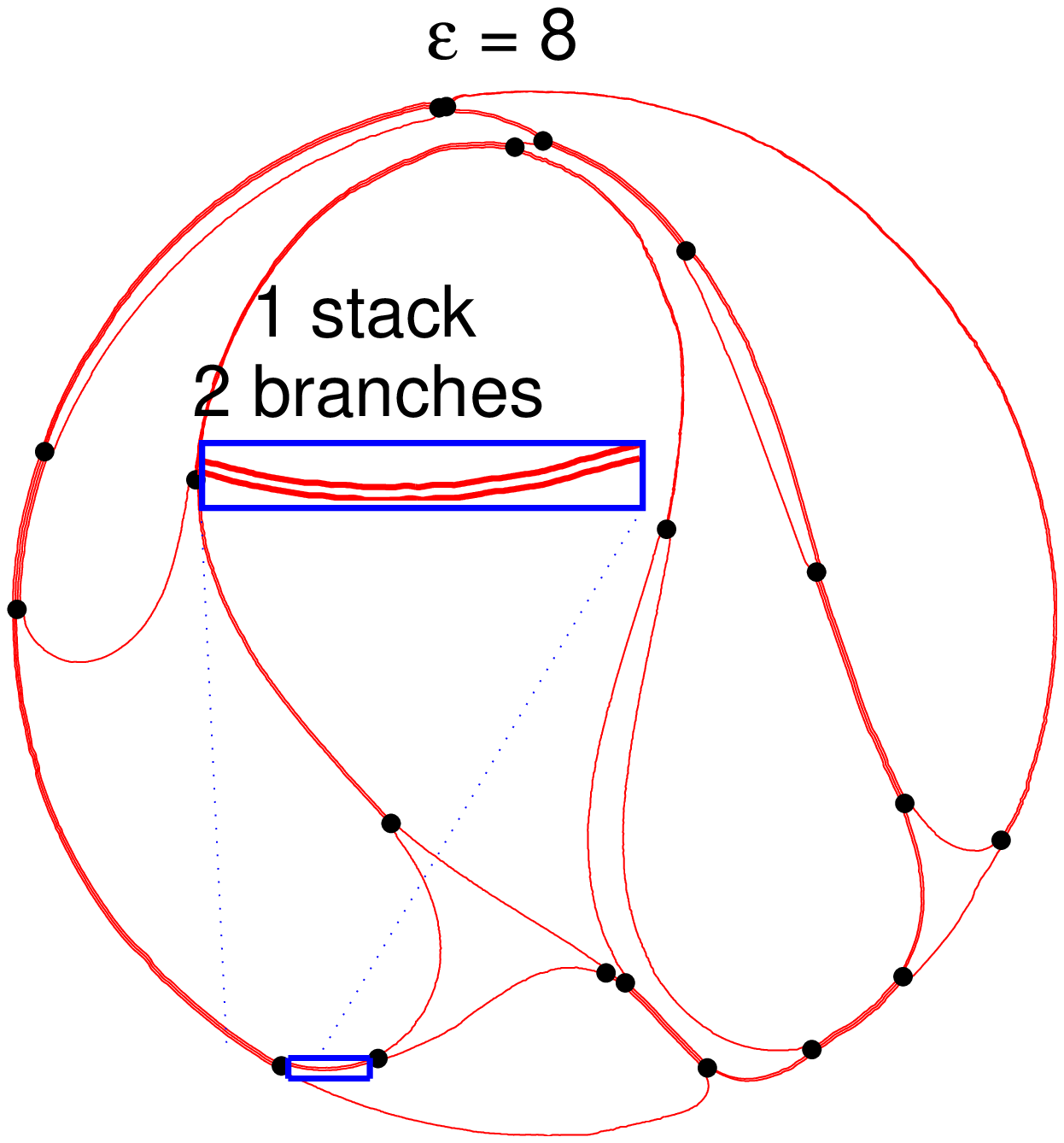}\qquad
\includegraphics[width=.25\linewidth]{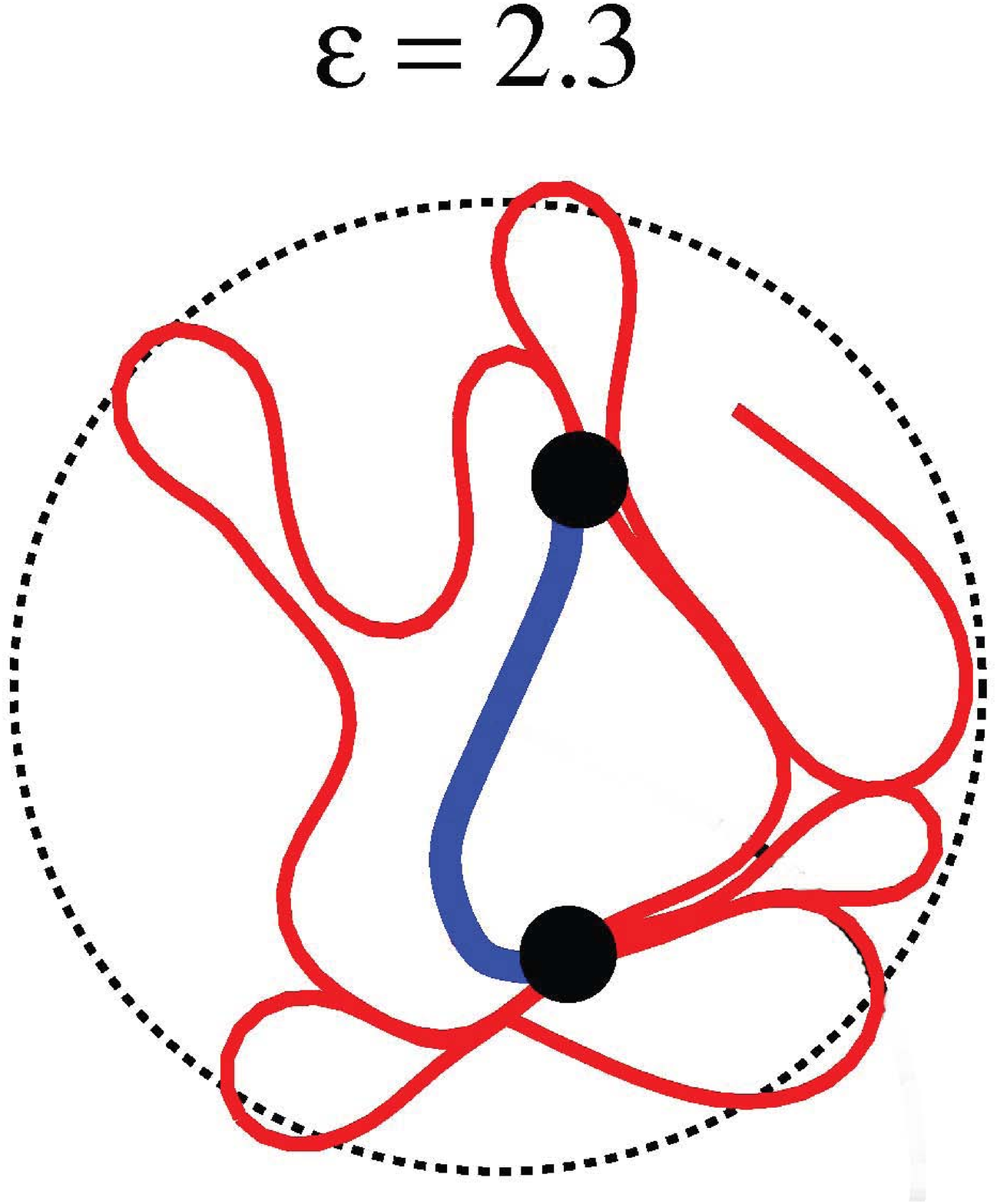} 
}  
\caption{{\it Definition of elementary branches. Parts of the rod delimited by two junction points ($\bullet$) define the elementary ``branches''. Some of them are in close contact and are delimited by the same junction points, defining a multi-branched stack. {\bf Experiments:} configuration from set $3$ at compaction rate~$\varepsilon=8$, with a snapshot showing a multi-branched stack.  {\bf Simulations:} the dashed circle represents the mean radius of gyration $\left\langle R_{\mbox{\tiny g}}\right\rangle$ at~$\varepsilon = 2.3$.}}
\label{branches_def}
\end{figure*}

A crude observation of examples of folded rods shows that, under constant control parameters, a wide variety of configurations are obtained in both the experiments (Fig.~\ref{fig_setexp}) as well as the numerical simulations (Fig.~\ref{fig_simu}).   This is an interesting first remark confirming that we do indeed span a large volume of phase space.  Although the shape of a folded rod looks very complicated, elementary parts of the rod, delimited by two neighbouring junction points (where the rod is locally in self-contact) can be identified.  These elementary supports are referred to as ``branches'', and could be relevant candidates for a ``microscopic'' definition and parametrisation of the macroscopic folded configuration.  From a mechanical point of view, branches are the natural elements, on which elastic equations are defined given the boundary conditions at their extremities.  Fig.~\ref{branches_def} shows two folded configurations coming from experimental and numerical sets, where junction points are shown by circles, that delimit branches.  Branches that are in close contact and delimited by the same junction points define a multi-branched stack.  We start our data analysis in section~\ref{sec_topo} and section~\ref{sec_geometro} by looking at the topological and geometrical properties.  While their behaviour are interesting, it is in fact the energetical properties studied in section~\ref{sec_energo} that turn out to be the desired general variable.

\section{General properties of the network}
\label{sec_topo}

\begin{table*} 
\begin{center}  
\begin{tabular}{l||l|l|l|l|l|l} 
& $\left\langle N_l\right\rangle$& $\left\langle N_{br}\right\rangle$ &$\left\langle n_{br}\right\rangle$& $\left\langle s\right\rangle/L^2$& 
$\left\langle \sqrt{s}/p\right\rangle$& $\left\langle l\right\rangle/L$  \\ \hline
$1$& 90& 380& 5& 6.6$\,10^{-6}$& 0.19& 2.5$\,10^{-3}$ \\  
$2$& 80& 310& 4& 1.4$\,10^{-5}$& 0.18& 3.4$\,10^{-3}$\\  
$3$& 40& 100& 3& 1.1$\,10^{-4}$& 0.18& 10$\,10^{-3}$ \\    
$4$& 80& 330&5& 2.6$\,10^{-5}$& 0.18& 4.4$\,10^{-3}$ \\ \hline
$5$& $\sim10$ & $\sim20$  & $\sim2$& 3.4$\,10^{-4}$& 0.19& 0.05
\end{tabular}    
\end{center}  
\caption{{\it Averages over all realizations of the number of multi-branched stacks $N_l$, number of branches $N_{br}$, number of branches  per stack $n_{br}=N_{br}/N_l$, surface area $s$, shape ratio $\sqrt{s}/p$ of elementary tiling voids and length of branches $\ell$.}}
\label{tab_macro}   
\end{table*}  

\subsection{Stacking}
\label{subsec_netw}

\begin{figure*}
\centerline{
\includegraphics[width=0.32\linewidth]{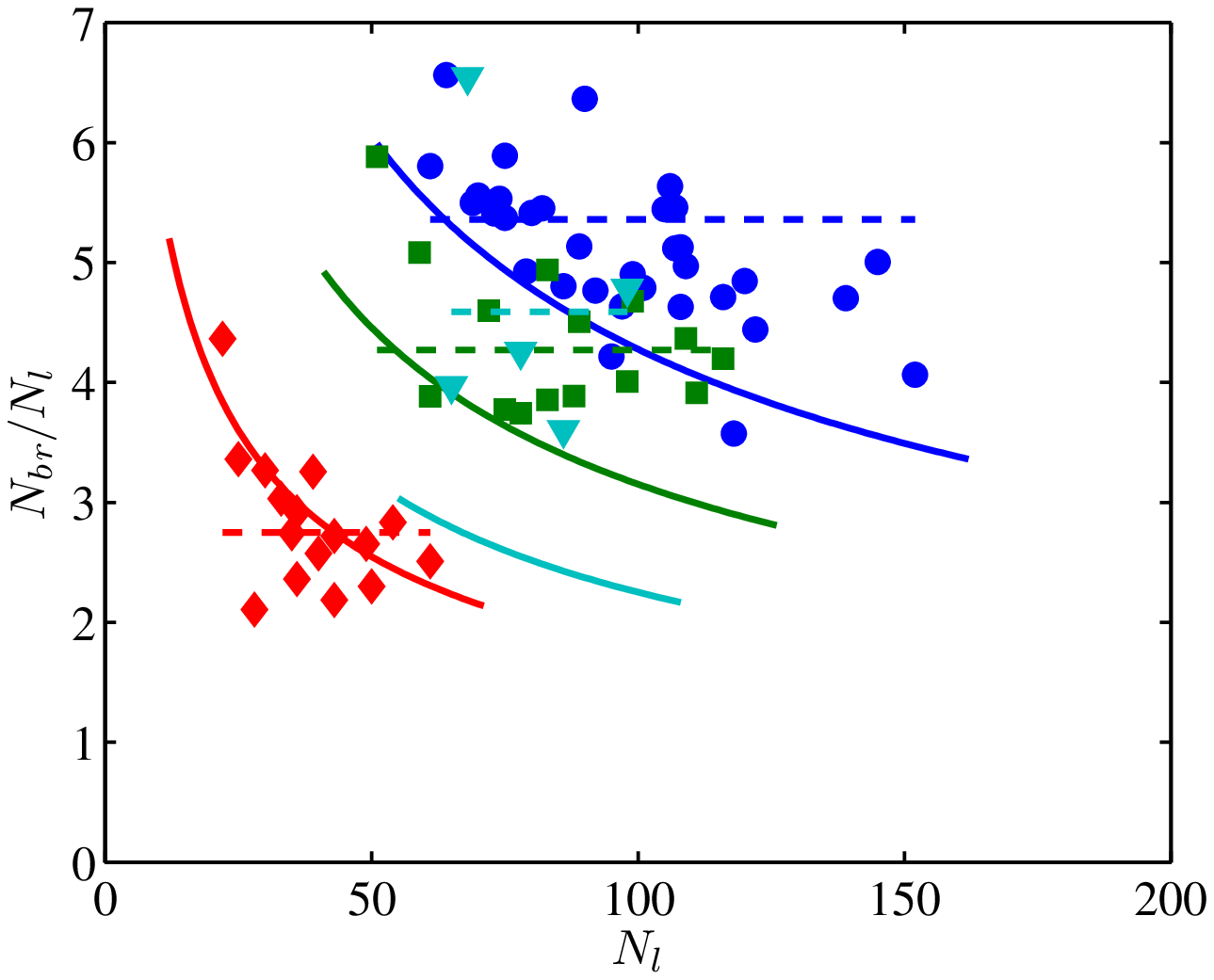}\hfill  
\includegraphics[width=0.32\linewidth]{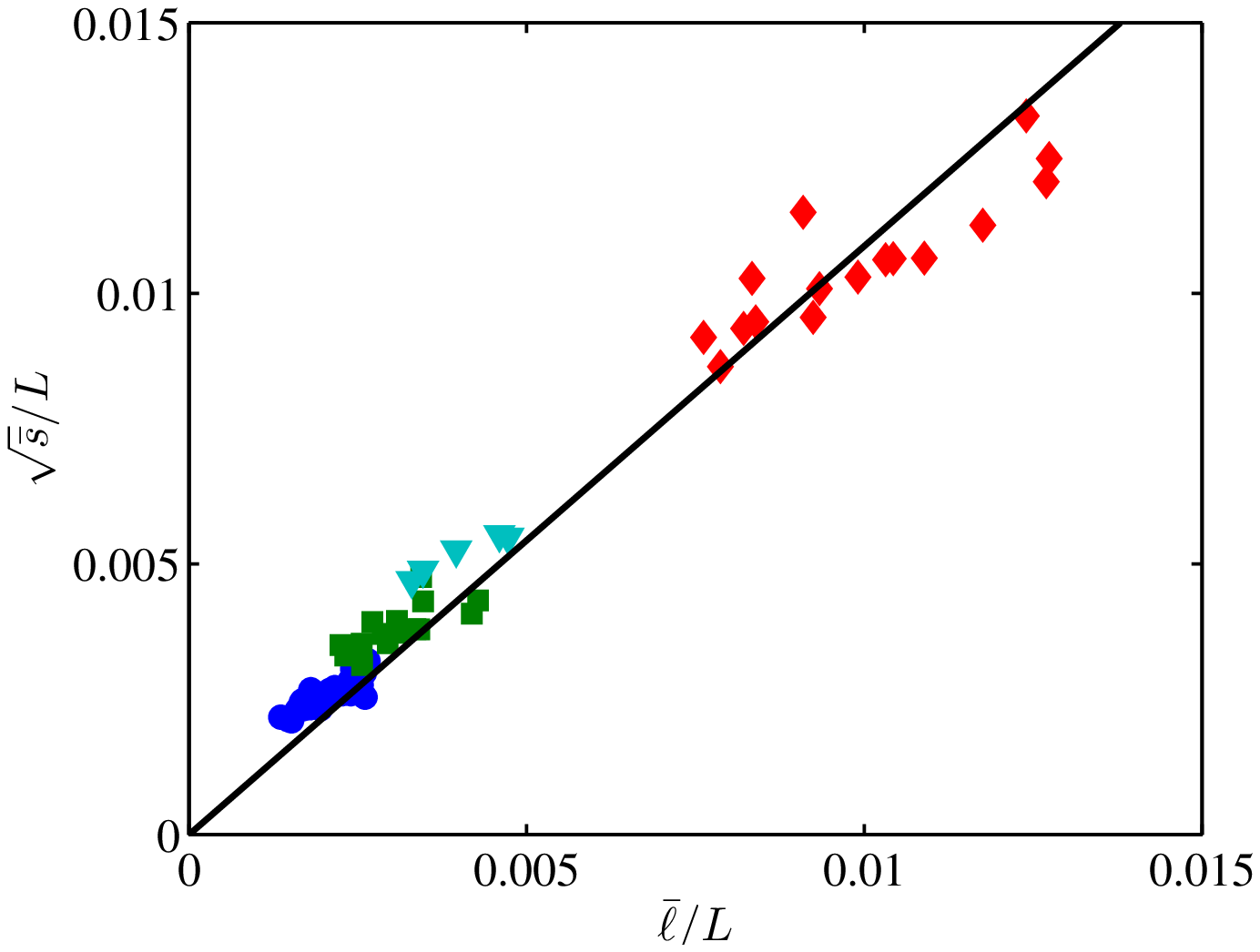} \hfill  
\includegraphics[width=0.32\linewidth]{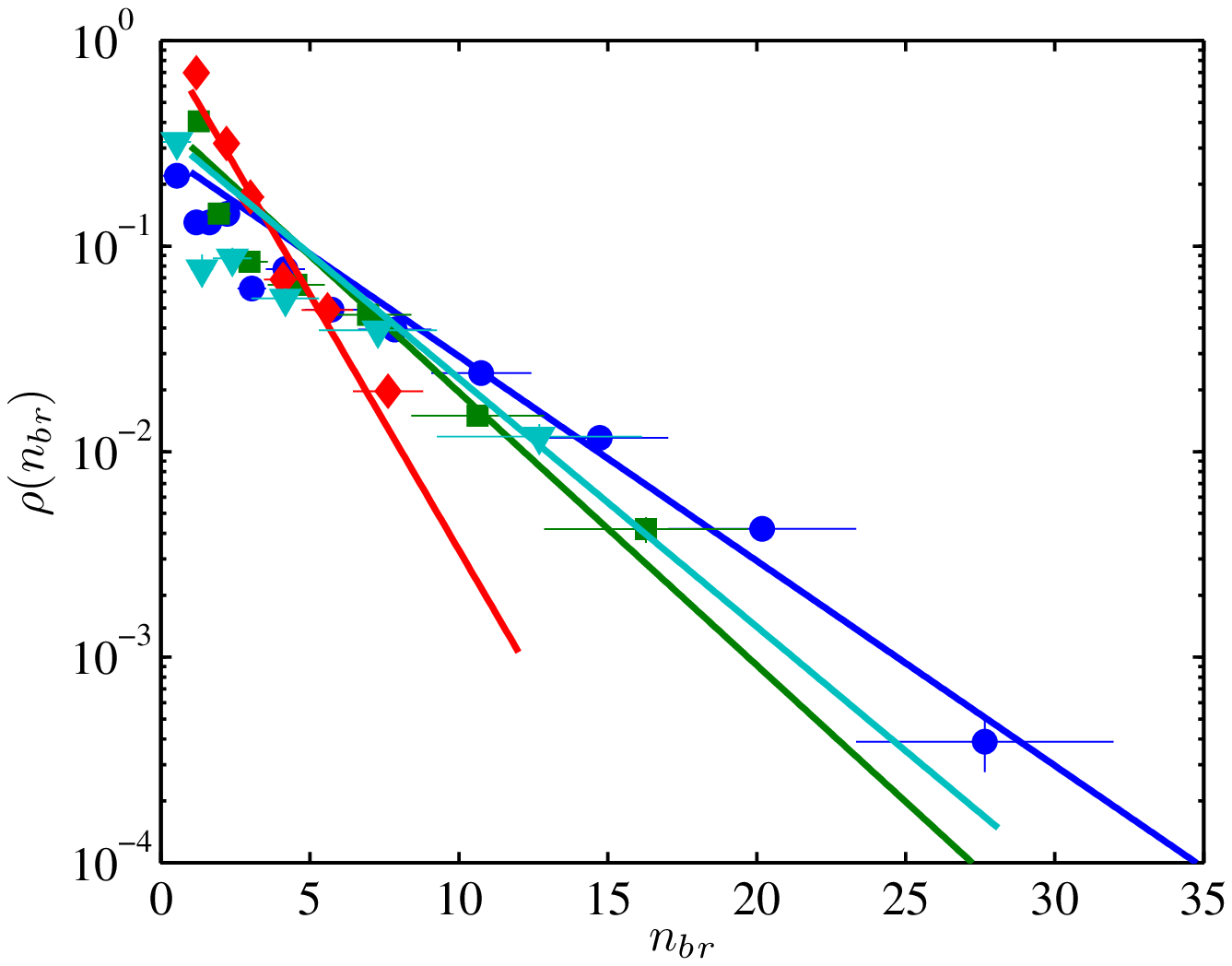}  
}
\caption{{\it {\bf Experiments.} {\bf a}) Mean number of branches per stack~$\overline{n}_{br}$ as function of the total number of stacks~$N_l$.  Continuous (resp. dashed) lines correspond to Eq.~(\ref{eq2}) (resp. Eq.~(\ref{eq4})).  {\bf b})  Square root of the mean area of voids $\sqrt{\overline{s}}$ as a function of the mean length of branches $\overline{l}$ (both quantities are non-dimensionalized by $L$). The continuous line corresponds to a linear fit of slope~$1.1$. {\bf c}) Illustration of the exponential distributions of the number of branches per stack~$n_{br}$ for the different sets of experiments. }}
\label{topolo_macro}
\end{figure*}

Here, we are interested in the number of different elementary units in a folded configuration: junction points, voids, branches, multi-branched stacks, and to their relative dependencies.  Junction points, multi-branched stacks and number of branches in close contact define respectively the nodes, the links and the values attributed to each link of a network~\cite{andresen07}.  Therefore, we start by investigating the topological properties associated with the network formed in the confined rods.

\paragraph*{{\bf Experiments.}}

A folded configuration is characterized by a number $N_l$ of multi-branched stacks. Each stack is formed by a number $n_{br}$ of multiple branches, the  sum of which is $N_{br}$ over the configuration. A natural question is how these network properties $N_l$  and  $N_{br}$ vary together?  The values of $\left\langle N_l\right\rangle$ and $\left\langle N_{br}\right\rangle$, reported in Table~\ref{tab_macro},  increase roughly with the compaction rate.  Fig.~\ref{topolo_macro}{\bf a} shows the mean number of branches per stack  $\overline{n}_{br}=N_{br}/N_l$, as a function of the total number of stacks $N_l$, for all experiments of all sets. Inside a single set, plotted points seem to follow a systematic trend (Fig.~\ref{topolo_macro}{\bf a}), showing that $N_{br}$ and $N_l$ are correlated.   
 
  Exactly $3$ multi-branched (or links) intersect at $99\%$ of the junction points, the value $3$ being the minimal possible by construction.  This allows to relate the total numbers of multi-branched stacks $N_l$ and of voids $N_v$ in a configuration: $N_l\approx 3 N_n/2\approx 3 N_v$. For a closed rod, the number of junction points (or nodes of the folded network) $N_n$ is equal to the number of voids $2N_v$ (or cells of the network). By definition, $N_v\simeq S/\overline{s}\approx N_l/3$ and $N_{br}\simeq L/\overline{\ell}$, where $\overline{s}$ and $\overline{l}$ are respectively the mean surface area of voids and the mean length of branches in a configuration. Indeed the total length of the rod $L$ is divided in the $N_{br}$ branches and the total available surface $S$ is tiled by the $N_v$ voids.  Moreover, $L$ and $S$ are related through the control parameter  $\varepsilon=L/\left(2\sqrt{\pi S}\right)$.  This allows to write the following relation between $N_{br}$ and $N_l$:  
\begin{eqnarray}
 N_{br} = \frac{2\sqrt{\pi}}{\sqrt{3}} \varepsilon \frac{\sqrt{\overline{s}}}{\overline{l}} \sqrt{N_l} \text{ ,} \label{eq1}  
\end{eqnarray}
through the parameter $\varepsilon$ and an a priori configuration dependent factor $\sqrt{\overline{s}}/\overline{l}$.  It is equivalent for the mean number of branches inside one stack $\overline{n}_{br}$ averaged over configuration to: 
\begin{eqnarray}
 \overline{n}_{br}=\frac{N_{br}}{N_l} = \frac{2\sqrt{\pi}}{\sqrt{3}} \varepsilon \frac{\sqrt{\overline{s}}}{\overline{l}} \frac{1}{\sqrt{N_l}} \text{ .} \label{eq2}  
\end{eqnarray} 
Formula~(\ref{eq1}) and~(\ref{eq2}) are valid for a single experiment, are these still verified for a whole set of experiments?  Can one write a relation between $N_{br}$ and $N_l$ through quantities averaged over sets of experiments (instead of configurations), or through scalar factors?  This raises the issue of the variations of $\sqrt{\overline{s}}/\overline{l}$ inside a set of experiments: is this ratio varying with $N_l$ and control parameters $\varepsilon$?  To this aim, we plot the square root of the mean voids area $\sqrt{\overline{s}}$ as a function of the mean branches length $\overline{l}$ for all experiments in Fig.~\ref{topolo_macro}{\bf b}. Even if these values change by factors $2$ inside a set and by a factor $5$ between different sets, these appear to be  proportional through the constant value $1.1$, independent of the precise configuration and of the compaction parameters, as shown by the linear fit in Fig.~\ref{topolo_macro}{\bf b}. So one can write from Eqs.~(\ref{eq1}) and~(\ref{eq2}): $N_{br}\propto\varepsilon\sqrt{N_l}$ and $\overline{n}_{br}\propto\varepsilon/\sqrt{N_l}$, that are plotted in continuous lines in Fig.~\ref{topolo_macro}{\bf a}, super-imposed on  experimental data.  We will see in the following sub-section that another relation between $N_{br}$ and $N_l$ can be predicted.   There is still an issue with the understanding of the universal value of $\sqrt{\overline{s}}/\overline{l}$, whatever the configuration, the compaction rate and the size of the sheet.

We found that the number of branches inside one stack $n_{br}$ follows an exponential distribution as shown on Fig.~\ref{topolo_macro}{\bf c}.  For the different sets of experiments, the mean $\left\langle n_{br}\right\rangle$ of the exponential distribution varies from $3$ to $5$ for increasing $\varepsilon$ (Tab.~\ref{tab_macro}).  Inside the sets of data, the relative fluctuations $\left\langle \overline{n}_{br}/\left\langle n_{br}\right\rangle-1 \right\rangle$ are only of the order of a few percents meaning that $\left\langle \overline{n}_{br}\right\rangle \simeq \left\langle n_{br}\right\rangle$.  A precise analysis of the pdf~$\rho(\overline{n}_{br})$ would require much more statistics.  Nevertheless, it is tempting to make the assumption that $\overline{n}_{br}\simeq \left\langle n_{br}\right\rangle$.  This implies that~$\overline{n}_{br}$ is not correlated with other configuration dependent variables such as~$N_l$.  Therefore, this reasoning leads to a prediction of the total number of branches~$N_{br}$ and the average number of branches per stack~$\overline{n}_{br}$ per configuration, as functions of the number of stacks~$N_l$ given by:
\begin{eqnarray}
 N_{br} = \left\langle n_{br}\right\rangle N_l \text{ ,} \label{eq3} \\
 \overline{n}_{br} = \left\langle n_{br}\right\rangle \text{ .} \label{eq4}  
\end{eqnarray}  
These linear predictions are plotted as dahes lines in Fig.~\ref{topolo_macro}{\bf a} for a comparison with Eqs.~(\ref{eq1}) and~(\ref{eq2}) and with the experimental data.  It turns out that both predictions~(\ref{eq1})-(\ref{eq2}) and (\ref{eq3})-(\ref{eq4}) describe well the data.  This suggests that $N_l$ and $N_{br}$ are in fact selected by the validity of both assumptions.  The intersection of the two relations corresponds to: 
\begin{eqnarray}
\frac{\left\langle n_{br}\right\rangle\sqrt{\left\langle N_l\right\rangle}}{\varepsilon} \simeq 2.2\sqrt{\pi/3}  \text{ .} \label{eq5}  
\end{eqnarray}
At the level of accuracy accessible in our experiments, the agreement is not so bad even though there seems to be a small dependence with~$\varepsilon$.

\paragraph*{{\bf Simulations.}}

\begin{figure*} 
\centerline{
\includegraphics[width=0.32\linewidth]{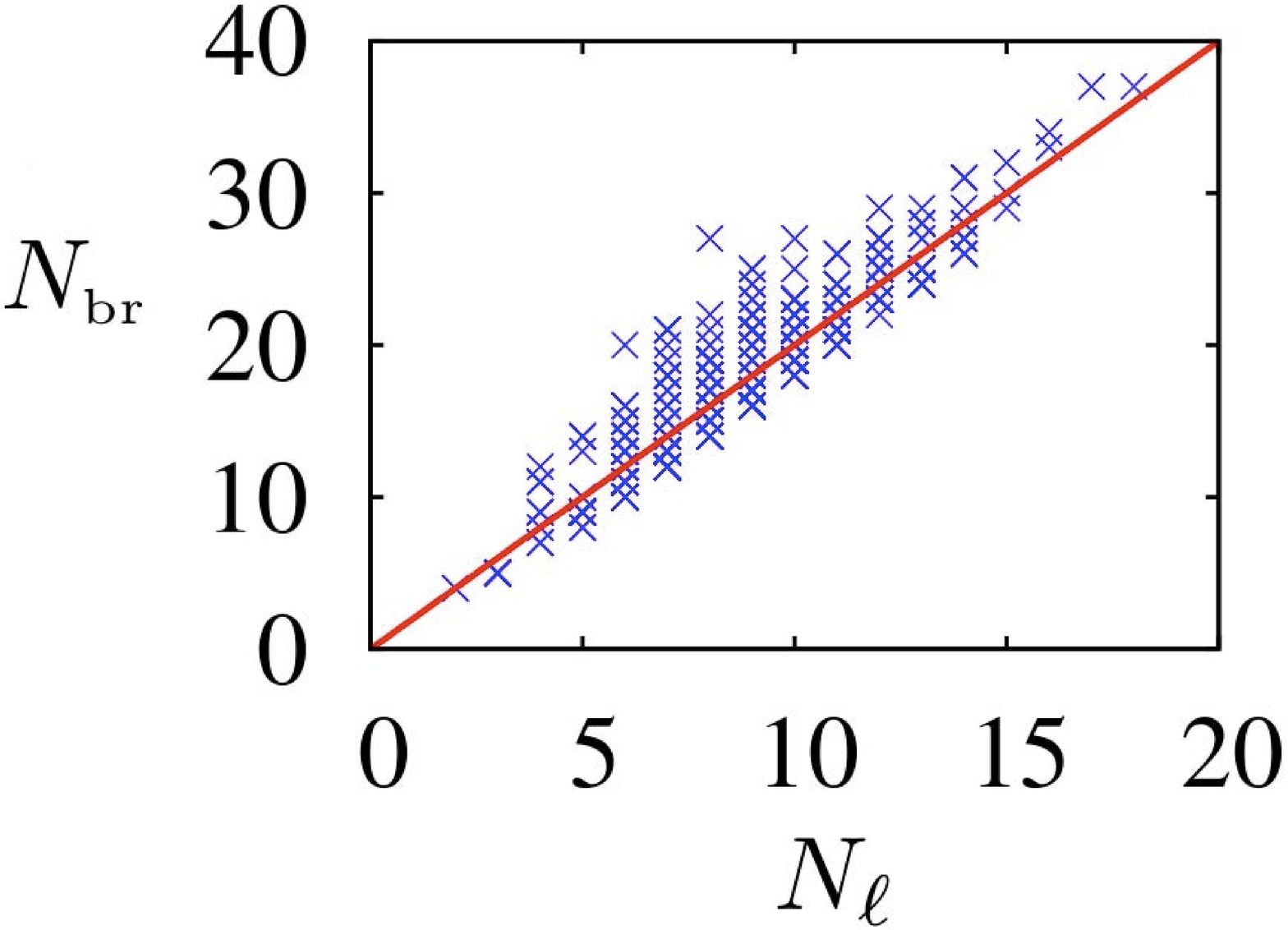}  
\includegraphics[width=0.32\linewidth]{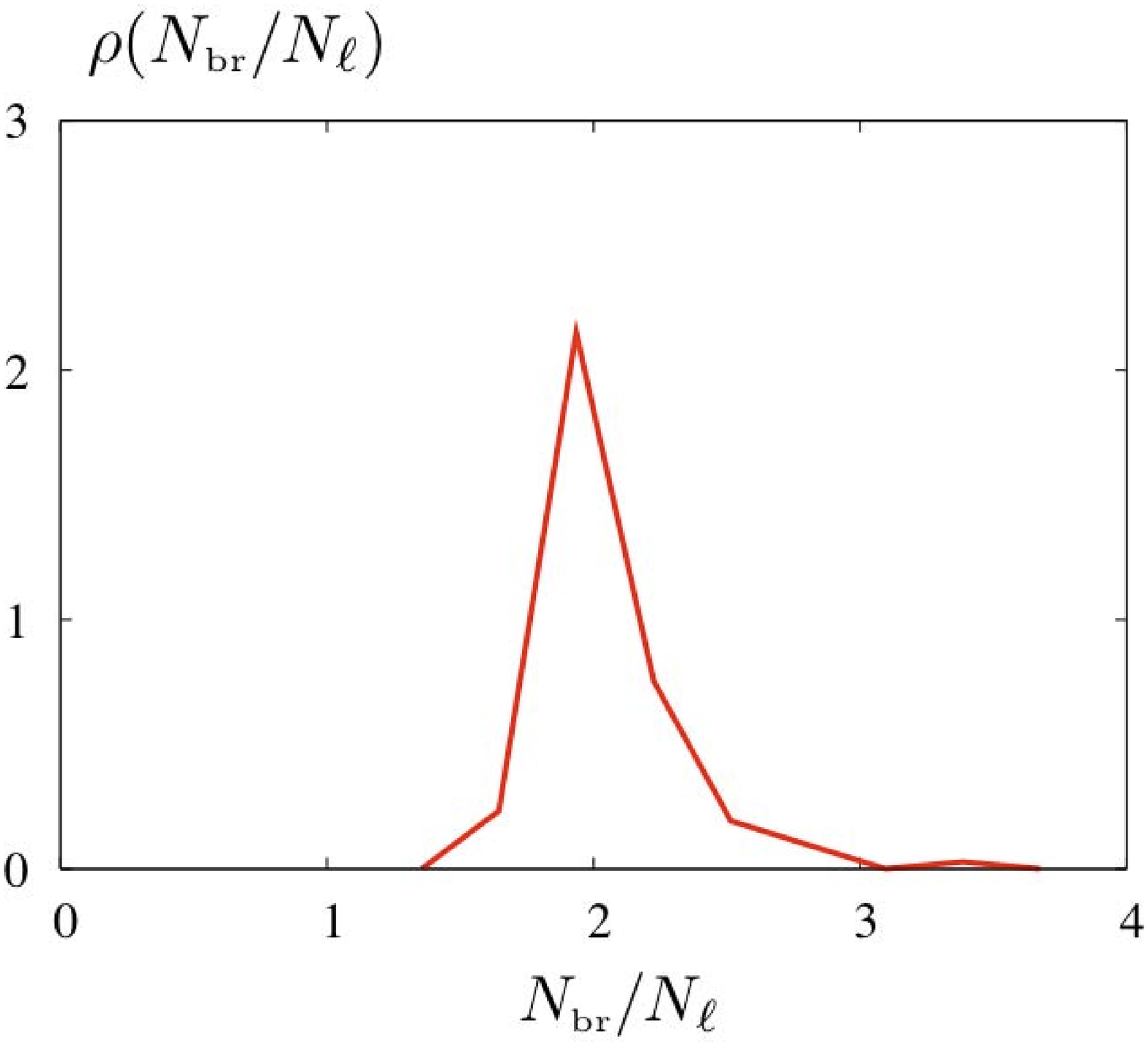}  
}
\caption{{\it {\bf Simulations.} {\bf a} Number of branches $N_{br}$ as a function of the number of multi-branched stacks $N_l$ in the simulations. Continuous line corresponds to a line of slope~$2$. {\bf b} Distribution of the ratio $N_{br}/N_l$.}}
\label{topolo_numerics}
\end{figure*}  

As the rod is open, the surface it encloses is not well-defined and not exactly constant from realizations to realizations.  The analysis presented for the experimental results cannot hold for the simulations.  However, Fig.~\ref{topolo_numerics}{\bf a} shows the number of branches $N_{\mbox{\tiny br}}$ as a function of the number of multi-branched stacks $N_l$ for all the numerical simulations: these verify the  relation~$N_{\mbox{\tiny br}} \simeq 2 N_{\mbox{\tiny l}}$; the distribution of~$N_{\mbox{\tiny br}}/N_{\mbox{\tiny l}}$ is indeed strongly peaked around~$2$ as shown on Fig.~\ref{topolo_numerics}{\bf b}.  Several configurations are characterized by the same pair ($N_{\mbox{\tiny br}}$, $N_l$).

Because the compaction rate achieved in the numerical simulations, $\left\langle \varepsilon\right\rangle=2.3$, is much smaller than those of the experiments, the number of branches in self-contact is only rarely more than$~3$.  Generically links contain just$~2$ branches. This explains the previous macroscopic relation~$N_{\mbox{\tiny br}} \simeq 2 N_{\mbox{\tiny l}}$, shown in Fig.~\ref{topolo_numerics}.

\subsection{Tiling}

\begin{figure*} 
\includegraphics[width=0.32\linewidth]{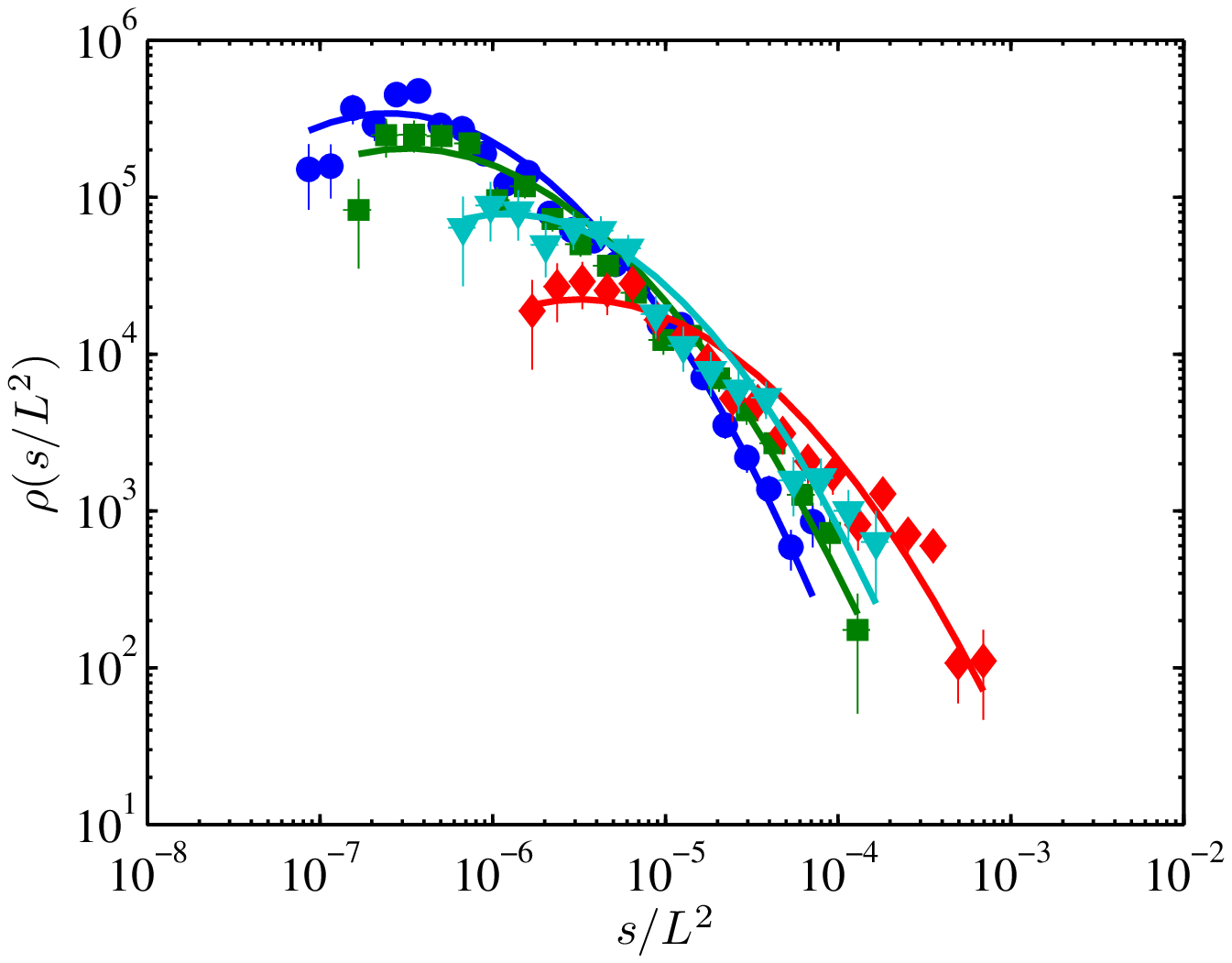} \hfill    
\includegraphics[width=0.32\linewidth]{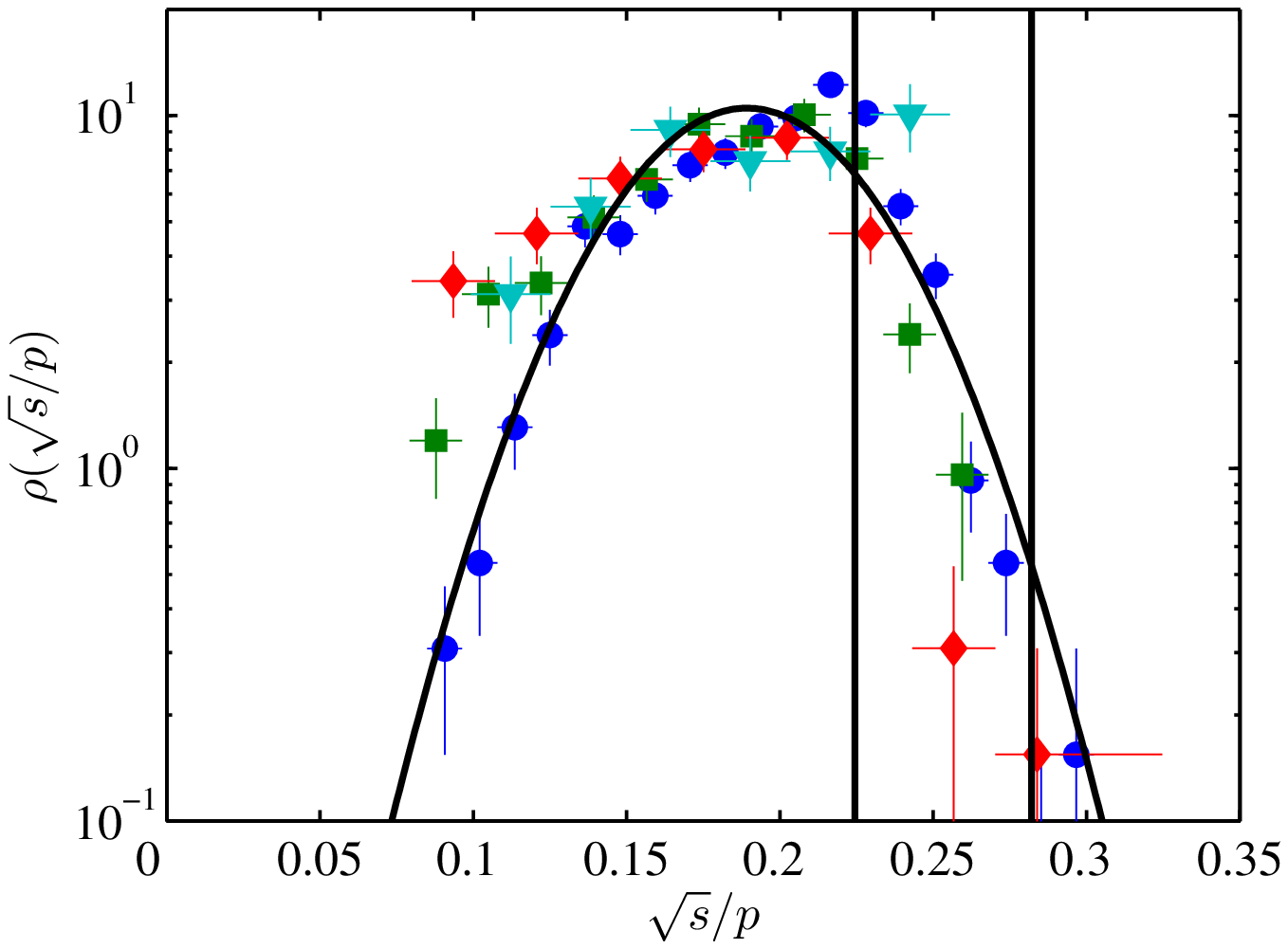} \hfill
\includegraphics[width=0.32\linewidth]{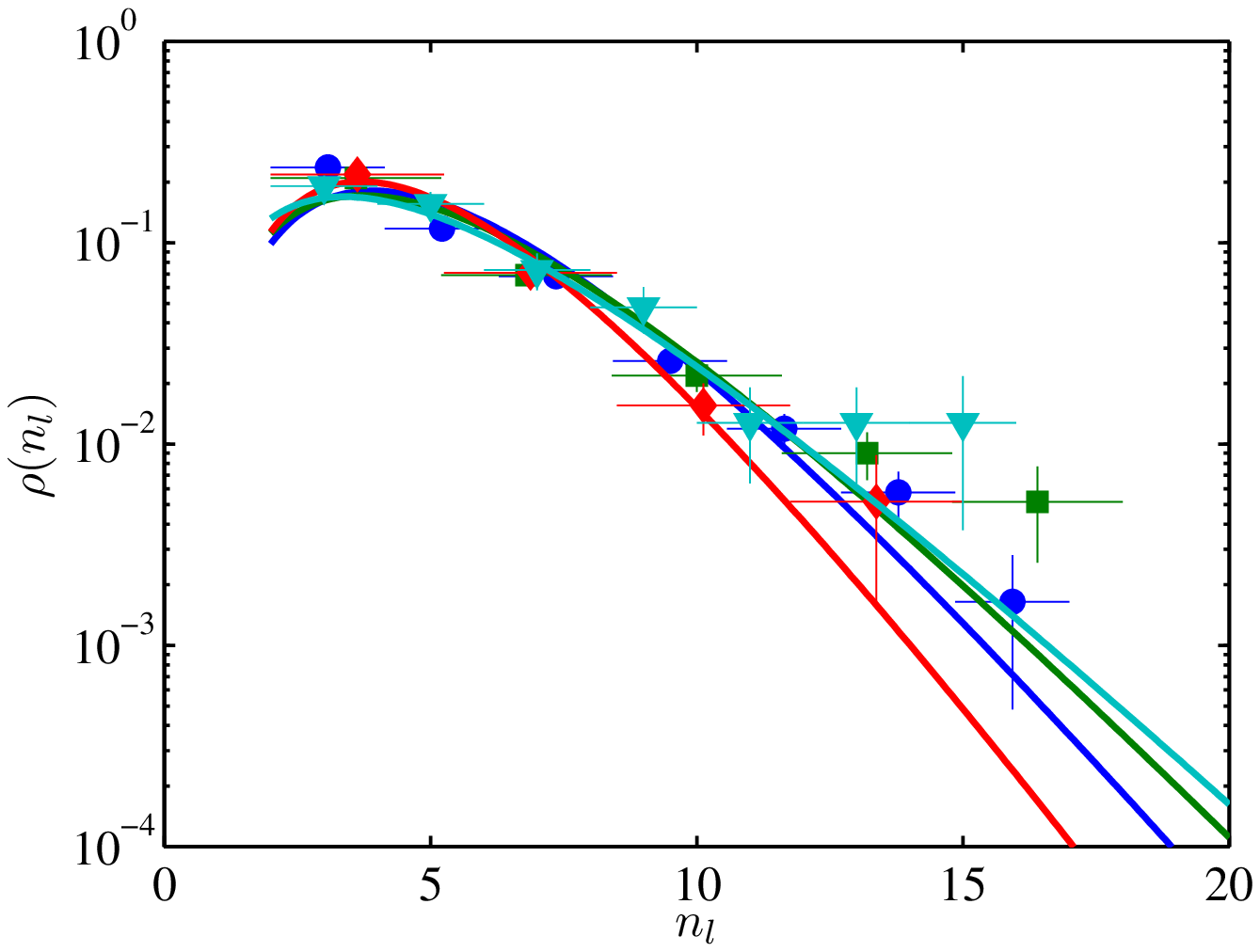}
\caption{{\it {\bf Experiments.} {\bf a}) Distribution of the area~$s$ of the voids.  Continuous lines correspond to fits by log-normal distributions.  {\bf b}) Distribution of the aspect ratio~$\rho(\sqrt{s}/p)$ where~$p$ is the perimeter of the voids.  It is well described by a Gaussian distribution.  The values $0.28$ for a circular shape and $0.22$ for a pinched closed elastic shape are drawn as vertical lines.  {\bf c})~Distribution of the number of links~$n_l$ surrounding a void.}}
\label{topolo_micro}
\end{figure*}

We focus here on the properties of the elements of tiling of the total available surface, such as their perimeter, surface and shape. Note that these elements correspond not only to loops (defined as an elementary ``closed'' part of the rod in~\cite{donato03}), but also to the voids that are delimited by such touching loops.  Does the spatial tiling of the available surface contain any information on how the folding might have happenned?

\paragraph*{{\bf Experiments.}} 

We found that the shapes as well as the sizes of the voids show a wide variability over~$2$ or~$3$ orders of magnitude.   The distribution of the area~$s$ of the voids shown in Fig.~\ref{topolo_micro}{\bf a} is well described by log-normal distributions.  The mean value $\left\langle s\right\rangle$, reported in Table~\ref{tab_macro}, decreases with the compaction rate.  The shape  of the voids can be characterized by the ratio of the square root of its surface over its perimeter $\sqrt{s}/p$, as done in~\cite{donato03}.  This aspect ratio is shown in Fig.~\ref{topolo_micro}{\bf b}.  The distributions are well described by an unique gaussian distribution independent on the control parameters:
\begin{eqnarray}
\rho_N^{\,\mu,\sigma}(x)= \frac{1}{\sqrt{2\pi}\,\sigma}\exp\left(-\frac{(x-\mu)^2}{2\sigma^2}\right) \text{ . } \label{eq_gaus}
\end{eqnarray}  
with~$\mu = 0.19$ and~$\sigma^2 = 14\ 10^{-4}$.  One can also look at the number of junction points belonging to each void contour, that is also the number of edges in terms of multi-branched stacks surrounding the void.  One should note that a junction point that belongs to a void contour does not correspond necessarily to a vertex where geometric properties (local tangent or local curvature) of the two adjacent stacks change discontinuously but rather to a change of the number of branches $n_{br}$ in the stacks.  The voids do not have a constant number of edges $n_l$ as shown by the distribution in Fig.~\ref{topolo_micro}{\bf c}.  The paucity of our data does not allow us to discriminate between Gamma, Log-normal or exponential distributions as they all describe well the data~$n_l\geq4$.

\begin{figure*} 
\centerline{
\includegraphics[width=0.32\linewidth]{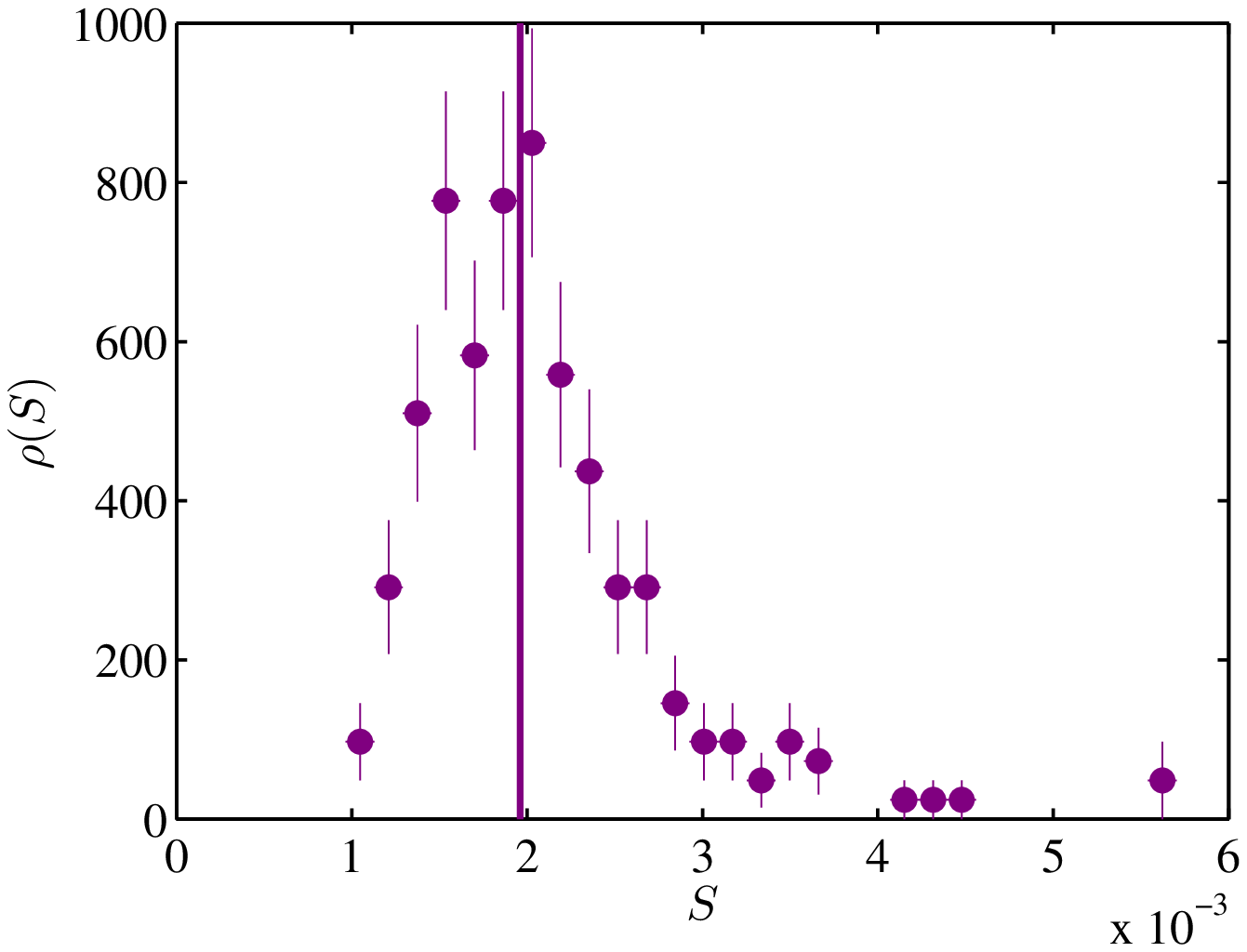} \hfill
\includegraphics[width=0.32\linewidth]{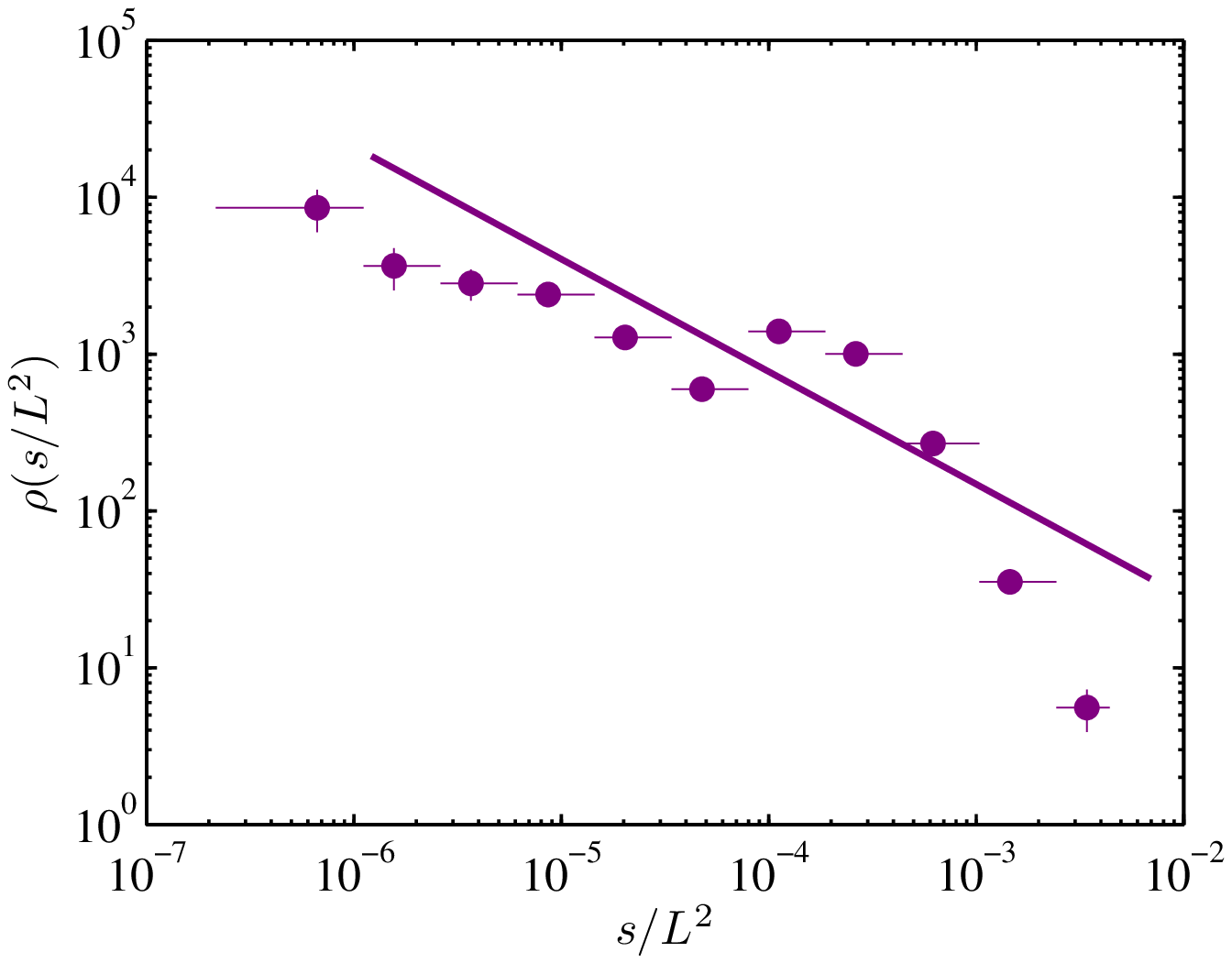} \hfill    
\includegraphics[width=0.32\linewidth]{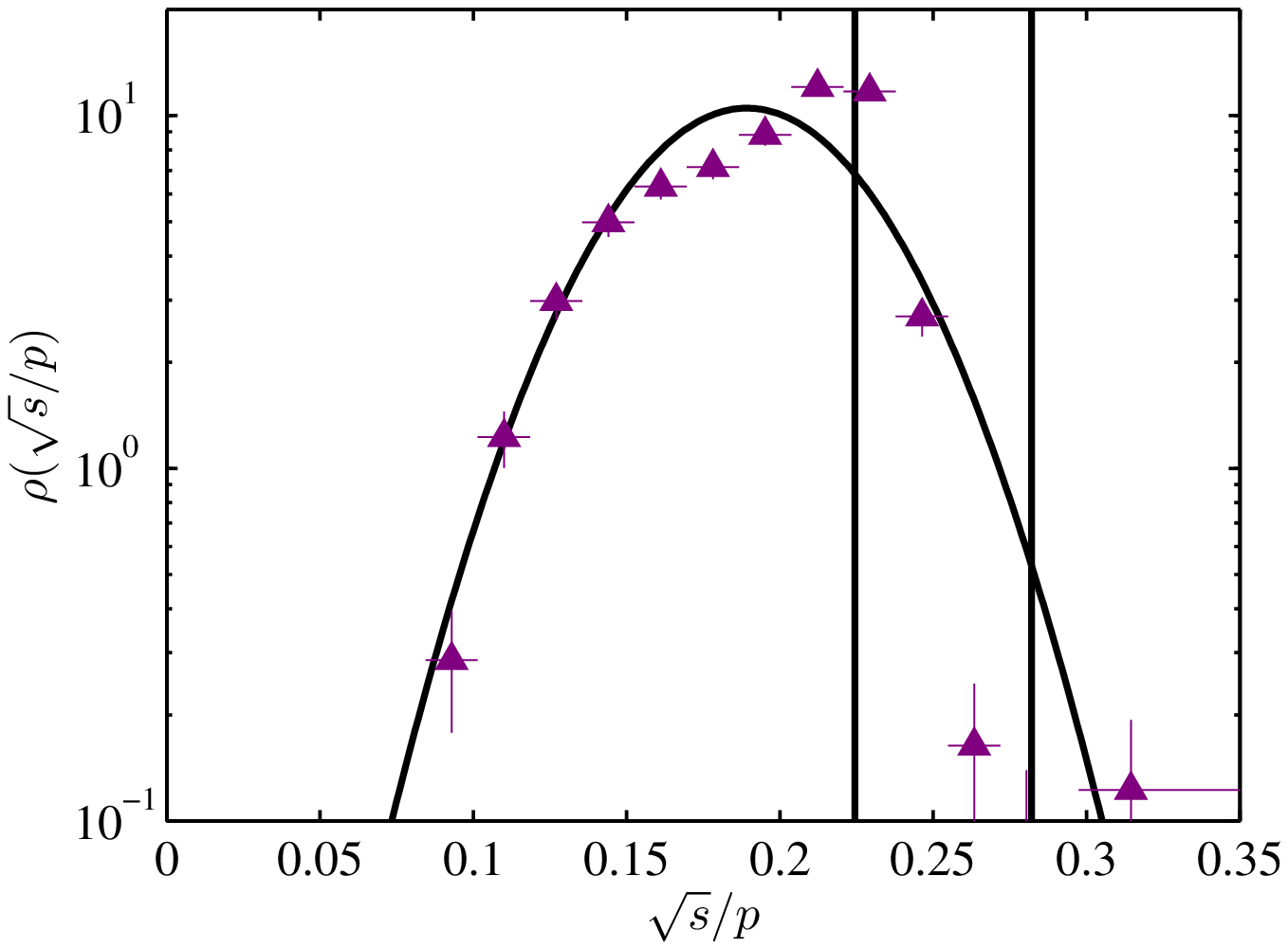} 
}
\caption{{\it {\bf Simulations.} {\bf a}) Distribution of the surface area $S$ occupied by the rod. {\bf b}) Distribution of the area $s$ of the elemtary voids. {\bf c}) Distribution of the aspect ratio~$\rho(\sqrt{s}/p)$ where~$p$ is the perimeter of the voids.  It is well described by a Gaussian distribution.  The values $0.28$ for a circular shape and $0.22$ for a pinched closed elastic shape are drawn as vertical lines.  These statistics are based on $1,435$ elementary~voids.}}
\label{topolo_micronum}
\end{figure*}

\paragraph*{{\bf Simulations.}} 

As already discussed, the simulations are not realized at constant available surface.  We defined the surface $S$ occupied by the open rod as the sum of surfaces of elementary voids; this distribution is shown in Fig.~\ref{topolo_micronum}{\bf a}.  The pdf~$\rho(S)$ is peaked around an average value of~$\left\langle S\right\rangle=4\ 10^{-3}L^2$ that has the same magnitude as that expected for a perfectly circular shape~$\pi \left\langle R_g\right\rangle^2=8\ 10^{-3}L^2$.  On the other hand, the distribution $\rho(s)$ of the surface of the elementary voids changes over~$3$ orders of magnitude and is closer to a power law distribution as shown on Fig.~\ref{topolo_micronum}{\bf b}.  The shape of this distribution is unchanged regardless of whether the surfaces of the voids are normalized by the occupied surface~$S$ or by the mean void surface~$\overline{s}$ of the configuration.  Fig.~\ref{topolo_micronum}{\bf c} shows the pdf of the aspect ratio $\sqrt{s}/p$ of voids: it is widely distributed of mean equal to~$0.17$, and well described by the same gaussian distribution than for the experiments. 

\subsection{Summary }

To summarize, we observed the difference of values of~$N_l$ and~$N_{br}$ between experiments ($\sim50$ and $\sim200$) respectively and simulations ($\sim10$ and $\sim20$) due to the different compaction rates~$\varepsilon$ ($\sim10$ and $\sim2$).  The number of elementary units growing with the compaction rate underlies the increasing complexity of the configurations.  Several predictions (Eqs.~(\ref{eq1})-(\ref{eq4})) are obtained on the relative variations of~$N_l$ and~$N_{br}$ either from geometrical relations or from observations at the scale of multi-branched stacks.  They both appear to be validated by the experimental data.  Eqs.~(\ref{eq1})-(\ref{eq2}) result of the independence of the ratio  $\sqrt{\overline{s}}/\overline{l}$ on the configuration, the compaction rate and the size of the sheet (Fig.~\ref{topolo_macro}{\bf b}): there is still an issue with the understanding of this constant value.  Note the large range and values accessible by $n_{br}$ (Fig.~\ref{topolo_macro}{\bf c}), the number of branches in close contact forming a multi-branched stack: it would be interesting to study what controls the value of $n_{br}$ inside one multi-branched stack.  However, whatever the degree of complexity of the folded configurations, some statistical characteristics stay unchanged, namely the shape ratio and the number of edges of the elementary voids.  The aspect ratio of the voids also appears to be independent on the compaction parameters $\varepsilon$ (varying from $2$ to $20$ between simulations and experiments) and on the compaction potential (hard-wall or quadratic).  The average value of the aspect ratio of the voids is $0.18$.  A perfectly circular shape provides an upper bound of $\sqrt{s}/p=0.28$.   For comparison, this ratio is $\sqrt{s}/p=0.22$ for a clamped elastic rod as can easily be shown by solving the Euler's elastica equations.  The values we report here can be explained by considering that the loops are confined:  their perimeter will not change but the area will be reduced due to the compression.  The experiments reported in~\cite{donato03} show values $\sqrt{s}/p\simeq0.17$-0.$20$, very close to~$0.22$, consistent with their injection method that tends to create a layered loops geometry.  Finally, the distributions of voids surface have been found to be Log-normal in the experiments.  This hints that voids are generated through a hierarchical process.  A succession of bifurcations such as the one predicted in~\cite{boue06} leads to folding events that successively ``break'' voids.  A new fragmentation event concerns all the previously existing fragments.  However, this pdf is closer to a power law distribution in the case of the numerical simulations.  Due to the small range of accessible compaction ratio, it is difficult to assess the relevance of this observation.

\begin{table*} 
\begin{center}  
\begin{tabular}{l||l|l|l|l|l|l|l|l} 
& $\left\langle \kappa_m\right\rangle L$& $\left\langle \ell\right\rangle/L$& $\langle \tilde{E}\rangle$& $\left\langle e\right\rangle L$& $\alpha_e$& $\chi_e L$  &$\alpha_E$&$\chi_EL$\\ \hline
$1$& 330& 2.5$\,10^{-3}$& 13$\,10^4$& 330& 0.16& 2000&10.8&1.2 $10^4$ \\  
$2$& 260& 3.4$\,10^{-3}$& 11$\,10^4$& 340& 0.19& 1800&12.7&8.3 $10^3$ \\  
$3$& 100& 10$\,10^{-3}$& 1.3$\,10^4$& 130& 0.41& 330&15.9&8.4 $10^2$ \\    
$4$& 200& 4.4$\,10^{-3}$& 7.3$\,10^4$& 220& 0.31& 720 &5.72&1.3 $10^4$\\ \hline
$5$& 37 & 0.05& 0.43$\,10^4$& 230& 0.31& 730&61 & 70
\end{tabular}    
\end{center}  
\caption{{\it ``Microscopic'' properties averaged over experiments from the same set. Geometry: mean curvature  $\kappa_m$ and length of branches $\ell$. Energy: Total energy $ \tilde{E}$, energy of branches $e$, and parameters of the Gamma distributions  $\alpha_e$ and $\chi_e$ (resp. $\alpha_E$ and $\chi_E$) for branches energy (resp. total energy), according to Eq.~(\ref{eq_gam}).}}
\label{tab_micro}   
\end{table*}  

\section{Geometrical properties of branches}
\label{sec_geometro}

In the following, we will characterize the geometry of the folded rod at two scales.  On the one hand, we will investigate the geometry of branches, namely theirs lengths and mean curvatures.  On the other hand, we will look at the local curvature along the rod.  However, let us first define two sub-systems in the experimental set-up.

\begin{figure*} 
\centerline{   
\includegraphics[width=0.2\linewidth]{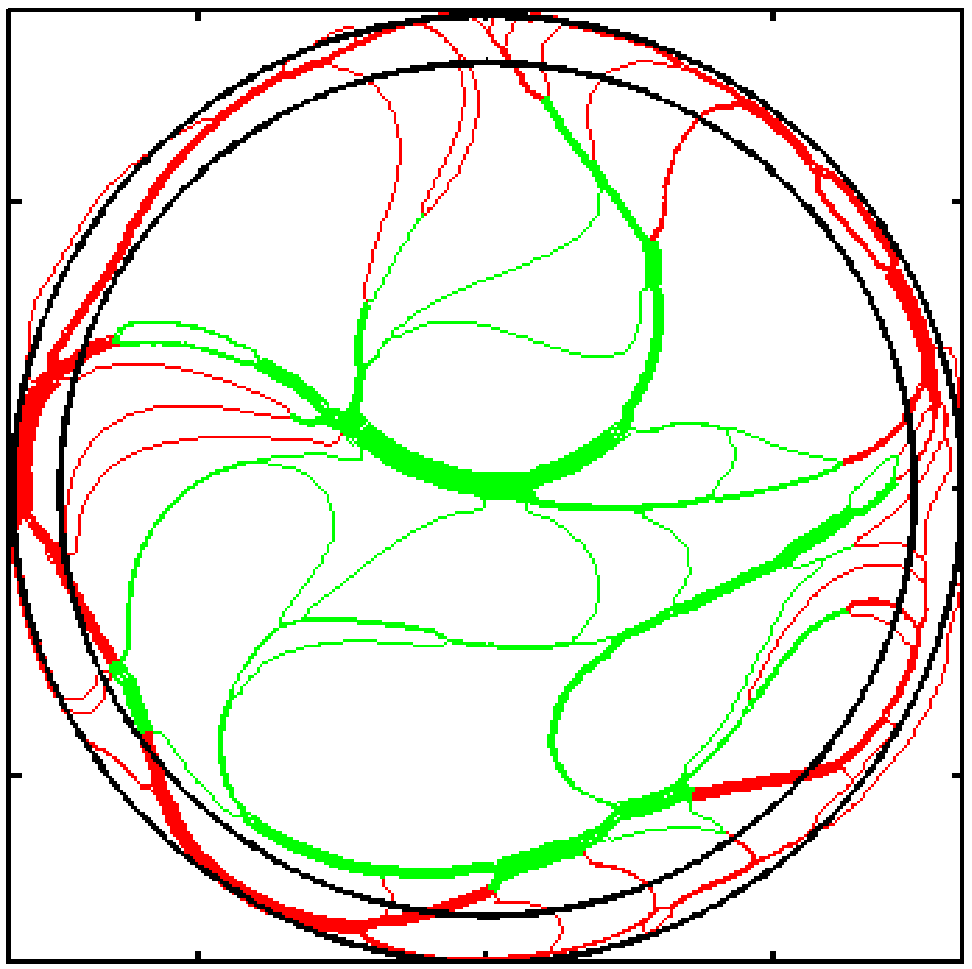}
\includegraphics[width=0.2\linewidth]{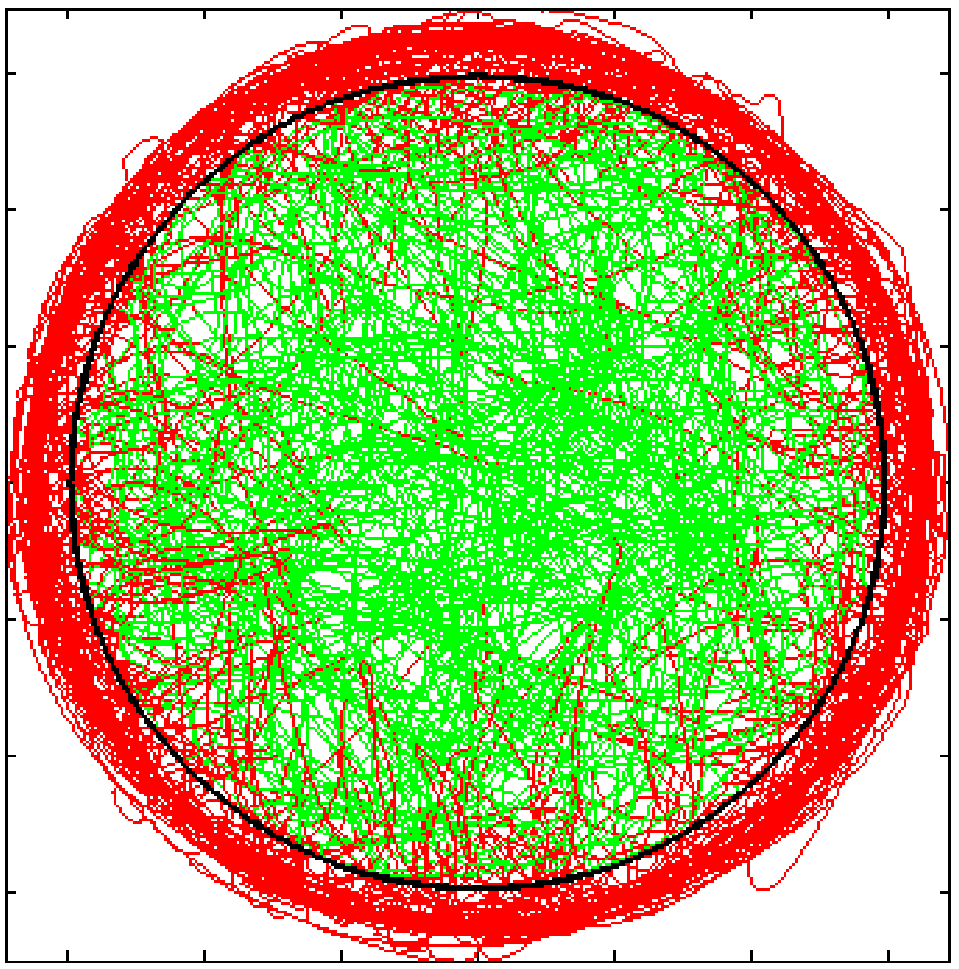}    }
\caption{{\it {\bf Experiments.} Definition of two sub-systems. {\bf a}) Branches that have one of their extremities in direct contact with the container wall are colored in red to separate them from the other branches that form the bulk of the system in green. {\bf b})~Super-imposition of the two sub-systems for experiments of set~$1$.  Note that because there is no well defined boundaries in the numerical simulations, this subdivision is unique to the experimental system.}}
 \label{fig_expbulkperiph}
\end{figure*}

\subsection{Two sub-systems in the experiments. }

The branches composing the folded rod are differentiated into two sub-systems: branches \emph{with} (resp. \emph{without}) an extremity in contact with the container, which we will refer to as \emph{periphery} (resp. \emph{bulk}).  Fig.~\ref{fig_expbulkperiph} shows branches in the bulk and in the periphery on one example of configuration and for all super-imposed configurations from set~$1$.  Statistical properties of branches can be studied either in the whole system, or separately between the periphery and bulk sub-systems.  We found that there is systematically more branches located at the periphery than in the bulk.  On average, the periphery (resp. the bulk) is composed of~$60\%$ (resp.~$40\%$) of branches.  Defining branches in the periphery as the ones for which both extremities are in contact with the wall would have resulted in a slightly different outcome.  However, our choice of definition allows to prevent any boundary effect on the bulk properties.  For example, we found that the average number of branches per stack is the same in the bulk and periphery; thus it is uniform in the whole system.

\subsection{Length of branches }

\begin{figure*} 
\centerline{  
\includegraphics[width=0.32\linewidth]{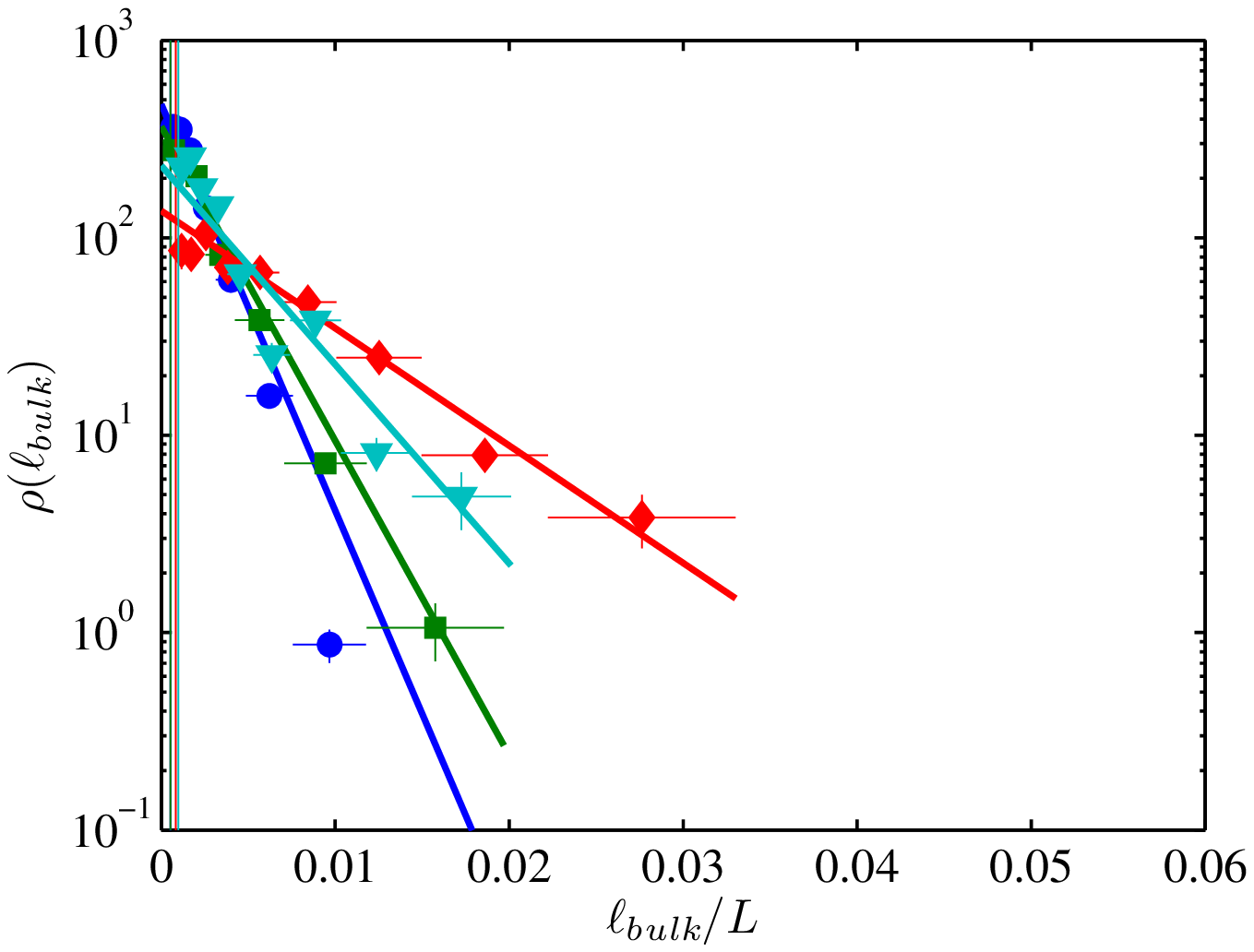} \hfill
\includegraphics[width=0.32\linewidth]{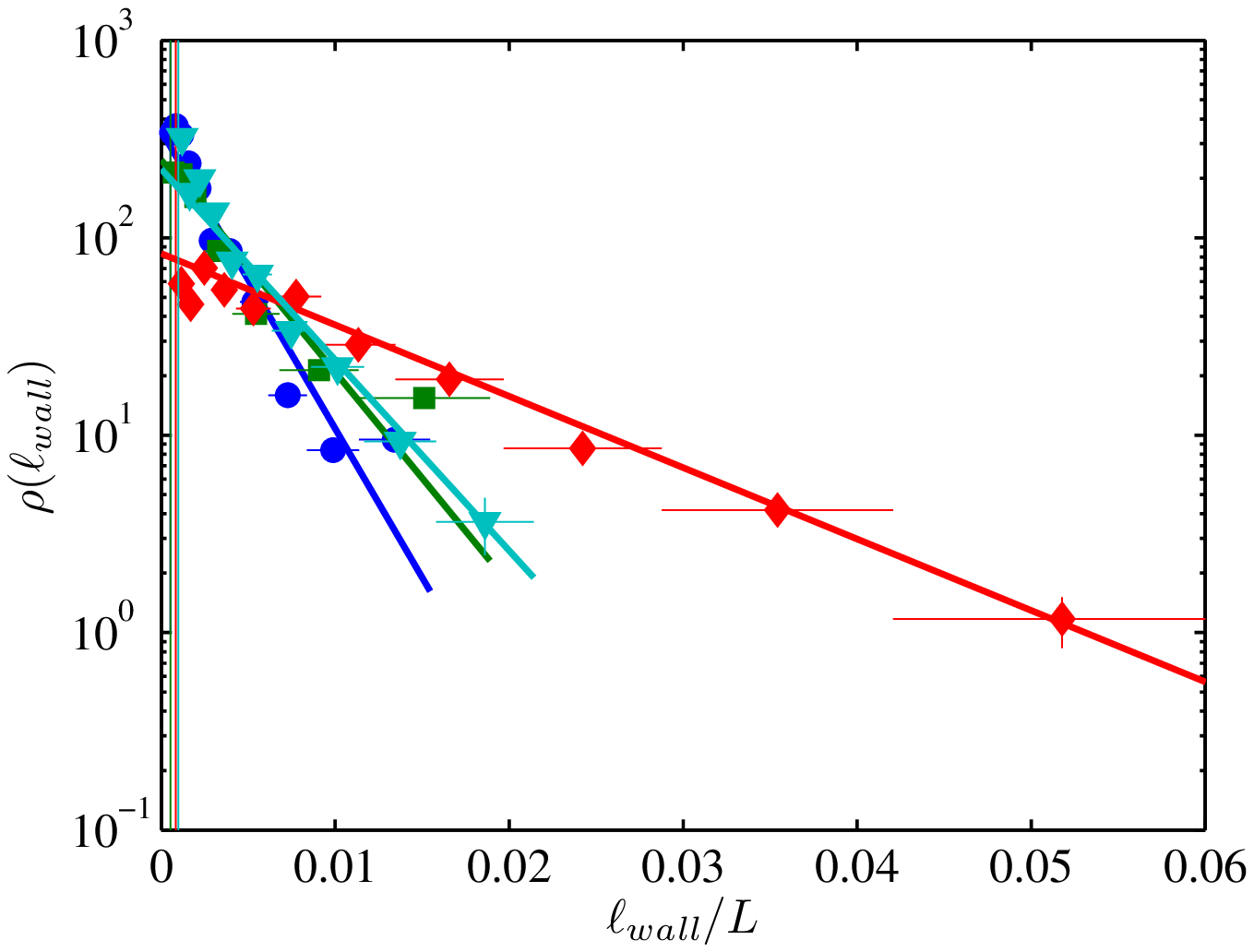} \hfill
\includegraphics[width=0.32\linewidth]{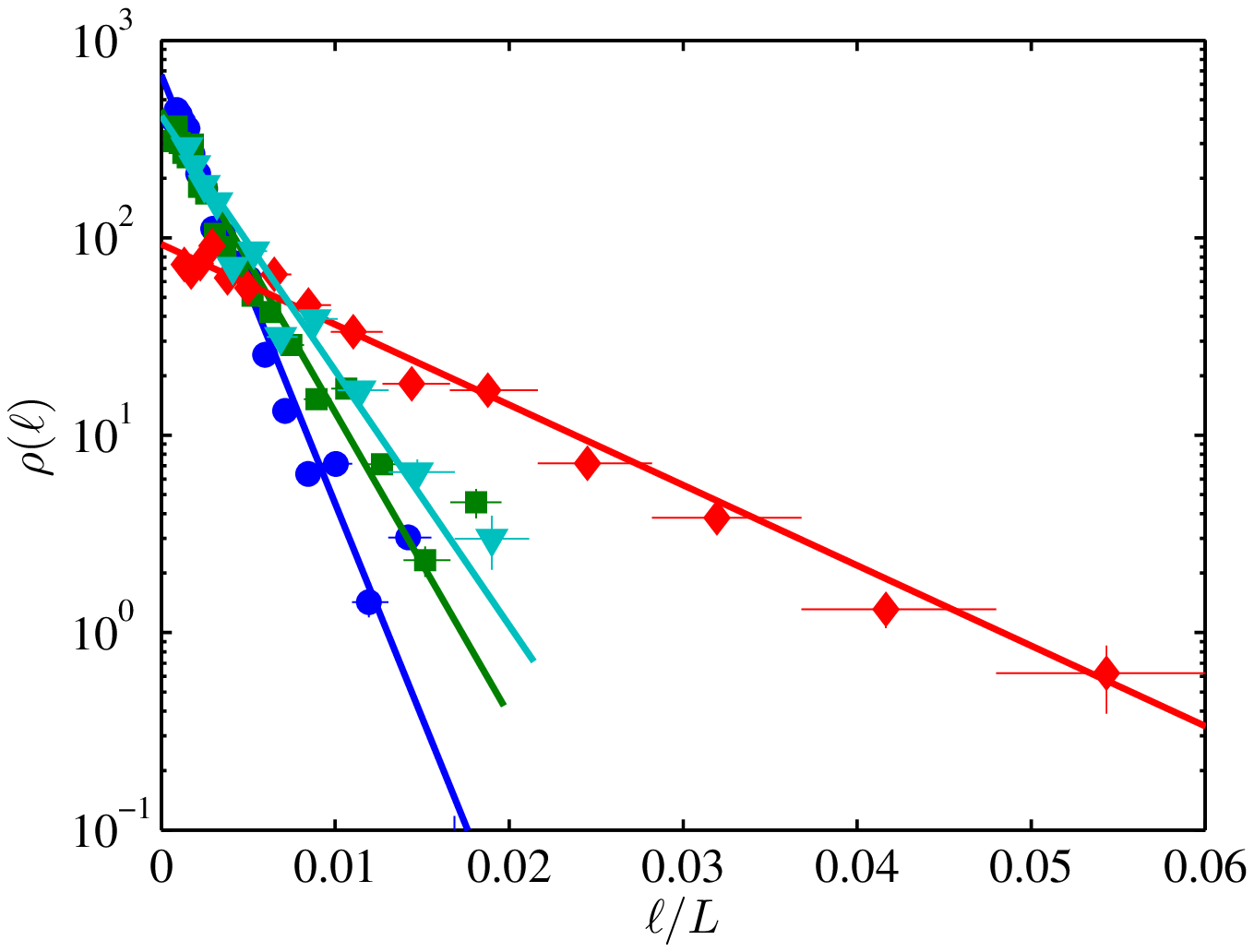} 
}
\caption{{\it {\bf Experiments.} Geometrical properties. Distribution of the length $\ell$ of branches {\bf a)}:  In the bulk sub-system, and exponential distributions of means $\mu=2.1\,10^{-3}$, $2.7\,10^{-3}$, $7.3\,10^{-3}$ and $4.3\,10^{-3}$  for sets $1$-$4$. {\bf b)}: In the periphery sub-system, with means~$\mu=2.9\,10^{-3}$, $4.0\,10^{-3}$, $12\,10^{-3}$ and $4.5\,10^{-3}$. {\bf c)}: For all branches in the whole system, with means~$\mu=2.5\,10^{-3}$, $3.4\,10^{-3}$, $10\,10^{-3}$ and $4.4\,10^{-3}$.  Vertical lines show the minimal length $\ell_{min}=1$~mm experimentally detected.}}
\label{geometro}
\end{figure*}

\paragraph*{{\bf Experiments.}}

The distribution of the length of branches is found to be exponentially distributed in both sub-systems as shown in Fig.~\ref{geometro}{\bf a}-{\bf b}.  It appears that the mean length is slightly larger for branches at the periphery than in the bulk.  The ratio of mean lengths between the two sub-systems $\left\langle \ell_{periph}\right\rangle/\left\langle \ell_{bulk}\right\rangle$ is between $1$ and $1.6$ depending on the set of experiments.  Using the fact that approximately $40\%$ (resp. $60\%$) of the branches are located in the bulk (resp. at the periphery), the mean length in the whole system verifies $\left\langle \ell\right\rangle\simeq0.4\left\langle \ell_{bulk}\right\rangle+0.6\left\langle \ell_{periph}\right\rangle$.  The resulting pdf in the whole system is therefore exponential as shown in Fig.~\ref{geometro}{\bf c}.  Quantitatively, the average length of branches $\left\langle \ell\right\rangle$, reported in Table~\ref{tab_micro}, decreases with the compaction rate. The values for branches in the sub-systems are written in the legend of Fig.~\ref{geometro}.

Note that the existence of a minimal value $\ell_{min}$, related to experimental detection limits, may introduce a small shift of the average $\left\langle \ell\right\rangle$ towards larger values, in comparison with the characteristic decreasing exponential length $\mu$: $\left\langle \ell\right\rangle\geq\mu$. Strictly, experimental biased data are distributed according to: $\left\langle \ell\right\rangle\exp(-\ell/\mu)/\mu^2$. However, as the  value  $\ell_{min}$ is very small in comparison with $\left\langle \ell\right\rangle$, this later well approximates $\mu$.  This method has the advantage to be insensitive to the choice of bins for the construction of the pdf, contrary to a simple minded fit.
 
\begin{figure*} 
\centerline{
 \includegraphics[width=0.64\linewidth]{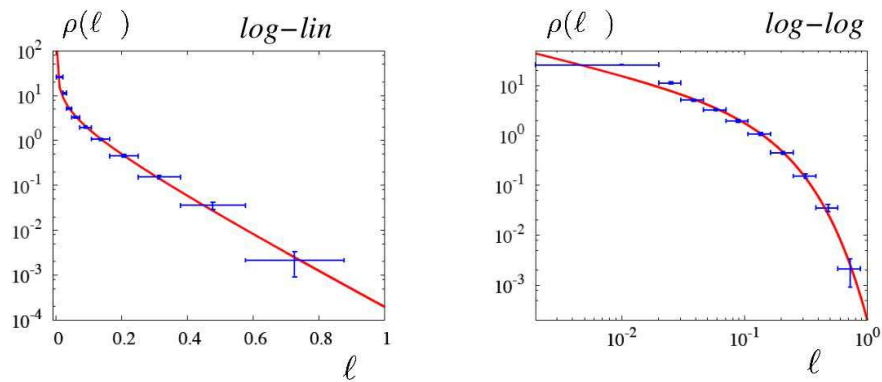} 
}
\caption{{\it {\bf Simulations.} Geometrical properties.  Probability distribution $\rho(\ell)$ of the length of branches $\ell$ in log-lin and log-log scales.  The continuous line represents a Gamma-law with parameters:~$\alpha=0.4$ and~$\chi=0.12$.  Control parameters are:~$\Lambda = 7\,10^5$, $D=750$ and there are~$4676$ branches.}}
 \label{fig_longsimu}
\end{figure*}

\paragraph*{{\bf Simulations.}}  The distribution of the length of the branches follows a Gamma law (Eq.~(\ref{eq_gam})) with shape parameter $\alpha$ smaller than~1: $\alpha=0.4$ and~$\chi=0.12$ as shown in Fig.~\ref{fig_longsimu}.  This observation is important because it indicates an accumulation of branches of small lengths, in comparison with the case of a pure exponential pdf which was observed in the experiments.

\subsection{Curvature of branches}

Several definitions are possible for the characterization of the branch curvature: 
\begin{equation}
 \kappa_{\pm} = \int_{0}^{\ell} \kappa(s)   \mbox{d}s /\ell \ \text{, } \ 
 \kappa_{+}  =  \int_{0}^{\ell} |\kappa(s)| \mbox{d}s /\ell \ \text{, and } \ 
 \kappa_{m}  = |\kappa_{\pm}| \text{ , }
\end{equation}
with $\kappa(s)=\frac{\mbox{d}^2 {\bf R}}{\mbox{d}s^2}$, the local curvature along the branch at  curvilinear abscissa $s$. When a branch does not exhibit an inflexion point the three definitions above are equivalent, with $\kappa_{\pm}$ having possibly a different sign. The statistical study of mean curvature of branches $\kappa_m$ allows for a representative sampling of all of the values of local curvature $\kappa(s)$.

\begin{figure*}   
\includegraphics[width=0.32\linewidth]{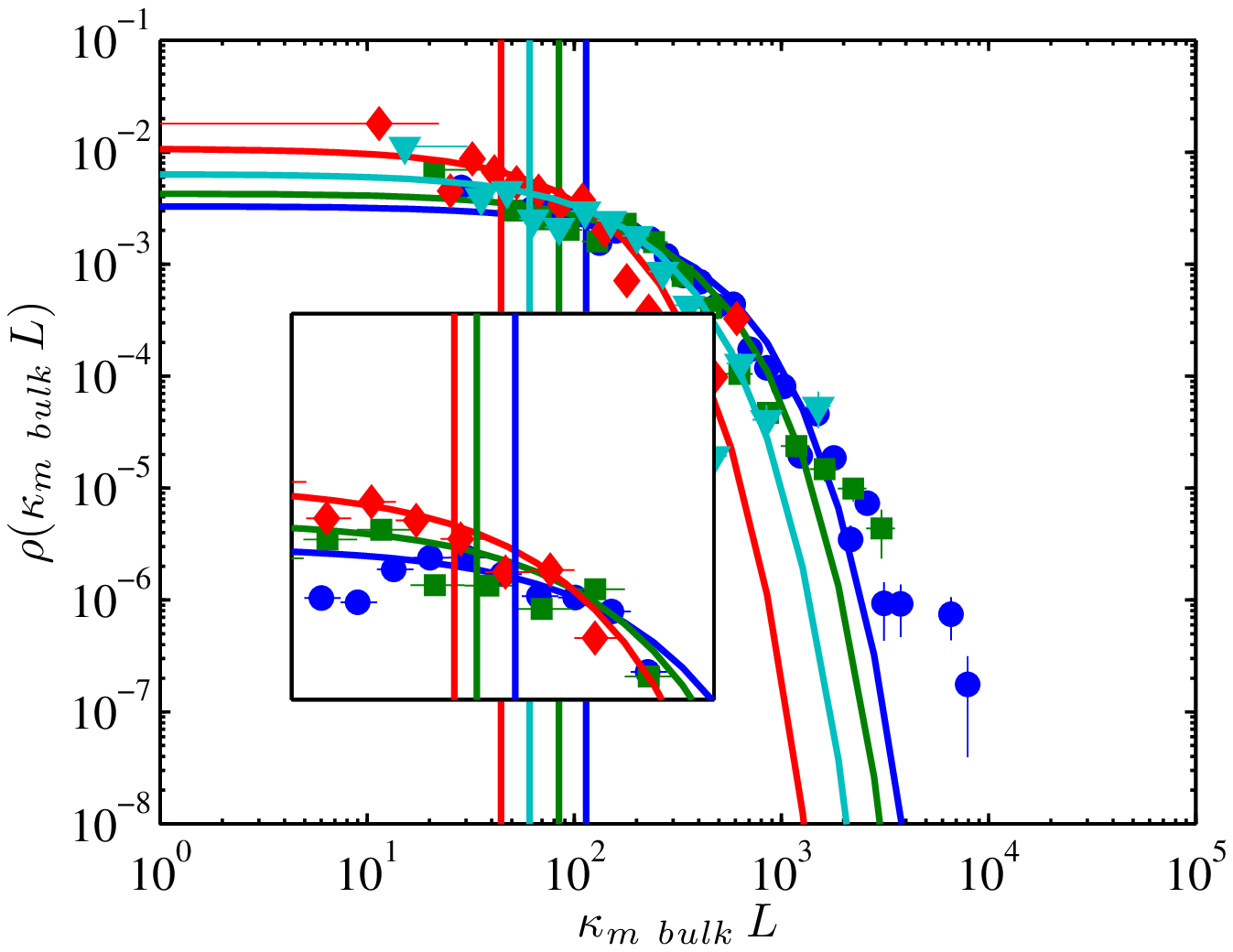} \hfill
\includegraphics[width=0.32\linewidth]{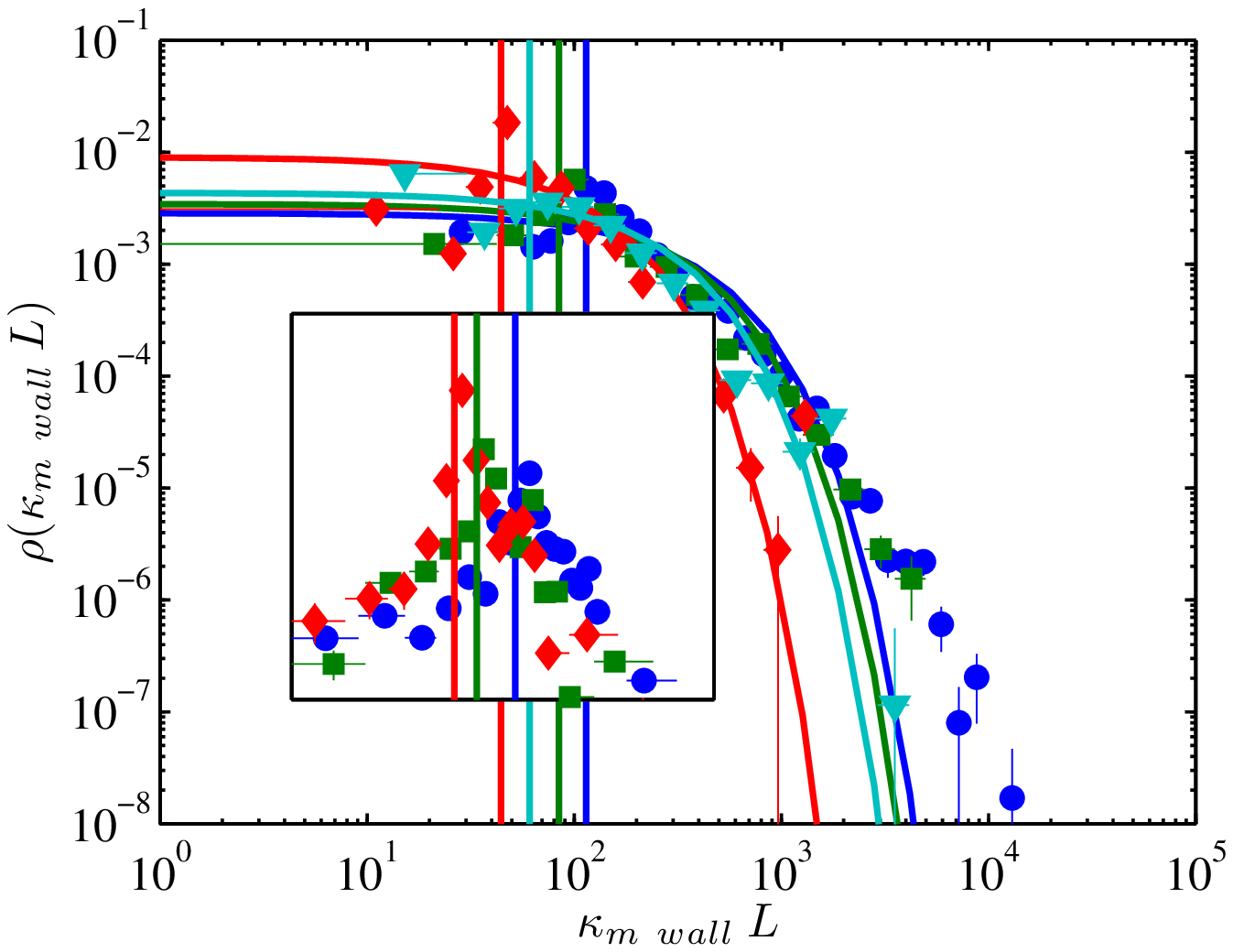} \hfill
\includegraphics[width=0.32\linewidth]{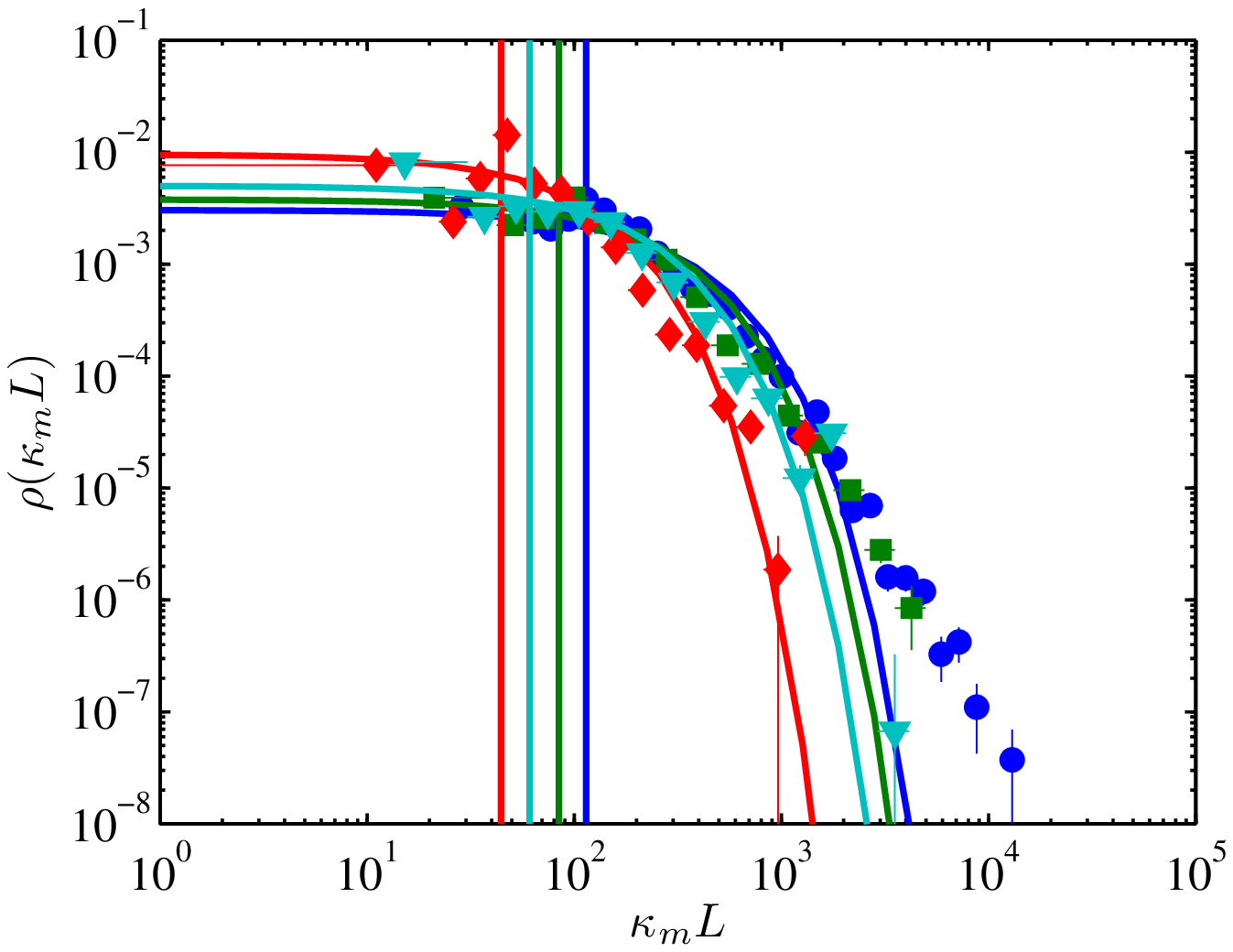} 
\caption{{\it {\bf Experiments.} Geometrical properties. Probability distributions $\rho(\kappa_m)$ of mean curvature of branches $\kappa_m$ for sets 1-4. {\bf a}: In the bulk sub-system and exponential distributions of mean $\left\langle \kappa_{m}\right\rangle=304$, $230$, $93$, $160$. {\bf b}: In the periphery sub-system: $\left\langle \kappa_{m}\right\rangle=350$, $290$, $110$, $230$. {\bf c}: For all branches in the whole system: $\left\langle \kappa_{m}\right\rangle=330$, $260$, $100$, $200$. The exponential pdf in {\bf b-c}Êdo not have the same mean than the data, but comes from {\bf a}.}}
\label{fig_courbexp2}
\end{figure*}

\paragraph*{{\bf Experiments.}}

We found that the whole rod has an increasing number of inflexion points as~$\varepsilon$ increases.  However, because inflexion points ususally coincide with junction points connecting adjacent branches, the local curvature does not change change sign at the branch level.  Indeed, only less than~$1\%$ of the branches have an inflexion point where $\kappa_+\neq \kappa_m$. Thus, the statistical distributions of $\kappa_{\pm}$ appear to be exactly symmetric around $0$, so that the pdf of $|\kappa_{\pm}|$ is equivalent to that of $\kappa_{m}$.  The pdf of the mean curvatures in the bulk are approximately distributed according to exponential laws as shown in Fig.~\ref{fig_courbexp2}{\bf a}.  By contrast, the same pdf in the periphery is characterized by a symmetric and sharp peak around the curvature of the hole~$\kappa_h$ which is clearly visible in the inset of Fig.~\ref{fig_courbexp2}{\bf b}.  The presence of this peak means that, even though both pdf's are well described by exponential laws, their means are significantly different.  As a result, the pdf $\rho(\kappa_m)$ of mean curvatures $\kappa_m$ for all branches in the whole system, verifying $\rho\simeq0.4\rho_{bulk}+0.6\rho_{periph}$, is also qualitatively a global  exponential law plus a symmetric peak around the hole curvature $\kappa_h$ (Fig.~\ref{fig_courbexp2}{\bf c}).

In addition, we found that lengths $\ell$ and mean curvature radius $\kappa_m^{-1}$ are slightly correlated (linear correlation coefficient $r\simeq0.45$).  Note that describing the branches as circles arcs, of perimeter smaller than for the corresponding whole circle, gives the inequality $\ell\leq2\pi/\kappa_m$, that is experimentally verified.

\paragraph*{{\bf Simulations.}} 

\begin{figure*} 
\centerline{
\includegraphics[width=0.32\linewidth]{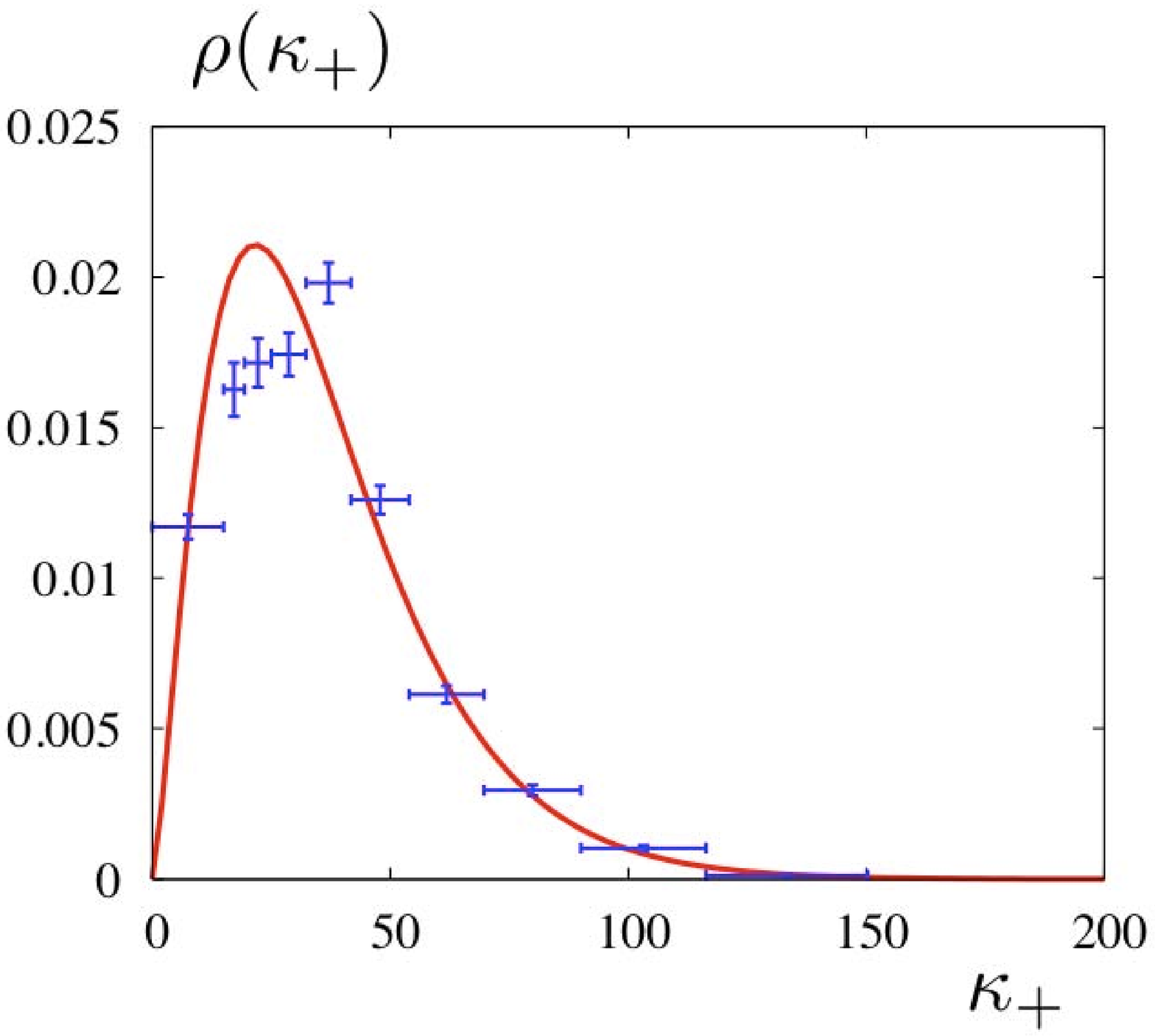} 
\includegraphics[width=0.32\linewidth]{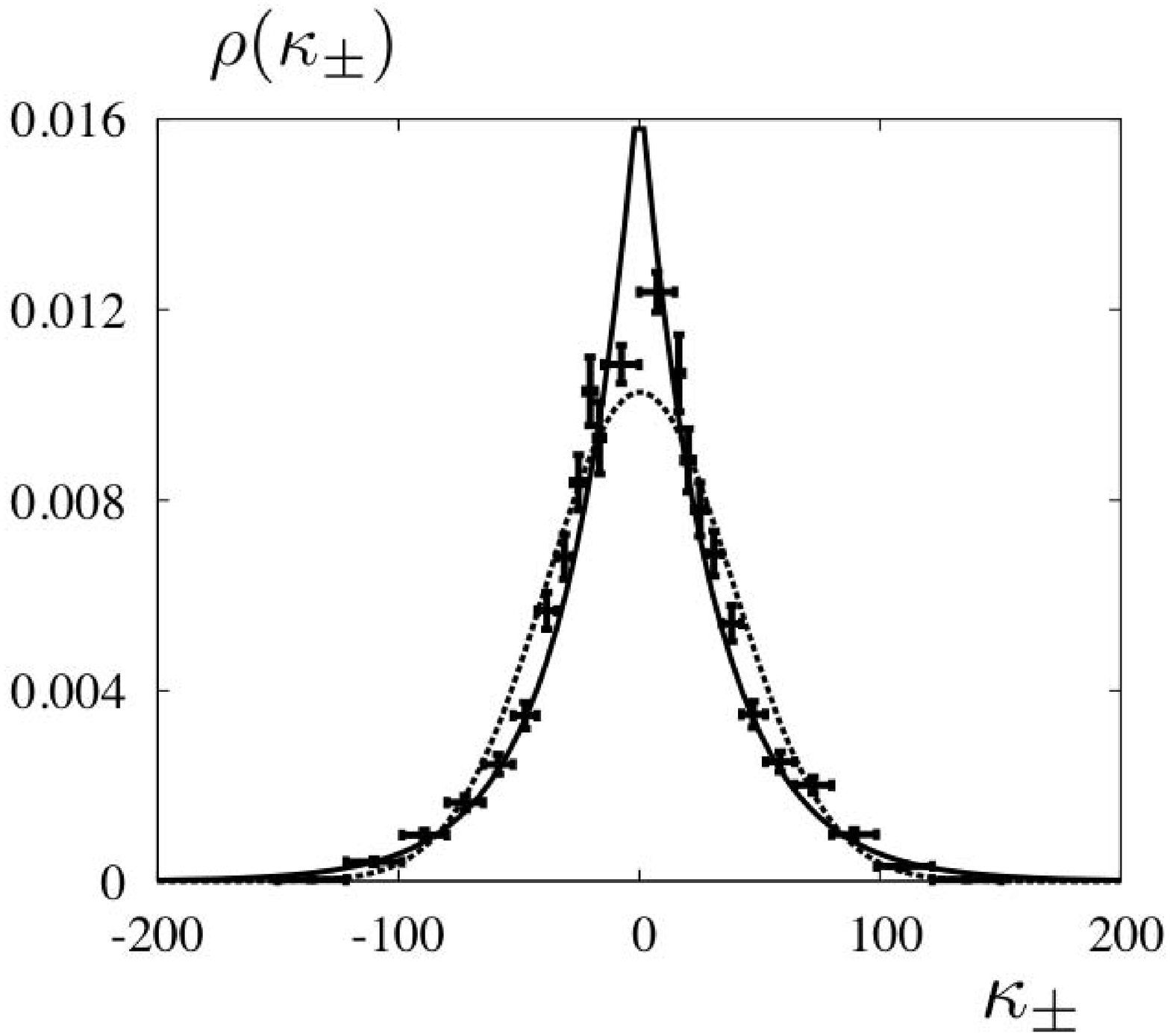} 
}
\caption{{\it {\bf Simulations.} Geometrical properties.  {\bf a}: Probability distribution of the absolute curvature per branch $\kappa_+$ and Gamma distribution.  Parameters for the Gamma-law are:~$\alpha=2.5$ and~$\chi=14.8$.  {\bf b}: Probability distribution of the curvature of branches $\kappa_{\pm}$ with a normal distribution (dashed line) with:~$\mu=0$ and~$\sigma = 39$, and an  exponential distribution (continuous line) with~$\mu = 30$. Statistical test:~$\chi^2=93$ for the exponential distribution and~$160$ for the normal law.  Data is averaged over 4,676 branches.}}
\label{pdf_num_abs_curv}
\end{figure*}

The distribution of~$\kappa_+$ is well described by a Gamma distribution with parameters~$\alpha=2.5$ and~$\chi=14.8$, as shown in Fig.~\ref{pdf_num_abs_curv}{\bf a}.  The average over all configurations is $\left\langle \kappa_+\right\rangle=\alpha\chi=37/L$.  As the linear scales plot shows (and corroborated by the shape parameter~$\alpha$ greater than~1), there is a power law drop-off for small values.  Otherwise, away from the peak the probability falls off with an exponential behaviour.  This peak is interpreted as the result of ``effective'' walls located at distance $R_g$ from the center which leads to an accumulation of curvature at~$\left\langle k_+\right\rangle=37 \simeq1/R_g= 20$.

As the compaction ratio is much smaller than in the experiments, many branches have inflexion points in the numerical simulations.  Therefore~$\kappa_+$ and $\kappa_\pm$ have very different statistical distributions.  Note that whereas $\kappa_+$ gives direct information on the curved state of the considered branch, $\kappa_\pm$ alone is not obvious to interpret.  A branch can have a small value of $\kappa_\pm$, because it is either straight or because it contains an inflexion point.  Therefore, a branch of curvature $k_+$ is bounded by $|\kappa_\pm|\leq \kappa_+$.  Fig.~\ref{pdf_num_abs_curv}{\bf b} shows the pdf of $\kappa_\pm$.  While it is symmetric around~$0$, it can equally be fitted by an exponential distibution of mean $\mu=30$, or by a gaussian distribution of variance~$\sigma=39$ (and mean~$0$).

\subsection{Summary }

We interpret the emergence of exponential distributions for the branches length $\ell$ (pure exponential in the experiments and just an exponential tail in the numerics) as the random division of the whole rod.  The length averages can be related to the number of branches.  For each configuration we have~$\overline{\ell}=L/N_{br}$.  One expects that this is still approximately valid for averages over all experiments of a same set, leading to $\left\langle \ell\right\rangle/L=\left\langle N_{br}\right\rangle^{-1}$ (non-dimensionalized length).  Thus the larger number of branches $N_{br}$ observed systematically for larger compaction rates $\varepsilon$, explains the different quantitative exponential distributions. Another indication of the stochastic formation of the branches comes from the well representation of curvature statistics by the sampling made on branches. One common feature to experiments and simulations on branches curvatures statistics is their exponential distributions   --pure exponential in bulk for experiments and for $\kappa_\pm$ in simulations, or exponential tail for $\kappa_+$ in simulations--. These exponential distributions of 2d curvatures can be compared with the exponential distributions found in~\cite{blair05}, for 3d curvatures in unfolded crumpled sheets of paper. 

After having looked at the geometrical properties, we will now turn our attention to the energetical characteristics of both experimental and numerical systems.

\section{Energetical properties}
\label{sec_energo}

\subsection{Expression of energy }

In experiments, because of the self-similar conical shape of the folded sheet, related to pure bending deformations, its elastic energy  is given by:
\begin{equation}
E_s = \frac{BZ}{2} \ln \left( \frac{r}{R_c} \right) \int_0^{L} \left( \frac{\mbox{d}^2 {\bf R}}{\mbox{d}s^2} \right)^2 \mbox{d} s , 
\label{en_cone}
\end{equation}
with on the one hand, the bending energy $B$, a logarithmic prefactor based on the ratio of the sheet radius $r$ and a cut-off length $R_c\simeq 10$~mm,  and on the other hand,  ${\bf R}$ and $L$ the position of the rod and its length, observed in the cross-section at distance~$Z$ from the cone tip.   By rescaling~$E_s$ by the characteristic length scale due to the conical shape, $Z \ln \left( r/R_c \right)$, we obtain the energy of the rod per unit of transversal length:
\begin{equation}
 E_r = \frac{B}{2} \int_0^L \left( \frac{\mbox{d}^2 {\bf R}}{\mbox{d}s^2} \right)^2 \mbox{d}s.
\end{equation}
This expression for elastic energy per unit of length is the same than for elastic energy of the numerical rod (Eq.~\ref{eq_enum}), except that $B$ in experiments has different units than $B$ in simulations (energy and energy times length respectively).  We should mention that Eq.~(\ref{en_cone}) is exact only in the regime of pure elastic deformations.  It turns out that for the highest compaction rates studied here, some configurations sustained a few localized plastic deformations.  This happens when the absolute value of the local curvature~$|\kappa|$ goes beyond the plastic threshold.  Through independent experiments, we measured it as~$\kappa_{\mbox{\tiny c}} = 0.54$~mm$^{-1}$ for sets $1$, $2$ and $4$ and $\kappa_{\mbox{\tiny c}} = 0.24$~mm$^{-1}$ for set $3$.  In order to account for the subsequent plastic softening of the sheets in these areas,  we substituted the quadratic dependance of~$E_s$ on the curvature by a more appropriate linear dependance of the form~$\kappa_{\mbox{\tiny c}} ( 2 \kappa - \kappa_{\mbox{\tiny c}})$.  This substitution does not affect the distributions, but only the total energy through extreme events.

In simulations and contrary to the experiments, an important distinction has to be made right away about the expression of the energy of the rod in the numerical simulations.  There are~2 independent terms contributing to the total energy: the bending energy and the confinement energy due to the quadratic potential.  Because the numerical simulations are based on an energy minimizing principle, the total energy of the rod has already been discussed in Section~\ref{sec_num}, and is rewritten here (omitting the hard-core repulsion term which is~$0$ for ``equilibrium'' configurations):
\begin{equation}
E = \frac{B}{2} \int_0^L \left( \frac{\mbox{d}^2 {\bf R}}{\mbox{d}s^2} \right)^2 \mbox{d}s + \frac{\lambda}{2} \int_0^L {\bf R}^2 \mbox{d}s.
\end{equation}

In non-dimensionalized form, the total energy of the rod can thus be rewritten as :
\begin{equation}
\tilde{E} = \underbrace{\int_{\Omega} \left(  \frac{\mbox{d}^2\tilde{R}}{ \mbox{d}\tilde{s}}  \right)^2}_{\mbox{\small experiments/numerics}} + \underbrace{\Lambda \int_{\Omega} {\bf \tilde{R}}^2 \mbox{d} \tilde{s}}_{\mbox{\small numerics only}},
\end{equation}
where the first term (bending) is common to both the experiments and the numerical simulations, while the second one (confinement) is only present in the numerical simulations.  In the following, we will be interested in two different levels of description (branch and whole rod scales) and~$\Omega$ represents the domain of interest: ``microscopic'' ($\Omega =$~branch) or macroscopic ($\Omega =$~whole rod).  Notice that in order to distinguish between the energy of branches and that of the whole rod, the notation~$e$ will refer to branches energy and $E$ to  the whole rod energy.

\begin{figure*} 
\centerline{
\begin{minipage}[b]{1\linewidth}
\includegraphics[width=0.32\linewidth]{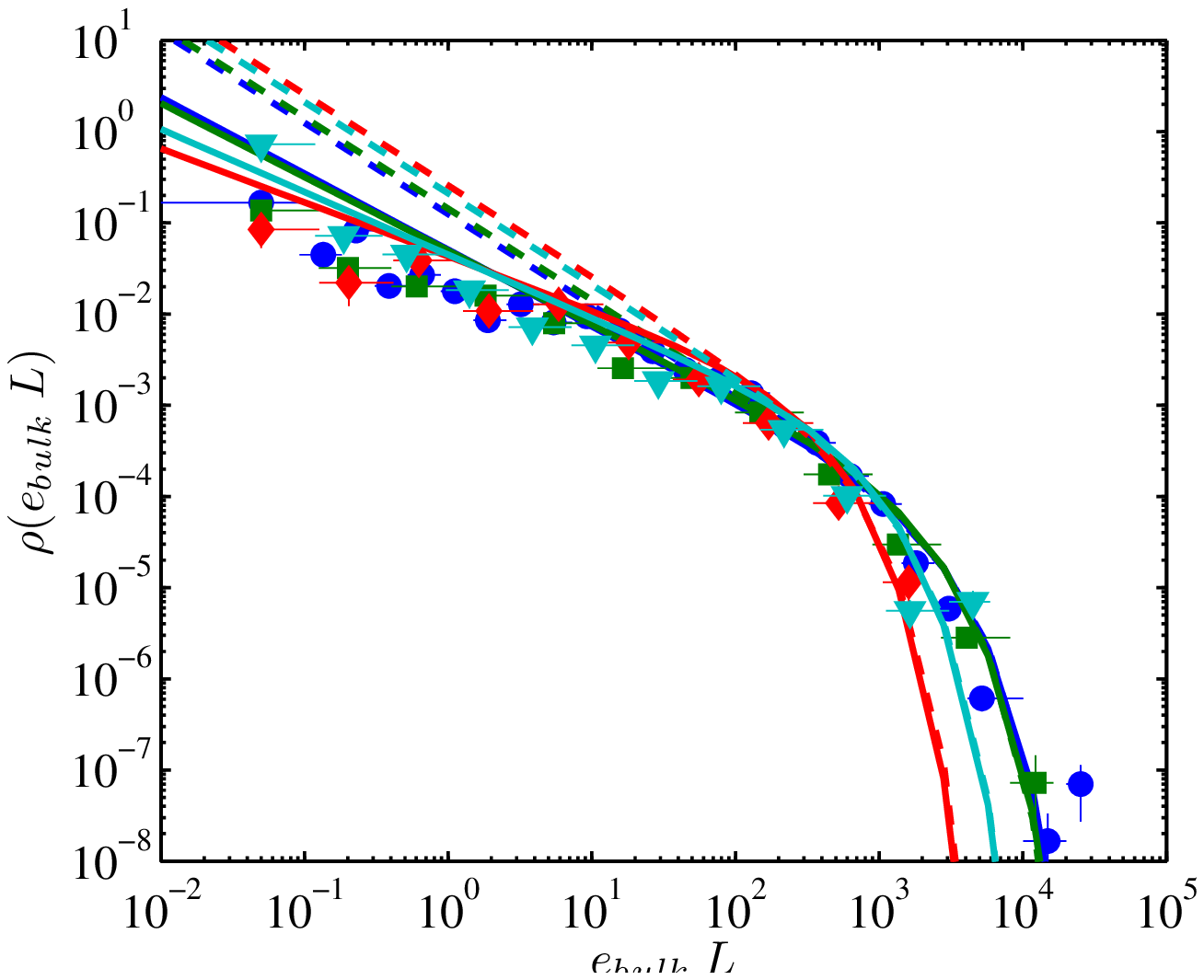} \hfill
\includegraphics[width=0.32\linewidth]{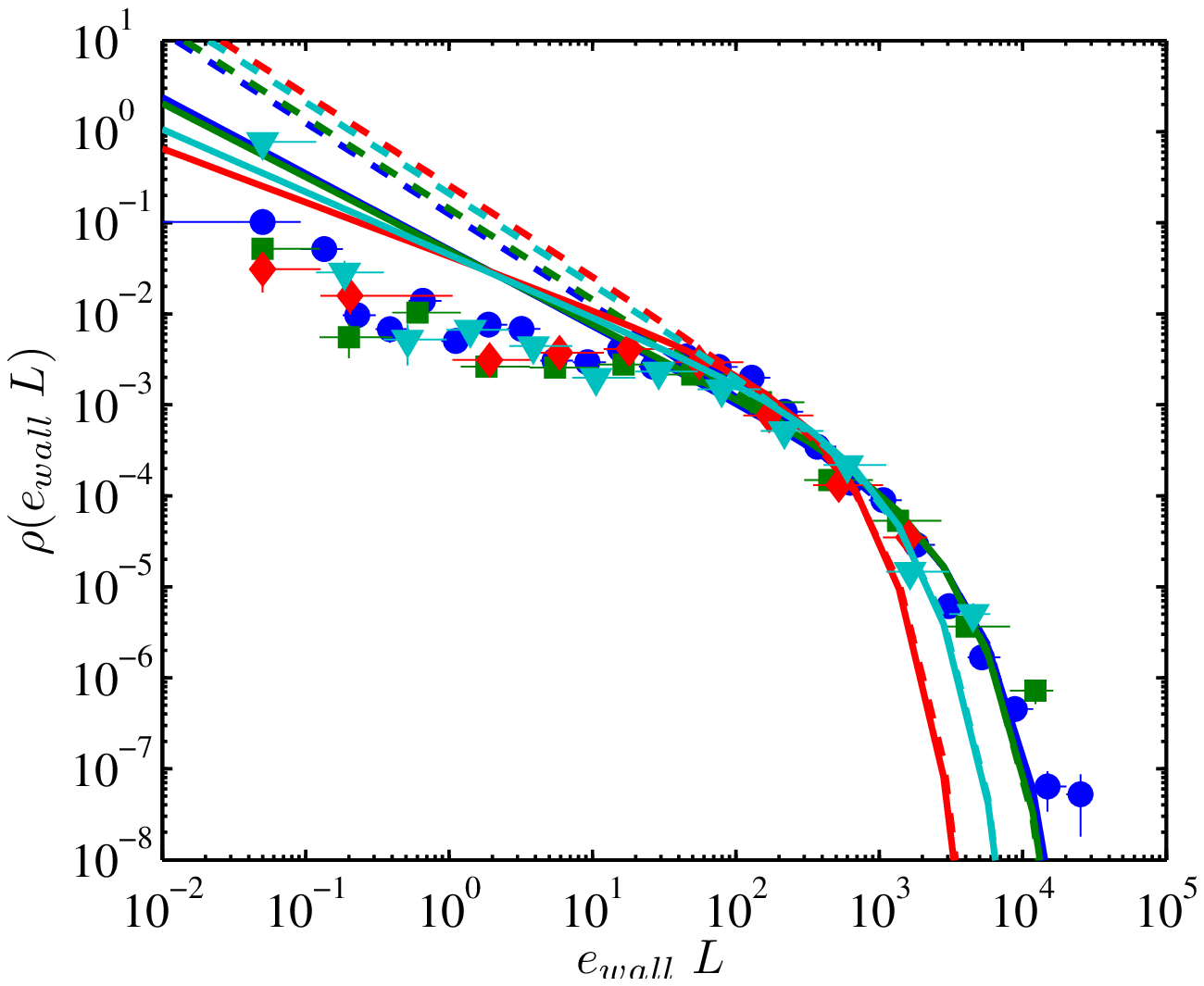} \hfill
\includegraphics[width=0.32\linewidth]{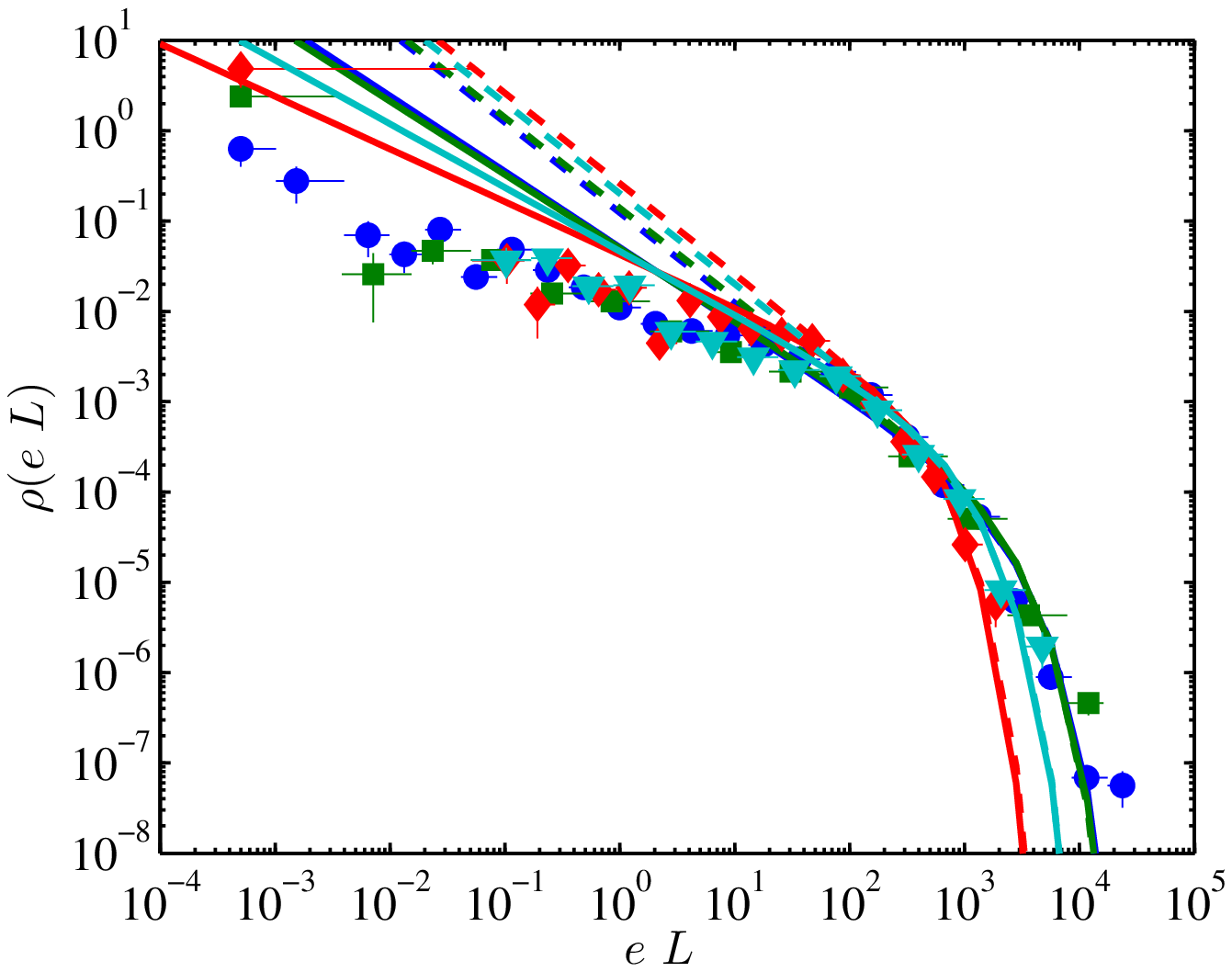} \\
\includegraphics[width=0.32\linewidth]{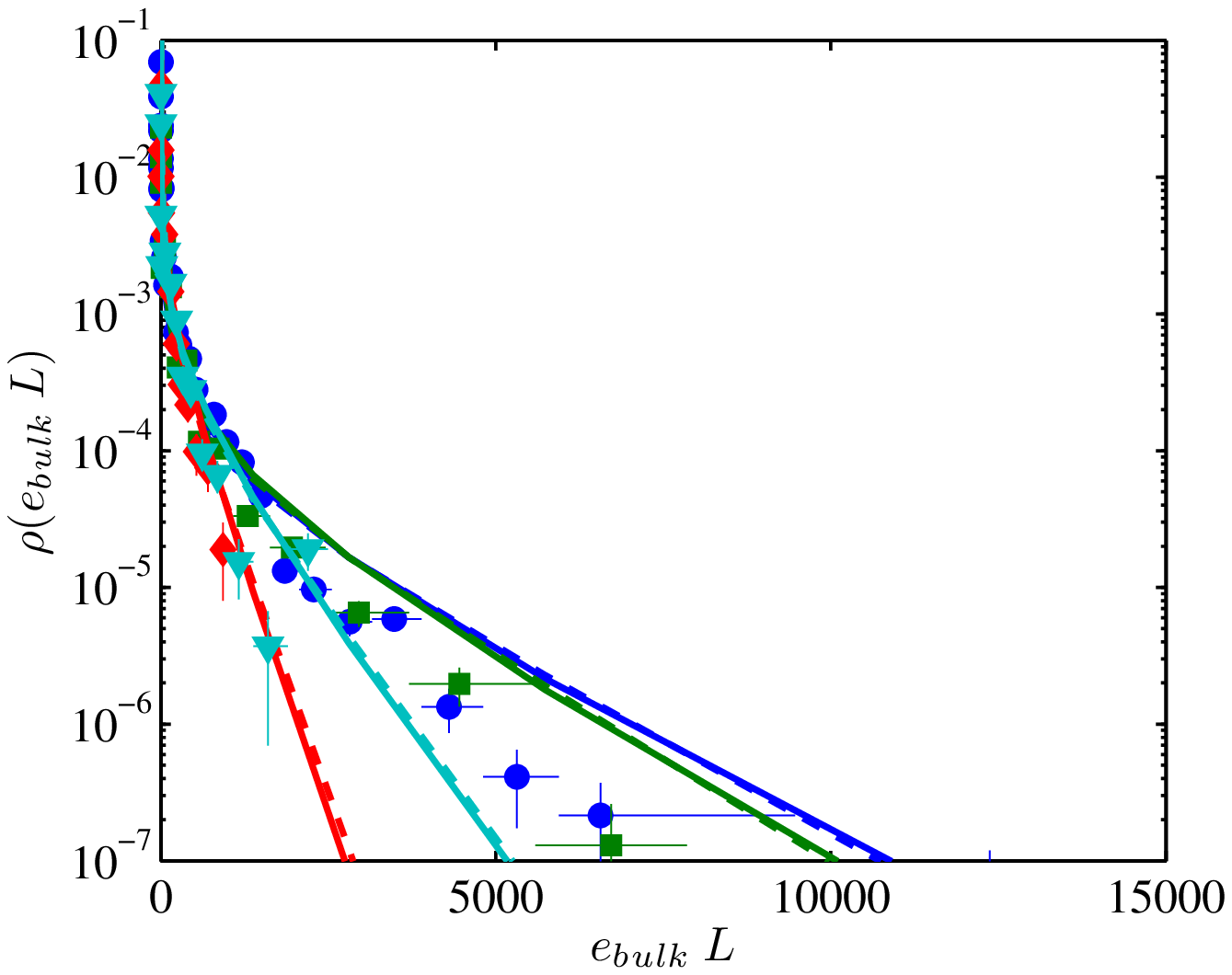} \hfill
\includegraphics[width=0.32\linewidth]{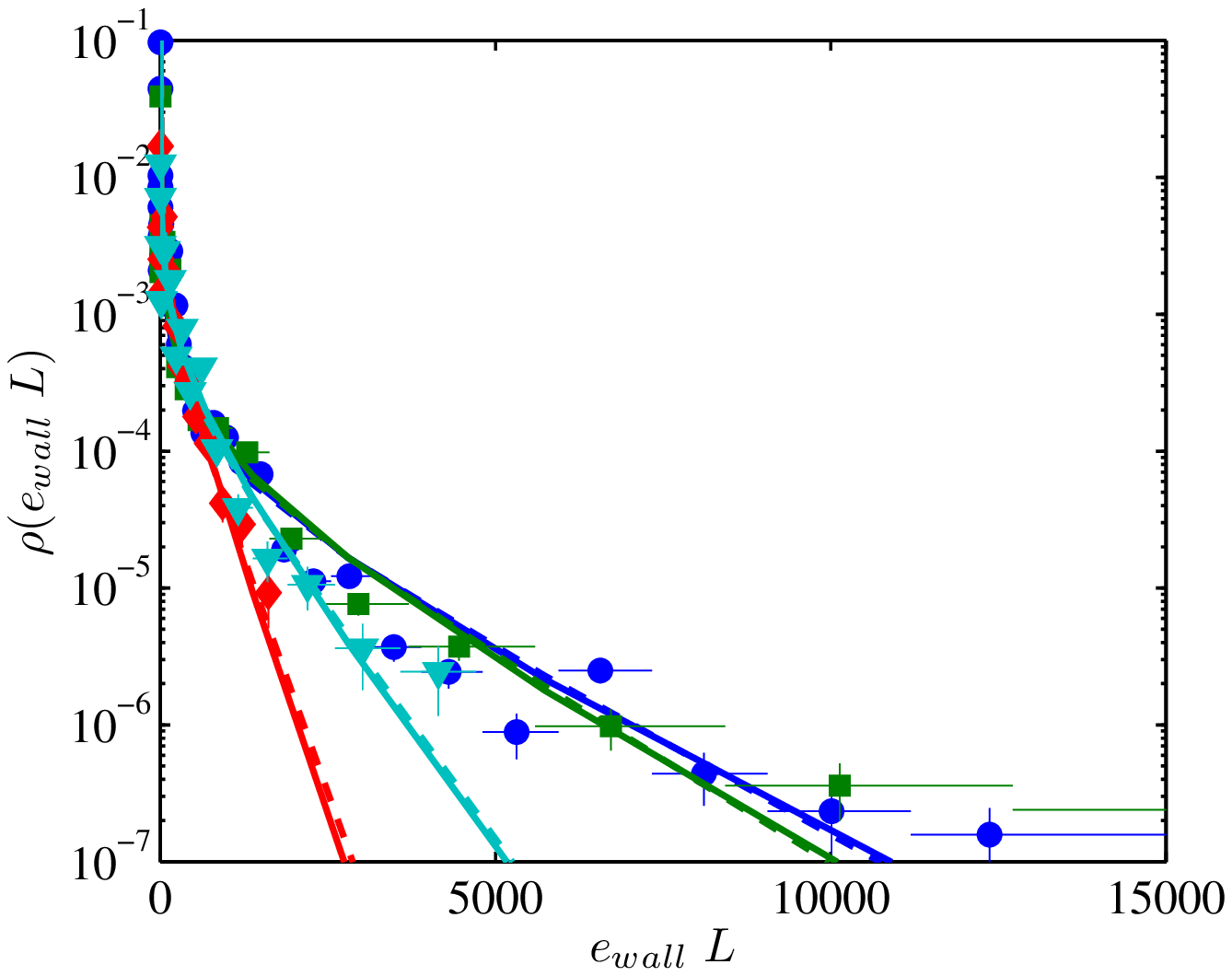} \hfill
\includegraphics[width=0.32\linewidth]{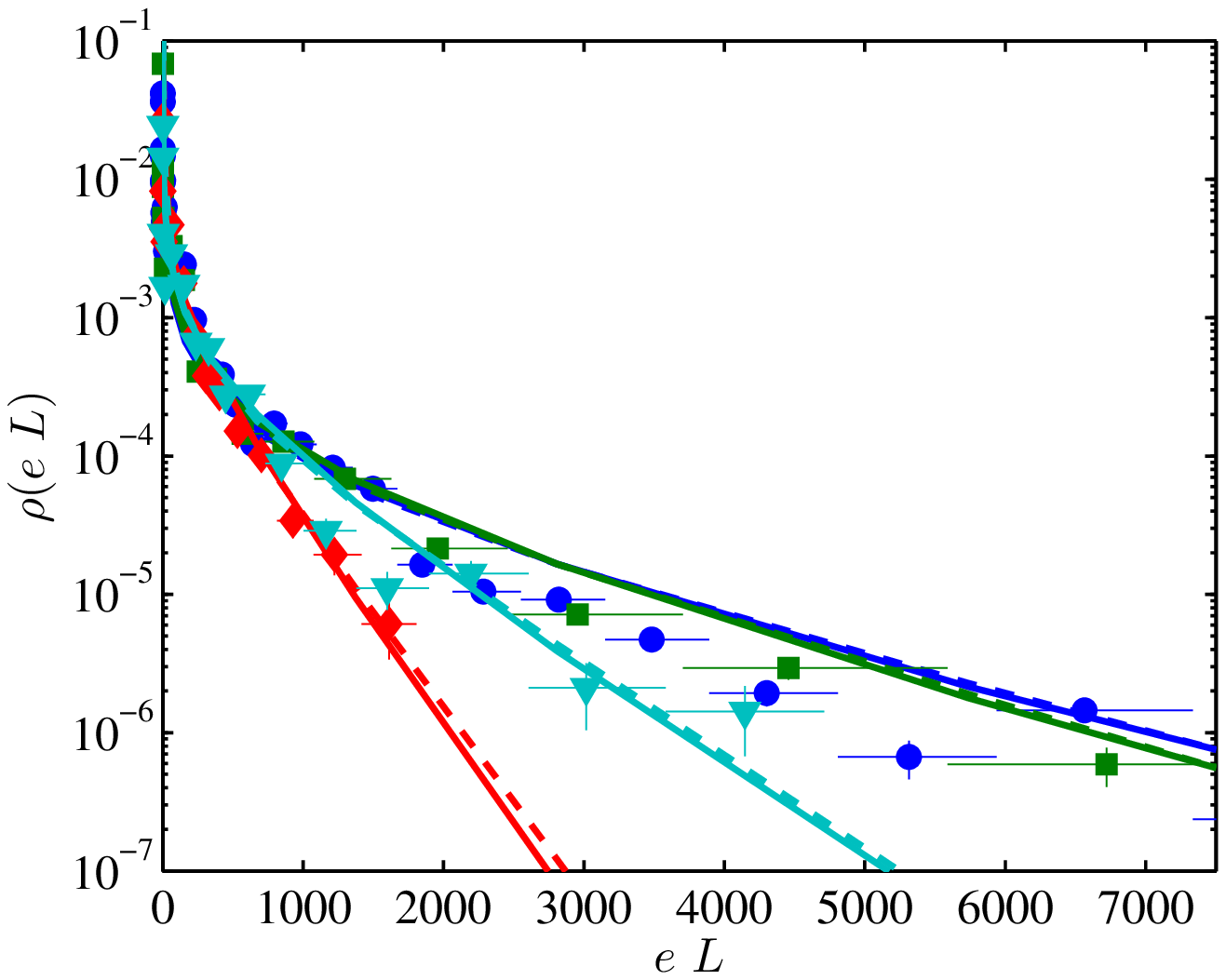} 
\end{minipage}
}
 \caption{{\it {\bf Experiments.} Probability distributions of the energy of the branches, Gamma and Bose-Einstein distributions  for sets 1-4.  {\bf a}: In the sub-system bulk: $\left\langle e\right\rangle=290$, $260$, $95$, $170$, $\alpha_{e}=0.17$, $0.23$, $0.41$, $0.24$ and $\chi_{e}=1700$, $1100$, $230$, $720$. {\bf b}: In the sub-system periphery: $\left\langle e\right\rangle=370$, $407$, $160$, $260$, $\alpha_{e}=0.17$, $0.20$, $0.44$, $0.37$ and $\chi_{e}=2200$, $2100$, $350$, $710$.   {\bf c}: In the whole system: $\left\langle e\right\rangle=330$, $340$, $130$, $220$, $\alpha_e=0.16$, $0.19$, $0.41$, $0.31$ and $\chi_e=2000$, $1800$, $330$, $720$. Lower figures: Same as Fig.~\ref{energo_micro} but represented in log-lin scale to better see the large exponential fall off of the distribution.}}
 \label{energo_micro}
\end{figure*}

\subsection{Microscopic level ($\Omega =$~branch)}

\paragraph*{{\bf Experiments.}}  

For all the four sets of experiments, the distribution of the energy of branches $\rho(e)$ is well described by Gamma laws whose parameters are given in the legend of Fig.~\ref{energo_micro}.  As the local curvature of a branch is smoothly distributed around~$k_m$/$k_+$ (section~\ref{sec_geometro}), we verified that the energy of a branch is roughly equal to the estimation~$\ell k^2_m$ (the linear correlation coefficient is~$\approx 0.6$).  Taking into account the rough linear correlation between $\ell$ and $k_m^{-1}$, it implies a roughly linear correlation between $k_m$ and $e$.

\paragraph*{{\bf Simulations.}}

\begin{figure*} 
\centerline{
\includegraphics[width=0.32\linewidth]{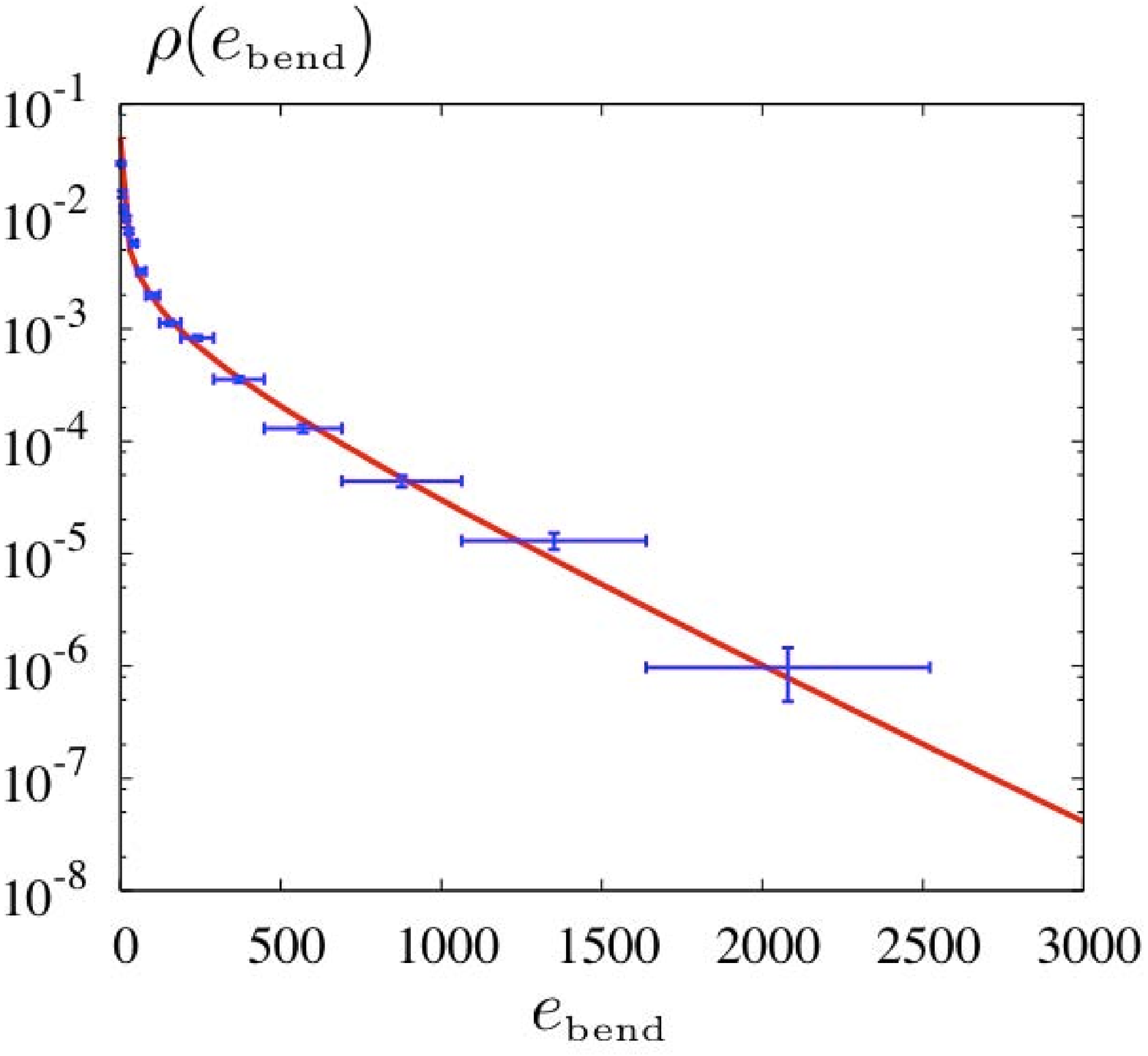}\hfill
\includegraphics[width= 0.32\linewidth]{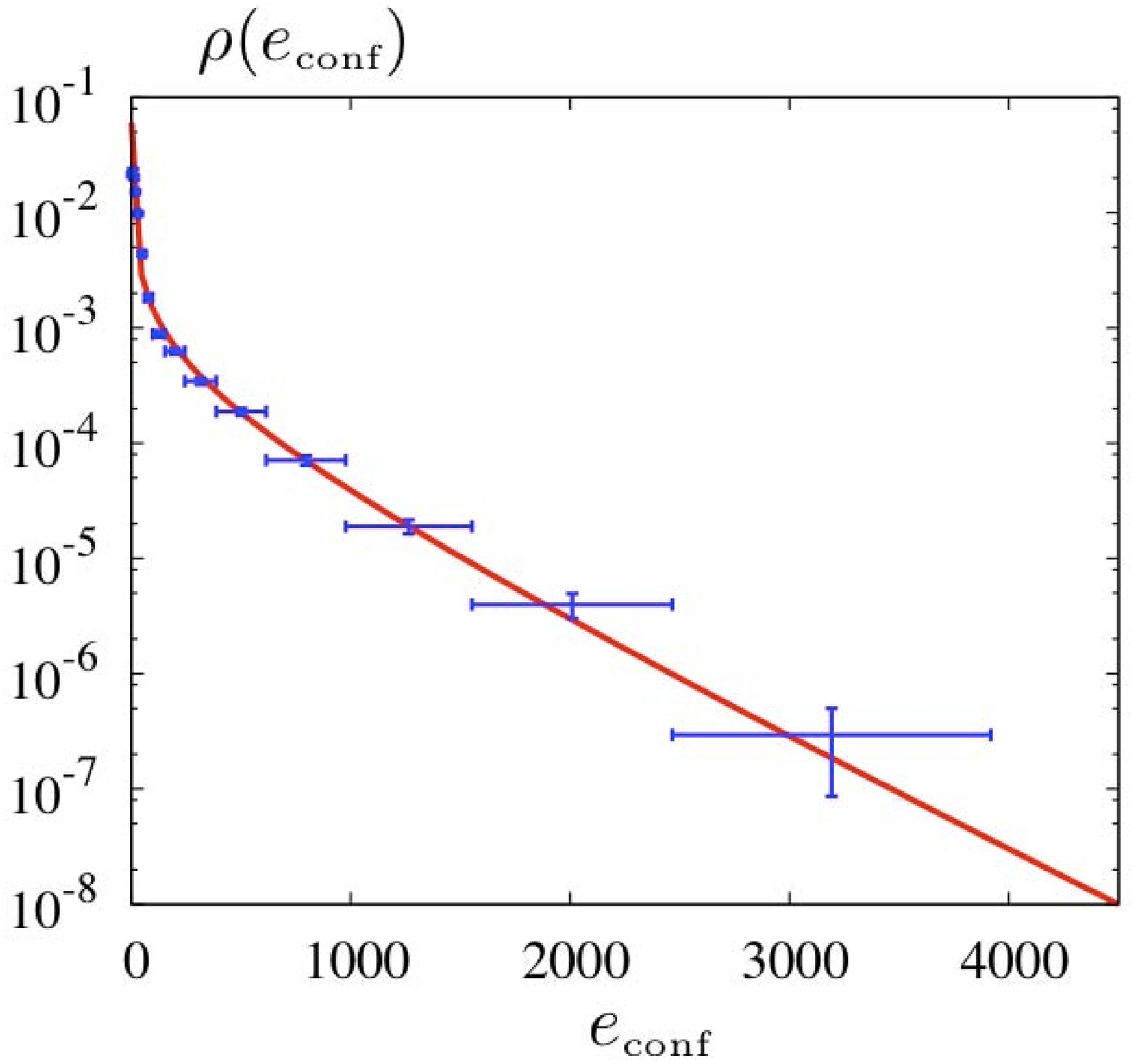} \hfill
\includegraphics[width= 0.32\linewidth]{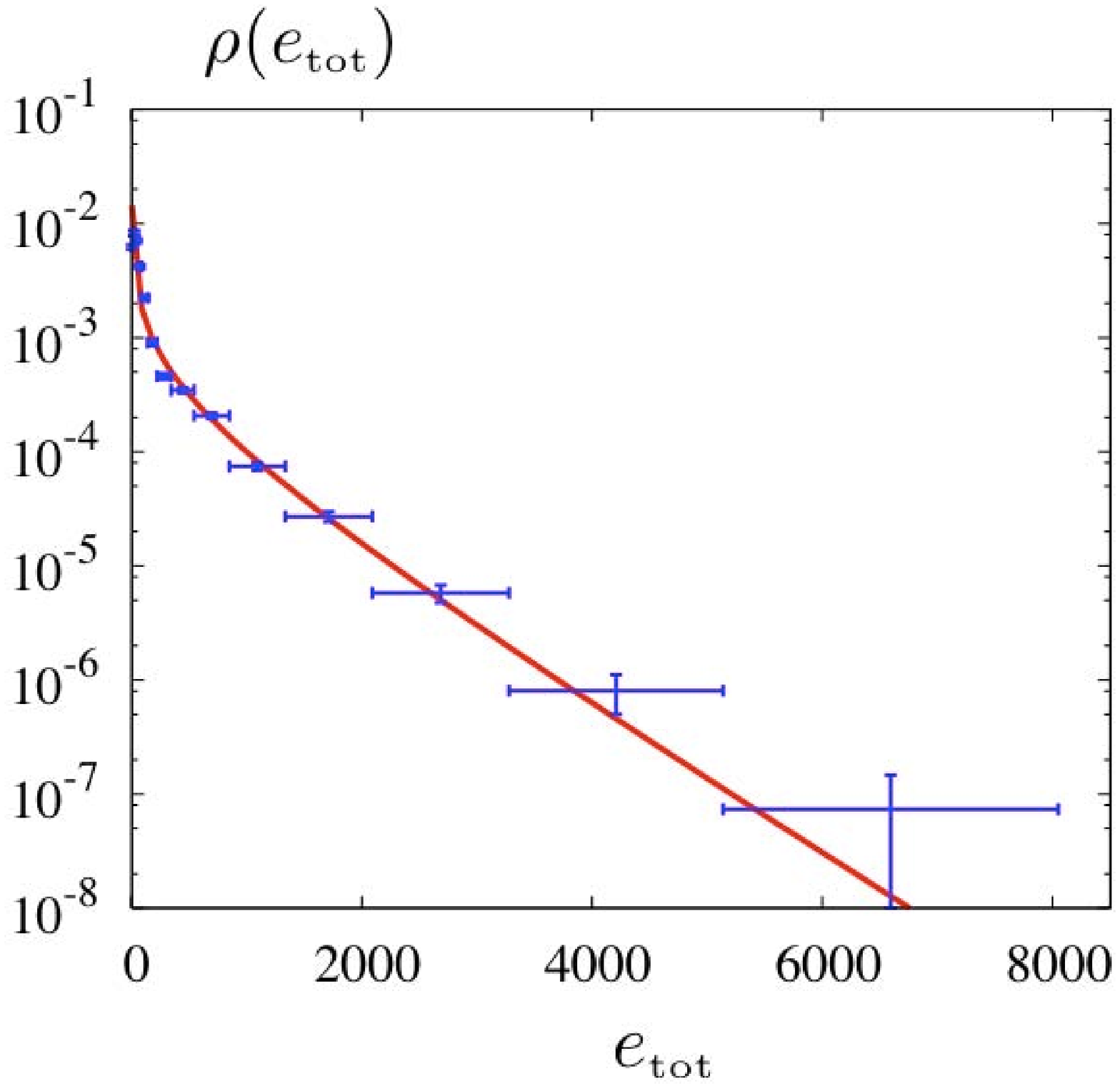}
}
\caption{{\it {\bf Simulations.} {\bf a} Probability distribution of the bending energy of branches.  The continuous line represents a Gamma-law with parameters:~$\alpha=0.35$ and~$\chi=340$. {\bf b} Probability distribution of the confinenement energy of branches.  The continuous line represents a Gamma-law with parameters:~$\alpha=0.23$ and~$\chi=492$.  {\bf c}
Probability distribution of the total energy of branches.  The continuous line represents a Gamma-law with parameters:~$\alpha=0.31$ and~$\chi=728$.
Other parameters are:~$\Lambda = 7\,10^5$, $D=750$ and there is~$4,676$ branches.}}
 \label{pdf_num_ener}
\end{figure*}

Because the total energy of branches is made up of~2 independent contributions, it is interesting to first analyze them separately.  It turns out however that both contributions, bending and confinement energies, are distributed according to a Gamma distribution with similar parameters (Fig.~\ref{pdf_num_ener}{\bf a-b}).  This is a confirmation that the numerical procedure indeed converges to true minima of the energy functional.  Summing up the bending and confinement energy yields the distribution of the energy of branches $e$ (Fig.~\ref{pdf_num_ener}{\bf c}) which is also well described by a Gamma law distribution.

\subsection{Macroscopic description ($\Omega =$~whole rod)}

Instead of looking at the level of branches (microscopic), let us now focus on the total energy distributions of the whole rod (macroscopic).

\paragraph*{{\bf Experiments.}}

\begin{figure*}  
\centerline{
\includegraphics[width=0.32\linewidth]{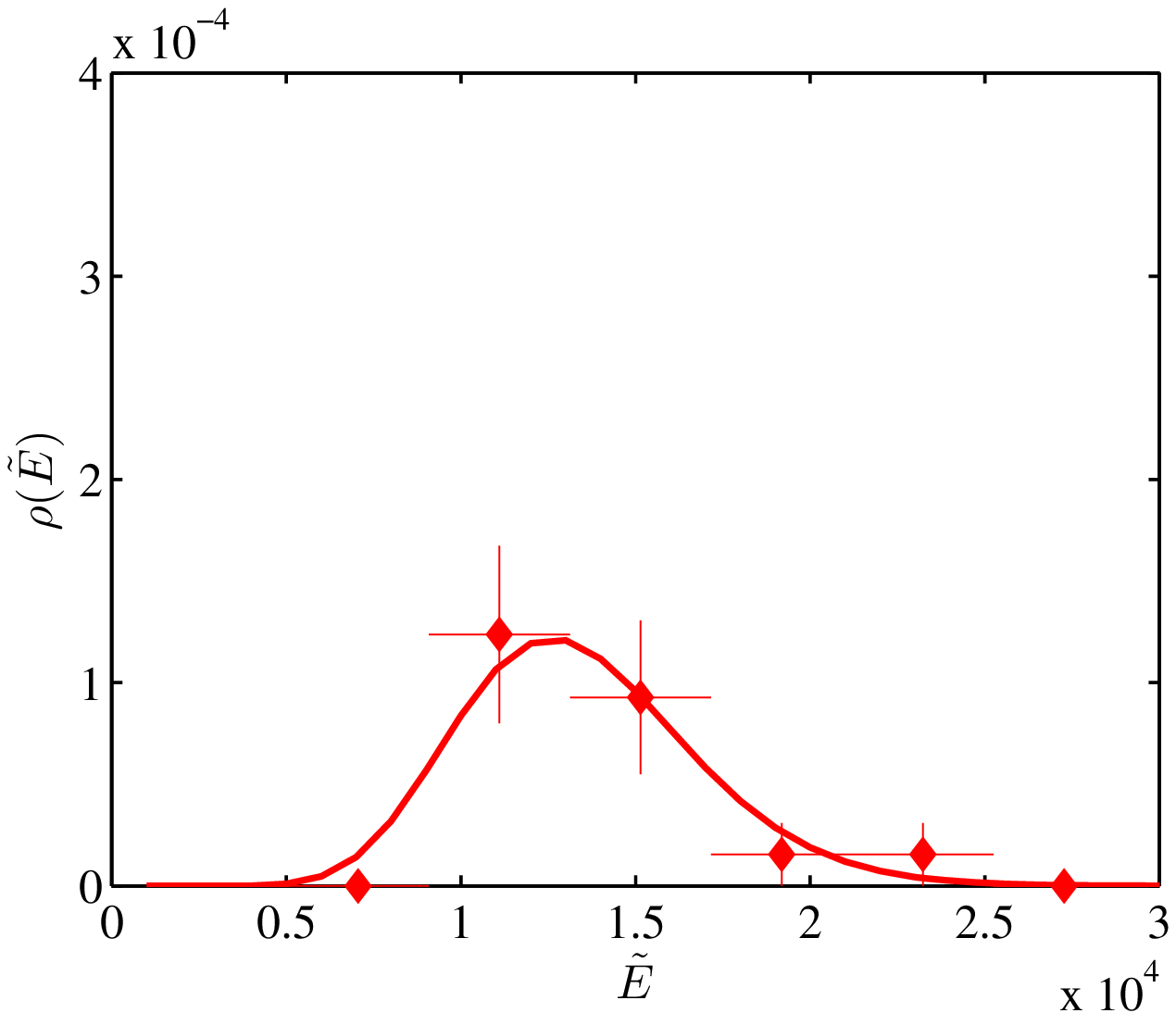} \hfill
\includegraphics[width=0.32\linewidth]{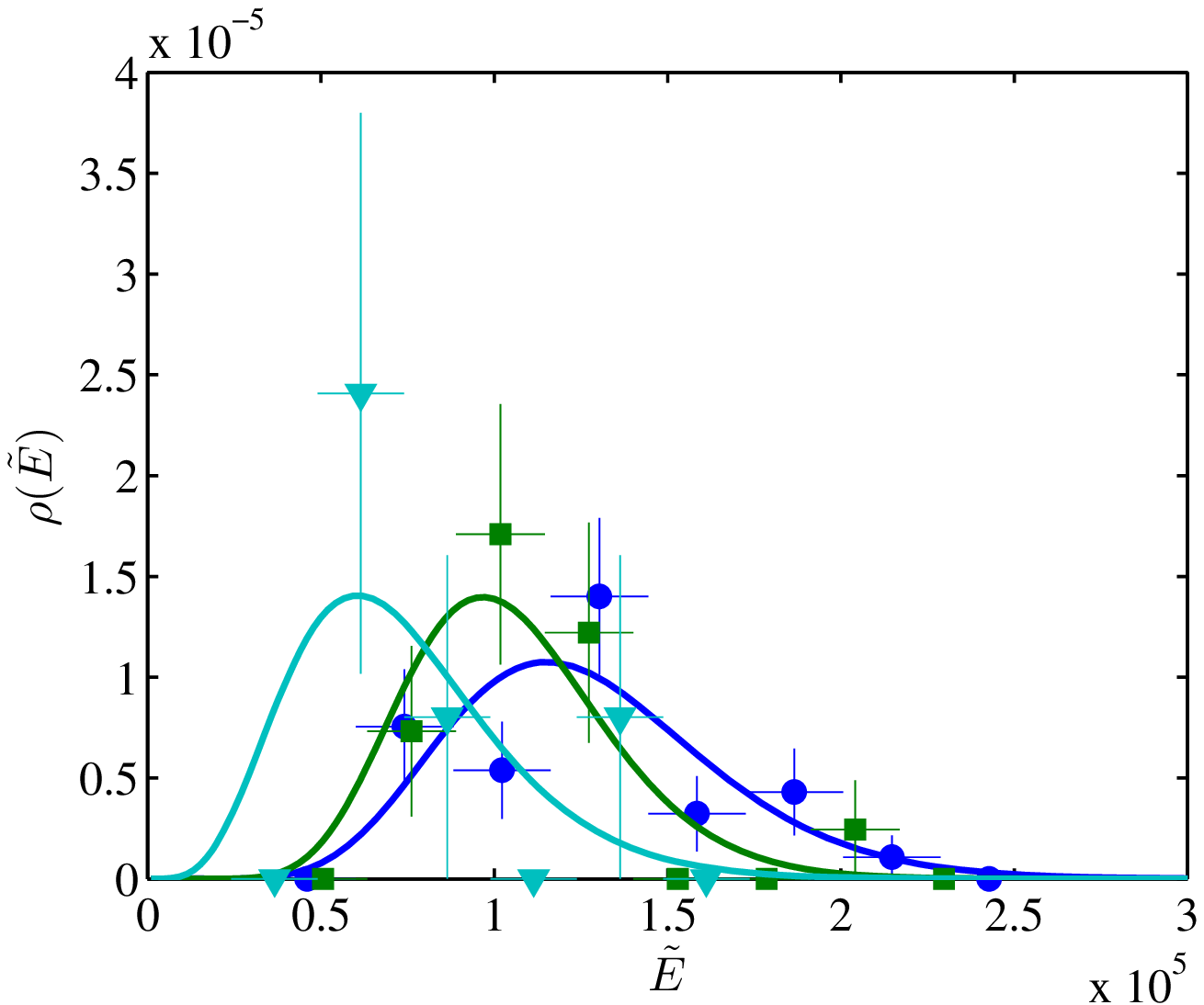} \hfill
\includegraphics[width=0.32\linewidth]{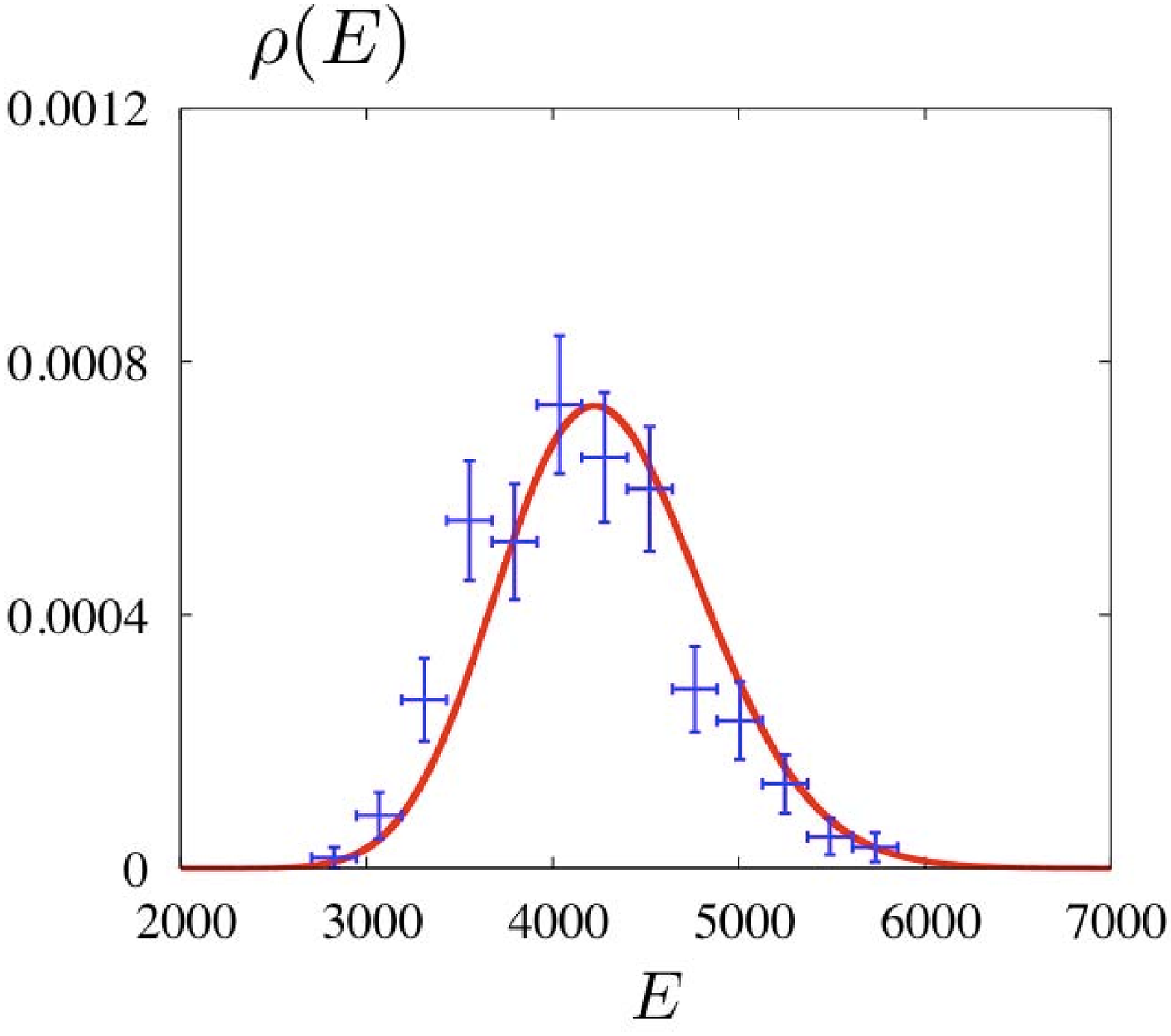}
}  
\caption{{\it  Distributions of total energy of the rod and comparison with Gamma distributions whose parameters $(\alpha,\chi)$ are. {\bf a-b) Experiments.}  $(10.8,1.2\,10^4)$, $(12.7,300)$, $(15.9,40)$, $(5.72,1.3\,10^4)$ and {\bf c) \bf Simulations.} ($\alpha=61,\chi=70$).}}
\label{discu}
\end{figure*}

\begin{figure*}  
\centerline{
\includegraphics[width=0.32\linewidth]{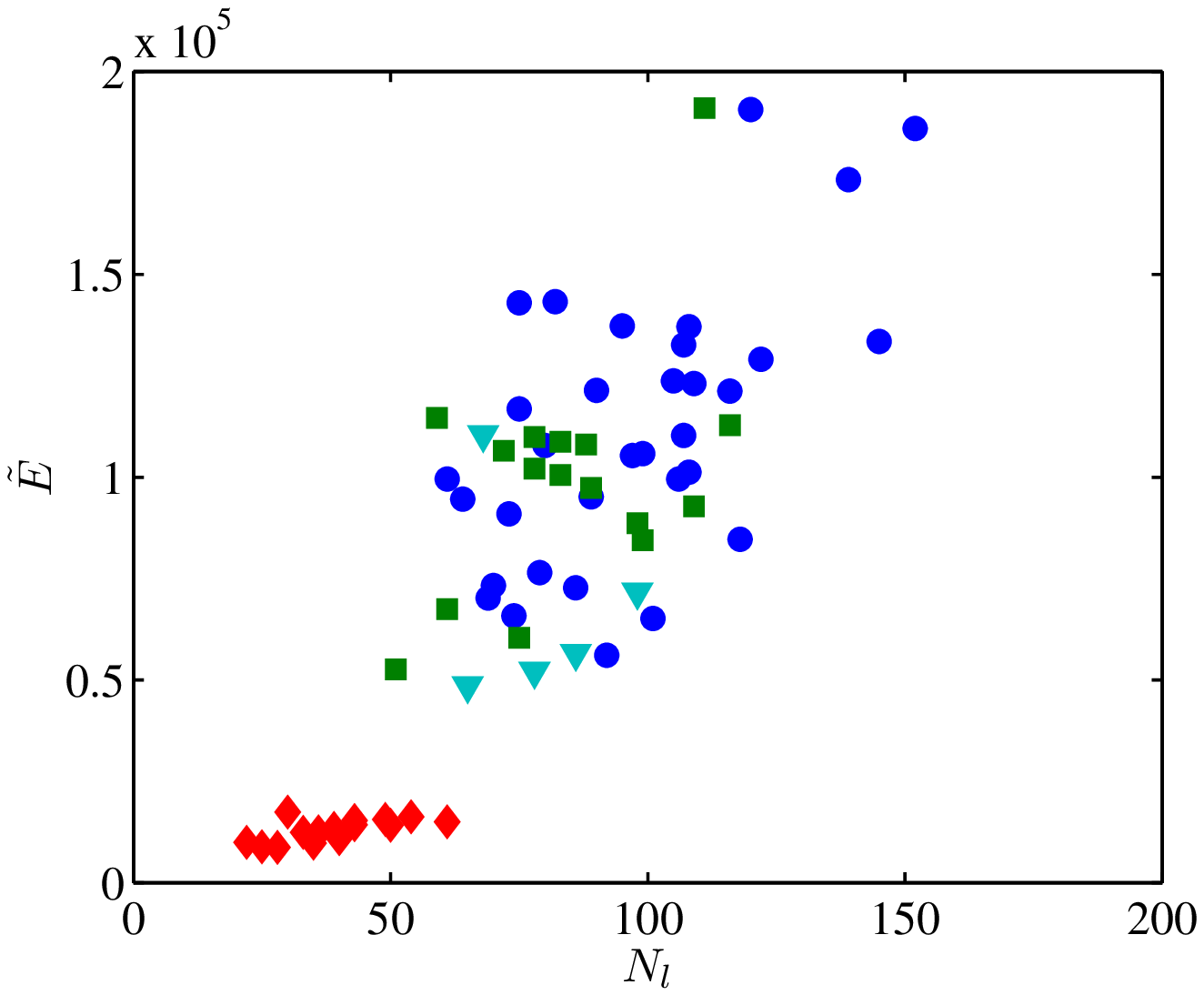} 
\includegraphics[width=0.32\linewidth]{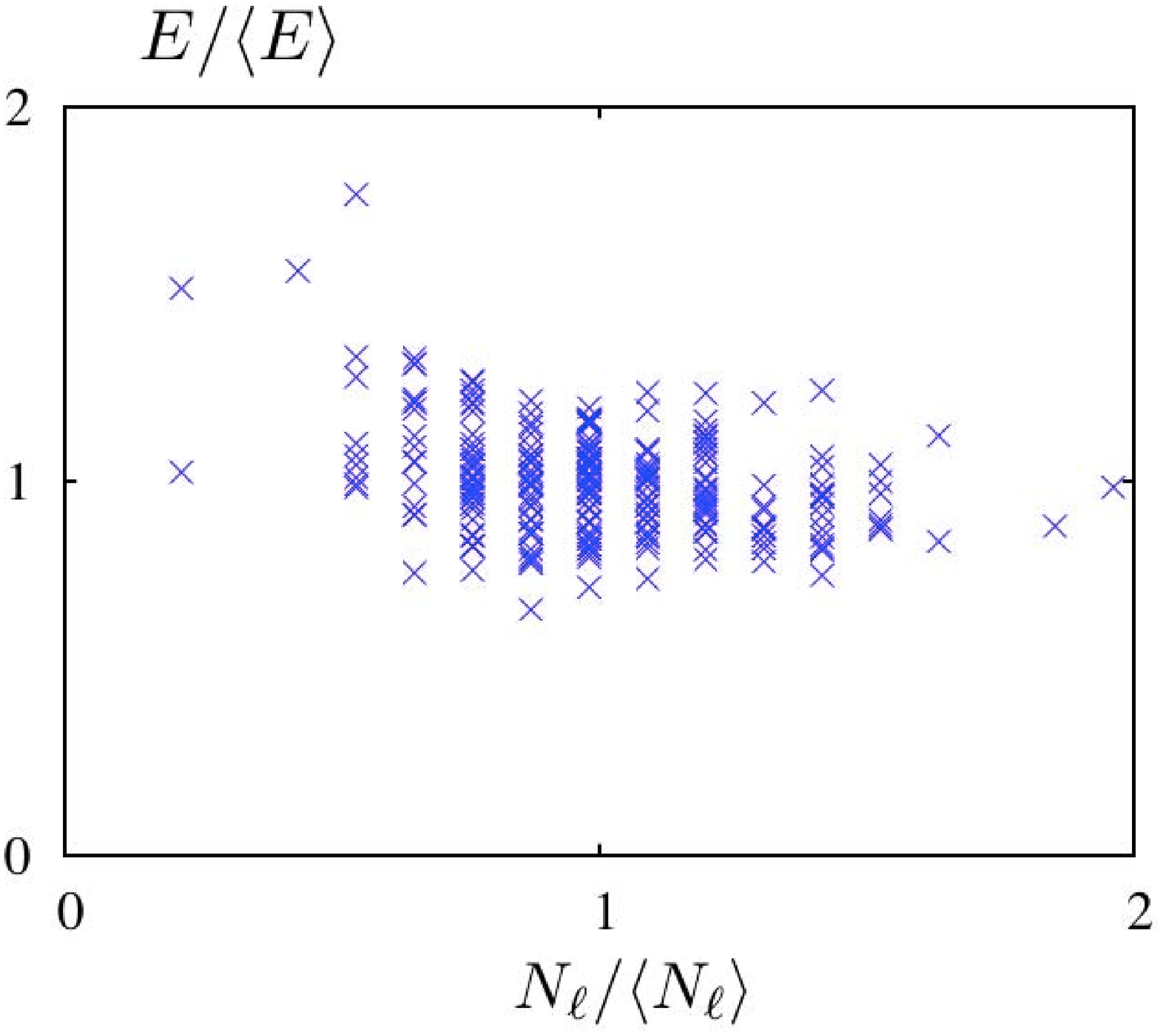}
}  
\caption{{\it Total energy versus topological quantities.  {\bf a)} {\bf Experiments.} Total energy $E$ of the rod as a function of the total number of branches $N_{br}$ and the total number of stacks $N_l$ for all experiments.  {\bf b)} {\bf Simulations.} Total energy of the rod (normalized by~$\left\langle E\right\rangle$) vs. the number of links (normalized by~$\left\langle N_{\mbox{\tiny links}}\right\rangle$).}}
\label{depend_nbr}
\end{figure*}

The pdf of the non-dimensionalized total energy $\tilde{E}$ of the rod seems asymmetric. In view of the comparison with the distribution of the branch energy, we also compare it to Gamma distributions as shown on Fig.~\ref{discu}.  Besides, Fig.~\ref{depend_nbr}{\bf a} shows that, individually, there is a significant correlation between the total energy of a folded configuration and its number of links/branches (linear correlation coefficient~$\approx 0.75$),  which confirms the central role of branches/links. 

\paragraph*{{\bf Simulations.}}

Just like we discussed for the microscopic description, the total energy is the sum of the bending energy and the confinement energy.  Again, the distributions are slightly asymmetric. They can be described by Gamma distributions with similar parameters: bending~$ \{ \alpha = 27 , \chi = 81 \}$ and confinement~$ \{ \alpha = 21 , \chi = 100 \}$.  The sum of these two distributions yields again a Gamma distribution~$ \{ \alpha=61, \chi=70 \}$ for the total energy of the rod as shown on~Fig.~\ref{discu}{\bf c}.  On the other hand, Fig.~\ref{depend_nbr}{\bf b} shows that for the small compaction rates obtained numerically there is no clear correlation between the energy of the rod and the number of branches that it contains.  We do however verify that the mean of the total energy is given on average by the average number of branches times their average energy:~$\left\langle E\right\rangle = \left\langle N_{\mbox{\tiny br}}\right\rangle\left\langle e\right\rangle$.

\section{Conclusion}
\label{sec_discu}

Despite some important differences beween the two systems (nature of confinement, compaction rates achieved, geometrical properties...), the major outcome of this comparison is the rather universal shape of the probability distribution of the energy of individual branches.  In both cases, this probability density is well described by a divergent Gamma distribution: it has an exponential decay at large energies, and turns into a power law divergence for small energies ($\alpha<1$).  This exponential fall off of the distribution is reminiscent of Boltzmann's law making it possible to extract a characteristic energy scale.  The parameter~$\chi$ in the Gamma distributions can then be considered as  ``effective'' temperature characterizing the system.  While we will come back below to the implications of this result, we can already say that it supports the notion of branches as the basic components of the folded rods (just like particles are in a real gas), interacting together through contacts forces at boundaries.  At the other end of the spectrum, we do notice an accumulation of branches with very small energy (soft power law divergence) hinting at some underlying energy condensation process.

\begin{table*}  
\begin{center}  
\begin{tabular}{l||l|l} 
& $\left\langle N_l\right\rangle$&  $\alpha_E/\alpha_e$  \\ \hline
$1$& 90&  68 \\  
$2$& 80&  55 \\  
$3$& 40& 56 \\    
$4$& 80&  23 \\ \hline
$5$ & $\sim10$ &  200   
\end{tabular}
\end{center}  
\caption{{\it Test of the branches as the independent variables.  While the experimental results suggest a convergence to a possible thermodynamic limit, the paucity of the numerical data cannot be interpreted in a simple way, see discussion in the text.}}
\label{tab_res} 
\end{table*}  

The sum of~$N$ {\it ``uncorrelated''} random variables distributed with a Gamma distribution of parameters~$\{\alpha_{\mu},\chi \}$ is another Gamma distribution with parameters~$\{ \alpha= N \alpha_{\mu}, \chi \}$.  Therefore the ratio~$\alpha/\alpha_{\mu}$ is a measure of the number of degrees of freedom present in the system.  Assuming branches are the basic elements composing the whole rod, we would expect this ratio to be related to the number of multi-branched stacks.  Table~\ref{tab_res} shows that the expected relation beween~$\alpha_E/\alpha_e$ and the number of links is only crudely observed for the compaction rates achieved here.  The experimental results show however that the higher the confinement and the better the agreement.  This trend indicates that due to the small number of branches (particles) present in the configurations there are still non-negligible finite size effects introducing non-trivial correlations.  As the number of branches increase, the system gets closer to the thermodynamic limit and the discrepancy indeed shrinks down a little.  As the shape parameter~$\alpha$ of the Gamma distributions, the probability density becomes more and more gaussian as a consequence of the central limit theorem.  On the other hand, the numerical results completely fail this analysis.  One possible interpretation is that, contrary to the experiments, one would expect branches rather than links to be the individual degrees of freedom in the numerical simulations.  This is because branches are determined to be in self-contact whenever their distance is smaller than some small cut-off distance.  This gives thick stack of branches more freedom to wiggle around and therefore behave somewhat more independently than in the experiments where there are true self-contact areas.  In that case, one should perhaps think in terms of a sum of~$N_{br}$ variables with~$N_l$ that are independent and~$N_{br}-N_l$ that are exactly correlated.  Our data can only hint in that direction but does not allow us any definite conclusion regarding this issue.  In any case, the simple minded analysis we presented here is still quite interesting and would deserve more work to understand if one really finally converges to the thermodynamic limit as the number of branches tend to infinity.

\begin{figure*} 
\centerline{
\includegraphics[width=0.32\linewidth]{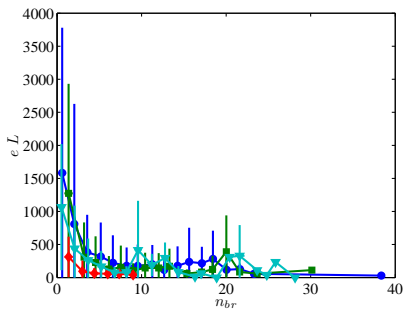} 
}
\caption{{\it {\bf Experiments.} Averaged values of the energy of one branch $e$ as a function of the number of branches in the multi-branched stack $n_{br}$.}}
 \label{energo_micro3}
\end{figure*}

\begin{figure*}  
\centerline{
 \includegraphics[width=0.4\linewidth]{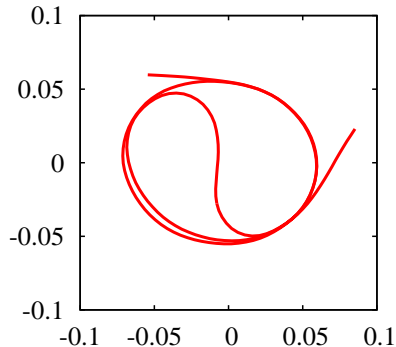} 
}  
 \caption{{\it {\bf Simulations.} Spiral configurations as the condensed phase of the system where all the branches are on top of each other.}}
\label{bose_condensate}
\end{figure*}

\paragraph*{{\bf Bose-Einstein condensation and annealing.}}

Another well-known statistical distribution that has the same asymptotic behaviours as Gamma laws is the celebrated Bose-Einstein distribution.  Using the analogy between branches and gas-particles, we compared our data with this distibution at null chemical potential because the number of branches (particles) is not pre-determined in either numerical simulations nor experiments:
\begin{equation}
\rho(e) = \frac{\alpha}{\exp \left( \beta e \right) - 1}.
\end{equation}
The agreement is qualitatively as good as the one obtained by using Gamma distributions.  For high confinements (in experiments), it is easy to see from the experimental pictures that many branches tend to accumulate close to each other, giving rise to thick stacks of heavily populated branches.  This phenomenon can be attributed to a transition from disordered configurations (isotropic phase) to ordered configurations (nematic phase) which was predicted theoretically.  In addition, Fig.~\ref{energo_micro3} reveals that these highly degenerate stacks (multiple branches close to of each other) do indeed carry very little elastic energy.  It is interesting to note that this behaviour holds for low confinements as well.  Because of their lower compaction rates, the configurations obtained numerically only rarely present stacks made up of more than~2 branches.  However the Bose-Einstein distribution describes equally well the data.  Since branches are not subjected to any exclusion principle (except for their vanishingly small thickness that prevents self-intersection), it is possible for them to be in the exact same state and effectively behave as integer spin particles explaining the observed good agreement of our data with the Bose-Einstein distribution.  While this comparison may seem surprising, we will give below a few more arguments justifying our position.

We saw that in addition to their exponential tail, the probability distributions of the branches display a power law divergence for small energies.  By closely looking at the experimental images one can see that most of the branches tend to accumulate close to each other creating thick stacks made up of several branches.  While this may not seem so obvious in the numerical simulations, one has to remember that they were obtained by quenching the rod to relatively much smaller values of compaction.  Indeed if instead of the quenching mechanism proposed here, we increase very slowly the control parameter~$\Lambda$ (annealing), a different behavior is observed.  We are no longer able to explore a wide phase space but instead converge to the true ground state.  A typical shape that is obtained by this process is the spiral pattern presented in~Fig.~\ref{bose_condensate}.  We expect this phenomenon of branch condensation to the exact same state to hold and even increase for higher confinements.  This is quite reminiscent of what happens during the condensation of integer spin particles (Bose-Einstein condensation) where particles are free to accumulate on the same energy levels.  The analog of the chemical potential in our situation would be the number of branches present in the rod.

In summary, we have described two different systems (experiment and numerical simulation) to study the statistical properties of confined rods.  The geometrical properties of the two systems are different due to the difference in the way confinement was applied.  We do find however that the statistical properties of the bending energy of the branches follow the same distribution.  Furthermore the parameters of this Gamma distribution share the same properties indicating a common behavior. 

Coming back to the question posed in the introduction, we can now answer to the existence of a statistical measure in this system.   The energy of the branches is in fact the relevant internal variable.  Moreover its distribution can be approximated over a wide range of energies by a Boltzmann distribution weighted by a characteristic energy reminiscent of the concept of ``effective'' temperature in the context of granular rheology.  This system could be used to test some recent theorems related to non-equilibrium statistical physics in a context different from granular (and colloid) matter.  An advantage of this system is that it is possible to measure the bending energy of branches which could be analogous to particles in a real gas.

\begin{acknowledgments}
We are grateful to \'Eric Sultan for his help and guidance during the initial stages of the numerical work.
\end{acknowledgments}

\bibliographystyle{abbrv}
\bibliography{PackingAugust2009}

\end{document}